\documentclass[11pt]{article}
\pdfoutput=1

\usepackage{amsmath,amssymb}
\usepackage{epsfig}
\usepackage{cite}

\def\Dslash{\hspace{3pt}\raisebox{1pt}{$\slash$} \hspace{-9pt} D}

\def\partialslash{\hspace{3pt}\raisebox{1pt}{$\slash$} \hspace{-6pt} \partial}
\def\SO{\textrm{SO}}

\numberwithin{equation}{section}

\title{
\vspace{-2cm}
\begin{flushright}
\small{CERN-PH-TH/2013-144}
\end{flushright}
\vspace{3cm}
\bf \LARGE
Light top partners and precision physics
\vspace{.2cm}}
\date{}
\author{
{\large Christophe Grojean$^{a,c}$, Oleksii Matsedonskyi$^b$, Giuliano Panico$^{c}$}\\
[10mm]
\normalsize\itshape $^a$ ICREA at IFAE, Universitat Aut\`onoma de Barcelona, Bellaterra, Spain\\
\normalsize\itshape $^b$ Dipartimento di Fisica e Astronomia and INFN, Sezione di Padova,\\
\normalsize\itshape Via Marzolo 8, I-35131 Padova, Italy\\
\normalsize\itshape $^c$ Theory Division, Physics Department, CERN, CH-1211, Geneva 23, Switzerland\\
}

\begin{document}
\maketitle
\begin{abstract}
\medskip
\noindent
We analyze the corrections to the precision EW observables in minimal composite Higgs
models by using a general effective parametrization that also includes the lightest
fermionic resonances.
A new, possibly large, logarithmically divergent contribution to
$\widehat S$ is identified, which comes purely from the strong dynamics. It can be
interpreted as a running of $\widehat S$ induced by the non-renormalizable Higgs
interactions due to the non-linear $\sigma$-model structure.
As expected, the corrections to the $\widehat T$ parameter coming from fermion loops
are finite and dominated by the contributions of the lightest composite states.
The fit of the oblique parameters suggests a rather stringent lower bound on the
$\sigma$-model scale $f \gtrsim 750$ GeV.
The corrections to the $Z \overline b_L b_L$ vertex coming from the lowest-order
operators in the effective Lagrangian are finite and somewhat correlated to the corrections
to $\widehat T$. Large additional contributions are generated by contact interactions with
$4$ composite fermions. In this case a logarithmic divergence can be generated and the
correlation with $\widehat T$ is removed.
We also analyze the tree-level corrections to the top couplings, which are expected to be
large due to the sizable degree of compositeness of the third generation quarks.
We find that for a moderate amount of tuning the deviation in
$V_{tb}$ can be of order $5\%$ while the distortion of the $Z \overline t_L t_L$
vertex can be $10\%$.
\end{abstract}

\newpage


\section{Introduction}

The discovery~\cite{:2012gk} by the LHC experiments of a scalar resonance with a mass around
$125\ {\rm GeV}$ and with production and decay properties compatible with the ones of the
Standard Model (SM) Higgs boson sets a landmark in the exploration of the sector responsible
for the breaking of the Electroweak (EW) symmetry. The value of the resonance mass together
with the absence of observation of any additonal new particles bring stringent constraints
on various models that were designed to address the naturalness problem.
For instance, the current results and in particular the lack of any signal
in the jets plus missing energy searches,
in addition to the indirect constraints from flavor physics, indicate that supersymmetric models
are in need for a non-minimal incarnation (see for instance Ref.~\cite{Hall:2011aa} and
references therein).
Important lessons can also be drawn for models
of strong electroweak symmetry breaking in which the Higgs boson emerges as a composite
particle~\cite{gk,Agashe:2004rs} (see also Ref.~\cite{DP}).
An interesting result is the fact that light fermionic top partners below $1\ {\rm TeV}$
are necessary to generate the correct
Higgs mass without too much fine-tuning~\cite{Matsedonskyi:2012ym,Redi:2012ha, Marzocca:2012zn, Pomarol:2012qf,Panico:2012uw, Pappadopulo:2013vca}.
This forces also the composite Higgs models into non-minimal territory with
some fermionic resonances below the expected typical mass scale of the resonances of the
strong sector. Moreover it motivates an extension of the effective description proposed
in Ref.~\cite{Giudice:2007fh} in order to include the appropriate dynamics and couplings of the
light top partners. 

It has been realized that light fermionic top partners offer nice distinctive collider signatures of composite Higgs models and the best search strategies at the LHC have been identified~\cite{Dennis:2007tv, Atre:2008iu, Anastasiou:2009rv, AguilarSaavedra:2009es,  Brooijmans:2010tn, Dissertori:2010ug, Han:2010rf, Bini:2011zb, Harigaya:2012ir, Berger:2012ec, Dawson:2012di, Okada:2012gy, DeSimone:2012fs, Aguilar-Saavedra:2013qpa, Gopalakrishna:2013hua} and are being applied by the experimental collaborations~\cite{Bounds53_CMS, Bounds53_ATLAS}.
Including the light fermionic resonances
in a general effective Lagrangian, as we will do in this work, can also
provide a model-independent tool to study these collider signatures. 

A third and essential motivation to consider an effective description of top partners is to reassess the status of the composite Higgs models regarding the EW constraints. The composite nature of the Higgs is indeed the source of an infrared-saturated contribution to the EW oblique parameters~\cite{Barbieri:2007bh} that, taken on its own, sets a stringent bound on the compositeness scale of the Higgs boson and inevitably raises the amount of fine-tuning~\cite{Contino:2010rs, Espinosa:2012im, Falkowski:2013dza, Ciuchini:2013pca}.
It is thus clear that a scenario with an acceptable amount of tuning can only be obtained
if further corrections to the EW parameters are present.

One possible source of additional contributions are the composite resonances and in
particular the fermionic ones. Even if they do not give tree-level corrections to the
EW oblique parameters, the top partners do contribute to them at one loop and these
contributions can be sizable if the partners are light. In this paper we extend previous analyses~\cite{Agashe:2003zs, Agashe:2005dk, Barbieri:2007bh, Gillioz:2008hs, Lodone:2008yy,  Pomarol:2008bh, Gillioz:2012se, Barbieri:2012tu} and we provide the first computation of the
fermion one-loop contribution to the $\widehat S$ parameter taking into account the Higgs non-linearities associated to its composite nature. The result of this computation is the
identification of a new logarithmically enhanced contribution that can be interpreted
as a running effect from the mass of the top partners to the scale of the EW vector resonances.
We also study the contributions of the top partners to the $\widehat T$ parameter which, though finite, can be large and positive, in particular in the presence of a light SU(2) singlet partner, and can compensate the Higgs contribution.
We also clarify the structure of the deviations
of the $Z\overline b_L b_L$ coupling which can become logarithmically divergent when $4$-fermion
interactions with a chirality structure LLRR are introduced in the composite sector. 

This paper is organized as follows. In section~\ref{sec:the_model} we present the effective Lagrangian describing a composite Higgs as Goldstone boson associated to the coset
$\SO(5)/\SO(4)$ together with the light top partners and their couplings to the SM fermions.
In section~\ref{sec:general_analysis} we present a general analysis of the corrections to the
EW observables. In particular we estimate the contributions of the top partners to
the EW oblique parameters and to the deviations of the couplings of the $Z$ gauge
boson to the $b$ quark. Section~\ref{sec:models} is devoted to the numerical
analysis of some explicit models.
In section~\ref{sec:compositetR} we repeat the previous analysis within an alternative set-up
in which the $t_R$ appears as a completely composite state.
And finally in section~\ref{sec:topcouplings} we compute the modifications of the couplings of
the top quark induced by the mixing with its partners. Afterwards we conclude
in section~\ref{sec:conclusions}.
The appendices resume our conventions and collect a few technical details.

\section{The model}\label{sec:the_model}

The first step in our analysis of the EW precision constraints is the identification
of a suitable parametrization of the composite Higgs models. As explained in the
Introduction, our strategy is to study the new physics effects on the EW parameters
from a low-energy perspective. The main advantage of this approach
is the possibility to capture the main features of the composite Higgs scenario and to describe
a broad class of explicit models in a unified framework.

The fundamental ingredient of the composite Higgs scenario is the identification of
the Higgs boson with a set of Goldstones coming from the spontaneous breaking
of a global symmetry of a new strongly-coupled dynamics.
For definiteness, in the following we will focus on the
case in which the Goldstone bosons are associated to the coset $\SO(5)/\SO(4)$.
This is the minimal choice that gives rise to only one Higgs doublet and contains
an $\SO(3)_c$ custodial symmetry. As we will see, the presence of a global symmetry
in the composite sector strongly constrains the structure of the effective Lagrangian
and in particular fixes the form of the Goldstones interactions.

In this paper we will be mainly interested in the corrections to the EW observables
that come from the presence of light fermionic resonances.
To analyze this aspect we will construct an effective description of the
composite models in which only the light fermionic states coming from the strong sector are
included, while the heavier fermionic states and the bosonic resonances are integrated out.
We associate to the heavy resonances a typical mass scale $m_*$, which can be interpreted
as the cut-off of our effective theory. In a generic
strongly coupled sector $m_*$ is connected to the coupling of the strong dynamics $g_*$ and
to the Goldstone decay constant $f$ by the relation $m_* \simeq g_* f$~\cite{Giudice:2007fh}.
Of course our effective description is valid as far as there is a mass gap between the light
and the heavy resonances $m_{light} \ll m_*$.

In the usual framework of composite Higgs models the SM fields do not come from the
strong dynamics, instead they are introduced as elementary states external
with respect to the composite sector. The elementary fermions are mixed to the composite dynamics
following the assumption of partial compositeness~\cite{Kaplan:1991dc}, which requires that they
have only linear mixing with the operators coming from the strong sector.
For simplicity we only include the top quark in our effective description. It is the field
that has the largest mixing with the composite states and induces the most important
corrections to the EW observables. The mixing of the elementary doublet $q_L$ and of the
singlet $t_R$ can be schematically written as
\begin{equation}\label{eq:elem_comp_mix_UV}
{\cal L}_{mix} = y_L \overline q_L {\cal O}_L + y_R \overline t_R {\cal O}_R + \textrm{h.c.}\,,
\end{equation}
where ${\cal O}_{L,R}$ are operators coming from the strong dynamics. An important
point is the fact that global $\SO(5)$ symmetry of the strong sector is unbroken in the UV
where the elementary--composite mixings are generated, thus the composite
operators ${\cal O}_{L,R}$ will belong to some liner representation of $\SO(5)$.
On the other hand, the elementary states transform only under
the SM gauge group and they do not fill complete $\SO(5)$ representations.
The mixing between elementary and composite states induces a (small) explicit breaking
of the global $\SO(5)$ invariance, making the Higgs a pseudo Goldstone boson and generating
an effective Higgs potential.
From a low-energy perspective, the mixing in eq.~(\ref{eq:elem_comp_mix_UV}) can be
reinterpreted as a linear mixing between the elementary states and some fermionic resonances
 $\Psi$ coming from the strong dynamics:
\begin{equation}
{\cal L}_{mix}^{eff} = y_L f\, \overline q_L \Psi_R + y_R f\, \overline t_R \Psi_L + \textrm{h.c.}\,.
\end{equation}

The assumption of partial compositeness also determines the coupling of the elementary
gauge fields with the composite sector.
The SM gauge fields are coupled to the strong dynamics via the weak gauging of a
subgroup of the global invariance. As well known, in order to accommodate the correct
hypercharges of the SM fermions, an extra Abelian subgroup must be added to the
global invariance of the composite sector, which becomes $\SO(5)\times\textrm{U}(1)_X$.
The SM $\textrm{SU}(2)_L$ group is identified with the corresponding factor of the
$\SO(4) \simeq \textrm{SU}(2)_L \times \textrm{SU}(2)_R$ subgroup of $\SO(5)$, while
the hypercharge generator corresponds to the combination $Y = T^3_R + X$, where
$T^3_R$ is the third generator of $\textrm{SU}(2)_R$ (see
appendix~\ref{app:ccwz} for further details and for our conventions).
The weak gauging induces another
small explicit breaking of the global $\SO(5)$ symmetry. This breaking is however typically
subleading with respect to the one induced by the top quark mixing.

\subsection{The effective Lagrangian}

We can now discuss in more details the structure of the effective theory and derive
the general form of the effective Lagrangian respecting our basic assumptions.
In our derivation we will follow the standard CCWZ approach~\cite{ccwz}, which
allows to build all the operators in the effective Lagrangian starting from elements
in irreducible representations of the unbroken global group $\SO(4)$.

The Higgs doublet is described by the set of $4$ Goldstone bosons $\Pi_i$ encoded
in the Goldstone matrix $U$,
\begin{equation}\label{eq:Goldstone_matrix}
U \equiv \exp\left[i \frac{\sqrt{2}}{f} \Pi_i T^i\right]\,,
\end{equation}
where $T^i$ ($i = 1,\ldots,4$) are the generators of the $\SO(5)/\SO(4)$ coset.
The operators in the effective Lagrangian can be written in terms of the $U$ matrix and of the
CCWZ operators $e_\mu$ and $d_\mu$, that come from the covariant derivative of the Goldstone matrix
(see appendix \ref{app:ccwz} for further details).
The $e_\mu$ symbol is used to build the covariant derivative of the composite fermions.
The $d_\mu$ symbol transforms as a $4$-plet of $\SO(4)$ and enters in the kinetic
terms for the Goldstones, which read
\begin{equation}
{\cal L}_{gold} = \frac{f^2}{4} d_\mu ^i d^\mu_i\,.
\end{equation}

The fermion sector of the theory depends on the quantum numbers we choose for the
composite sector operators ${\cal O}_{L,R}$. In the following we will concentrate on the
case in which the operators belong to the fundamental representation of $\SO(5)$.
With this choice we are able to parametrize the low-energy dynamics of several explicit
models proposed in the literature
(see for example Refs.~\cite{Contino:2006qr,Panico:2008bx,Anastasiou:2009rv,Panico:2010is,Panico:2011pw,DeCurtis:2011yx}).
The requirement of a mixing with the elementary top quark fixes the $\textrm{U}(1)_X$
charge of these operators to be $2/3$.

As mentioned before, in the effective theory we can describe the low-energy dynamics
of the strong sector through a set of fermionic states. For simplicity we include
only one level of composite fermions in our effective description and we identify
the cut-off with the mass of the lightest of the other resonances.
In the CCWZ approach the fields are introduced as irreducible representations
of the unbroken group $\SO(4)$ and transform non-linearly under the full $\SO(5)$ symmetry.
The quantum numbers of the ${\cal O}_{L,R}$ operators determine the representations
of the fields that can be directly coupled to the elementary fermions.
The fundamental representation of $\SO(5)$ decomposes under $\SO(4)$ as
${\bf 5} = {\bf 4} + {\bf 1}$. For this reason we include in our theory
two composite fermion multiplets corresponding to representations ${\bf 4_{2/3}}$ and ${\bf 1_{2/3}}$
of $\SO(4)\times \textrm{U}(1)_X$, which we denote by $\psi_4$ and $\psi_1$ respectively.

In order to estimate the size of the coefficients of the various terms in the effective
Lagrangian we need to use a suitable power-counting rule. Following the approach of
Refs.~\cite{Giudice:2007fh,DeSimone:2012fs} we adopt the following formula
\begin{equation}\label{eq:power-counting}
{\cal L} = \sum \frac{m_*^4}{g_*^2} \left(\frac{y\, \psi_{el}}{m_*^{3/2}}\right)^{n_{el}}
\left(\frac{g_* \Psi}{m_*^{3/2}}\right)^{n_{co}}
\left(\frac{\partial}{m_*}\right)^{n_d}
\left(\frac{\Pi}{f}\right)^{n_\pi} \left(\frac{g A_\mu}{m_*}\right)^{n_A}\,,
\end{equation}
where $\psi_{el}$ generically denotes the elementary fields $q_L$ or $t_R$, while
$\Psi$ denotes the composite fermions. Notice that each insertion of an elementary
fermion is accompanied by a corresponding factor of the elementary-composite
mixing $y$. We assume that the rule in eq.~(\ref{eq:power-counting}) has only two
exceptions~\cite{DeSimone:2012fs}.~\footnote{Notice that the power-counting rule can also be
violated in the presence of sum rules which forbid the generation of some operators.}
The first one is the kinetic term of the
elementary fermions, which we set to be canonical. This is justified by the fact that
the elementary fermions are external with respect to the strong
dynamics and their kinetic term is set by the UV theory.
The second exception is the mass of the fermion resonances included in our low-energy description,
which we assume to be smaller than the cut-off $m_*$. This is needed in order to write
an effective theory in which only a few resonances are present, while the other ones,
at the scale $m_*$, are integrated out.

The full effective Lagrangian can be split into three pieces which correspond to the
terms containing only composite states, the ones containing only elementary fields
and the elementary--composite mixings:
\begin{equation}
{\cal L} = {\cal L}_{comp} + {\cal L}_{elem} + {\cal L}_{mixing}\,.
\end{equation}
The leading order Lagrangian for the composite fermions is given by
\begin{equation}\label{eq:lagr_comp}
{\cal L}_{comp} = i \overline \psi_4 \Dslash \psi_4 + i \overline \psi_1 \Dslash \psi_1
- m_4 \overline \psi_4 \psi_4 - m_1 \overline \psi_1 \psi_1
+ \left(i\, c\, \overline \psi_4^i \gamma^\mu d_\mu^i \psi_1 + \textrm{h.c.}\right)
+ \frac{1}{f^2} (\overline \psi \psi)^2\,,
\end{equation}
where the index $i$ labels components of the $\SO(4)$ $4$-plets. Notice that the covariant
derivative of the $\psi_4$ field contains, in addition to the usual derivative
and to the coupling to the $\textrm{U}(1)_Y$ gauge boson $B_\mu$, the CCWZ $e_\mu$ symbol:
$D_\mu \psi_4 = (\partial_\mu - 2/3 i g' B_\mu + i e_\mu) \psi_4$. The presence of the
$e_\mu$ term is essential to restore the full $\SO(5)$ invariance of the Lagrangian and gives rise to
non-linear derivative couplings between the $4$-plet components and the Goldstones.
In addition to the usual kinetic and mass terms we can also write an additional term
using the CCWZ $d_\mu$ symbol. This operator
induces some interactions between the $4$-plet and the singlet mediated by the gauge fields
and by the Goldstones. In general two independent terms with the $d_\mu$ symbol can be
present, one for the left-handed and one for the right-handed composite fermions.
For simplicity, however, we assumed that the strong sector is invariant under parity,
which forces the two operators to have the same coefficient.

\begin{figure}
\centering
\includegraphics[width=.6\textwidth]{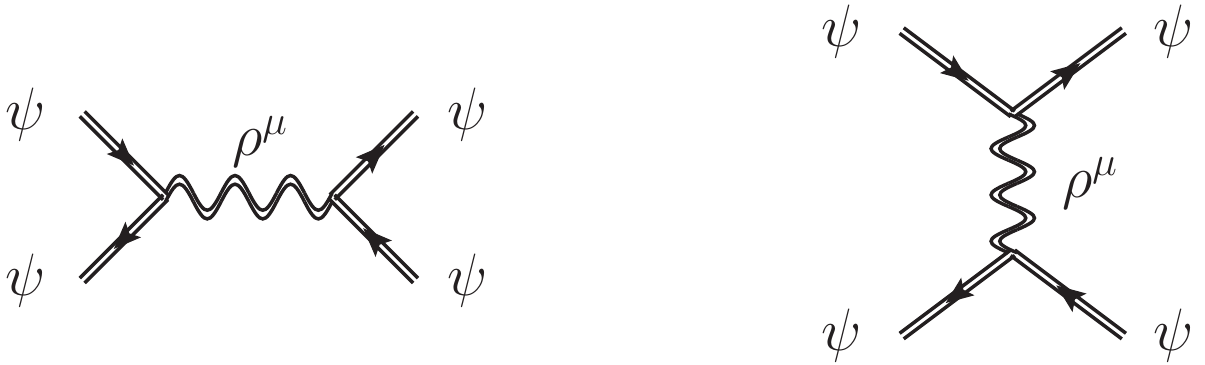}
\caption{Structure of the Feynman diagrams that generate $4$-fermions
operator through the exchange of heavy gauge resonances. In the diagrams
we represent the composite resonances with a double line.}
\label{fig:4-ferm}
\end{figure}

Finally we denote collectively by $(\overline \psi \psi)^2/f^2$
possible contact interactions with $4$ composite fermions.
In spite of having dimension $6$
these operators are not suppressed by the cut-off $m_*$, instead, their natural
coefficient is of order $1/f^2$. Operators of this kind are typically generated by the exchange
of heavy vector or scalar resonances (see diagrams in fig.~\ref{fig:4-ferm}).
The suppression due to the propagator of the heavy boson is compensated by the large
coupling, $g_* \simeq m_*/f$, thus explaining the order $1/f^2$ coefficient.

The Lagrangian involving the elementary fields includes the usual canonical kinetic terms
\begin{equation}
{\cal L}_{elem} = i \overline q_L \Dslash q_L + i \overline t_R \Dslash t_R\,,
\end{equation}
and the elementary--composite mixing
\begin{eqnarray}
{\cal L}_{mixing} &=& y_{L4} f \left(\overline q_L^{{\bf 5}}\right)^I U_{Ii}\, \psi_4^i
+ y_{L1} f \left(\overline q_L^{{\bf 5}}\right)^I U_{I5}\, \psi_1 + \textrm{h.c.}\nonumber\\
&&+\, y_{R4} f \left(\overline t_R^{{\bf 5}}\right)^I U_{Ii}\, \psi_4^i
+ y_{R1} f \left(\overline t_R^{{\bf 5}}\right)^I U_{I5}\, \psi_1 + \textrm{h.c.}\,,\label{eq:lagr_elem}
\end{eqnarray}
where $q_L^{{\bf 5}}$ and $t_R^{{\bf 5}}$ denote the embedding of the elementary
fermions in an incomplete $\bf 5$ of $\SO(5)$, namely
\begin{equation}
q_L^{{\bf 5}} = \frac{1}{\sqrt{2}}
\left(
\begin{array}{c}
i\, b_L\\
b_L\\
i\, t_L\\
- t_L\\
0
\end{array}
\right)\,,\qquad\qquad
t_R^{{\bf 5}} =
\left(
\begin{array}{c}
0\\
0\\
0\\
0\\
t_R
\end{array}
\right)\,,
\end{equation}
and $U$ is the Goldstone matrix defined in eq.~(\ref{eq:Goldstone_matrix}).
The form of the elementary--composite mixings is dictated by the $\SO(5)$ symmetry.
The assumption of partial compositeness encoded in eq.~(\ref{eq:elem_comp_mix_UV})
tells us that the elementary fields are mixed with operators that transform in a
\emph{linear} representation of $\SO(5)$. The $\psi_4$ and $\psi_1$ CCWZ fields, instead,
transform non-linearly under the global symmetry, so they can not be directly
mixed with the elementary fields. To write down a mixing term we thus need to compensate
for the non-linear transformation and this can be done by multiplying the CCWZ fields
by the Goldstone matrix.

Notice that the coefficients that appear in our effective Lagrangian are in general complex.
By means of chiral rotations of the elementary and composite fields one can remove only
$3$ complex phases, thus some parameters are still complex. In order to simplify the
analysis we assume that our Lagrangian is invariant under CP~\cite{DeSimone:2012fs}.
Under this hypothesis all the parameters in the Lagrangian in eqs.~(\ref{eq:lagr_comp})
and (\ref{eq:lagr_elem}) are real.~\footnote{The CP invariance fixes the coefficient of the
$d_\mu$ symbol term to be purely imaginary. Thus our parameter $c$ is real.}

\section{General analysis of the EW parameters}\label{sec:general_analysis}

In this section we provide a general analysis of the new physics corrections to the
EW observables, in particular we will focus on the oblique parameters, $\widehat S$
and $\widehat T$, and on the $Z \overline b_L b_L$ coupling.
As we will see, several effects can generate distortions of this parameters
and it is important to carefully study all of them.
The primary aim of this section is to estimate the size of the various corrections
and to determine which observables can be reliably computed
in our low-energy effective  approach.

\subsection{The oblique parameters}

We start our analysis by considering the oblique EW parameters, $\widehat S$
and $\widehat T$,~\cite{Peskin:1991sw,Barbieri:2004qk} that encode the corrections to the
two point functions of the EW gauge bosons.
The contributions to the oblique parameters come from three main effects: the Goldstone nature
of the Higgs, the presence of vector resonances and the presence of fermionic resonances.

The first effect is related to the non-linear Higgs dynamics which induces a modification
of the Higgs couplings with the EW gauge bosons. This distortion is present in any
composite-Higgs model and is fully determined by the symmetry breaking pattern that
gives rise to the Goldstones, in our case $\SO(5)/\SO(4)$.
In particular the leading logarithmically-enhanced contribution
is universal and is completely fixed by the IR dynamics~\cite{Barbieri:2007bh}.
As we will see, while the contribution to $\widehat S$ is small, the effect on $\widehat T$
is sizable and, without further corrections, would lead to very stringent bounds on
the Higgs compositeness scale $f$.

The second source of corrections is the presence of EW gauge
resonances. In our effective Lagrangian approach the gauge resonances have been integrated out,
thus this corrections arise as a purely UV effect. The most important contribution
is generated at tree level due to the mixing of the composite resonances with the
elementary gauge bosons and it gives a sizable correction to the $\widehat S$ parameter.

Finally the third class of contributions comes from loop effects
induced by the composite fermions. This is the class of contributions we will be
mainly interested in in the present analysis. As we will see, these corrections are
typically large and including them is essential in order to obtain
a reliable fit of the EW parameters.
Although these effects have been already considered in the literature, most of the
previous analyses did not take into account the full non-linear structure of the composite
Higgs Lagrangian. Our analysis will show that the non-linearities are relevant
and their inclusion can significantly affect the result and lead to new important
effects.

\begin{figure}
\centering
\includegraphics[width=.4\textwidth]{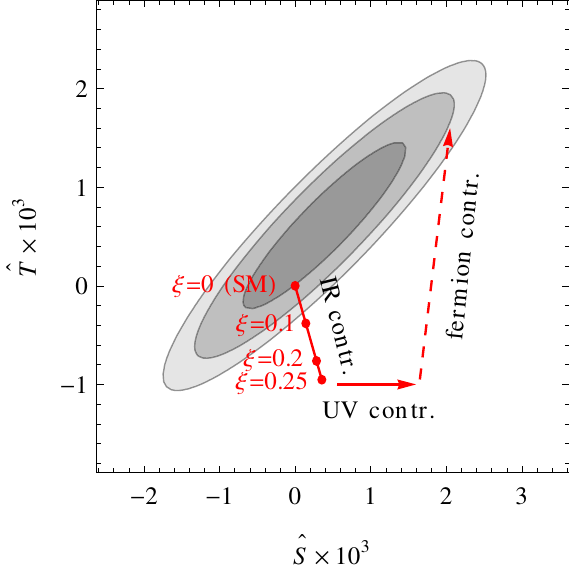}
\caption{Constraints on the oblique EW parameters $\widehat S$ and $\widehat T$~\cite{Baak:2012kk}.
The gray ellipses correspond to the $68\%$, $95\%$ and $99\%$ confidence level contours
for $m_h = 126\ {\rm GeV}$ and $m_t = 173\ {\rm GeV}$. The red lines show the contributions
that arise in composite Higgs models as explained in the main text. The IR contribution
corresponds to the corrections due to non-linear Higgs dynamics,
approximately given in eqs.~(\ref{eq:S_IR}) and (\ref{eq:T_IR}), and is obtained fixing
$m_* \sim 3\ {\rm TeV}$. The UV contribution is due to the EW gauge resonances (see eq.~(\ref{eq:S_UV})).}
\label{fig:ST_plane}
\end{figure}

\subsubsection*{The $\widehat S$ parameter}

At tree level the $\widehat S$ parameter receives a correction due to the
mixing of the elementary gauge fields with the composite vector bosons.
An estimate of this correction is given by~\cite{Giudice:2007fh}
\begin{equation}\label{eq:S_UV}
\Delta \widehat S \simeq \frac{g^2}{g_*^2} \xi  \simeq \frac{m_w^2}{m_*^2}\,.
\end{equation}
The UV dynamics can lead to deviations with respect to the above formula. However those
deviations are typically small and eq.~(\ref{eq:S_UV}) is usually in good agreement with
the predictions of explicit models. Assuming that the correction in eq.~(\ref{eq:S_UV})
is the dominant contribution to $\widehat S$ (or at least that the other contributions
to $\widehat S$ are positive), a rather strong upper bound on the mass of the EW gauge
resonances is found, $m_* \gtrsim 2\ {\rm TeV}$ (see the fit of the oblique parameters
in fig.~\ref{fig:ST_plane}).

The other contributions to the $\widehat S$ parameter arise at loop level due to
the non-linear Higgs dynamics and to the presence of fermion resonances.
The leading contribution due to the non-linear Higgs dynamics is given by~\cite{Barbieri:2007bh}
\begin{equation}\label{eq:S_IR}
\Delta \widehat S = \frac{g^2}{192 \pi^2} \xi \log\left(\frac{m_*^2}{m_h^2}\right)
\simeq 1.4\cdot 10^{-3}\, \xi\,.
\end{equation}
where $g$ denotes the SM $\textrm{SU}(2)_L$ gauge coupling.
In the above formulae we identified the cut-off with
the mass scale of the EW gauge resonances and we chose $m_* \sim 3\ {\rm TeV}$
and $m_h = 126\ {\rm GeV}$ to derive the numerical estimate.

The contribution in eq.~(\ref{eq:S_IR}) arises from one-loop diagrams with gauge bosons and Goldstone
virtual states. The diagrams contributing to $\widehat S$ are
superficially logarithmically divergent. However, in the SM the logaritmic divergence
exactly cancels due to the physical Higgs contribution. This is no longer true when
the Higgs couplings are modified and in composite Higgs models a residual logarithmic
dependence on the cut-off scale is present.~\footnote{A more detailed analysis
of the corrections to the $\widehat S$ parameter related to the Goldstone nature of the Higgs
has been presented in Ref.~\cite{Orgogozo:2012ct}.}
As can be seen from the numerical estimate the contribution in eq.~(\ref{eq:S_IR}) is
much smaller than the absolute bounds on $\widehat S$ (compare fig.~\ref{fig:ST_plane})
and is typically negligible.

Let us finally consider the contribution due to loops of fermionic resonances. The general
expression for the corrections to $\widehat S$ due to an arbitrary set of new vector-like
fermion multiplets has been derived in Ref.~\cite{Lavoura:1992np}. The final formula
contains a divergent contribution to $\widehat S$ given by
\begin{equation}\label{eq:div_S}
\Delta \widehat S^{div}_{ferm} = \frac{N_c g^2}{96 \pi^2} {\rm Tr}\left[U_L^\dagger Y_L
+ U_R^\dagger Y_R\right] \log(m_*^2)\,,
\end{equation}
where $U_{L,R}$ and $Y_{L,R}$ are the matrices of the couplings of left- and right-handed fermions
to the $W^3_\mu$ and to the $B_\mu$ gauge bosons respectively and $N_c$ is the number
of QCD colors. In a renormalizable theory
in which the couplings of the gauge bosons to the fermions are just given by the usual
covariant derivatives it is easy to see that the trace appearing in eq.~(\ref{eq:div_S})
vanishes, so that no logarithmically divergent contribution to $\widehat S$ is
present.~\footnote{To prove this one can notice that the sum of the $W^3_\mu$
couplings to the fermions in each $\textrm{SU}(2)_L$ multiplet is zero. After EWSB the gauge
couplings of the fermion mass eigenstates are obtained by unitary rotations of the initial
coupling matrices. These rotation clearly cancel out in the trace in eq.~(\ref{eq:div_S}),
so that the divergent term vanishes.}
This is no longer true when the Higgs is a Goldstone boson. In this case
higher order interactions of the gauge bosons mediated by the Higgs
are present in the Lagrangian.
Interactions of this kind are contained in the $e_\mu$ term in the covariant derivative
of the composite $4$-plet $\psi_4$ and in the $d_\mu$-symbol term.
After EWSB a distortion of the gauge couplings to the fermions is induced by these operators
and a logarithmically divergent contribution to $\widehat S$ is generated.
The presence of a logarithmically enhanced contribution can be also understood in simple
terms as a running of the operators related to the $\widehat S$ parameter. We postpone a discussion
of this aspect to the end of this subsection.

The logarithmically divergent correction can be straightforwardly computed:
\begin{equation}\label{eq:S_div_explicit}
\Delta \widehat S^{div}_{ferm} = \frac{g^2}{8 \pi^2} (1 - 2 c^2)\, \xi \log\left(\frac{m_*^2}{m_4^2}\right)\,.
\end{equation}
It is important to notice that this contribution is there only if at least one
$\SO(4)$ $4$-plet is present in the effective theory. In fact, as we said, the only terms in the
effective Lagrangian that can lead to relevant distortions of the gauge couplings are
the $4$-plet kinetic term and the $d_\mu$-symbol term, which are clearly absent if only
singlets are present.
The connection of the divergence with the $4$-plets justifies the identification of the
argument of the logarithm in eq.~(\ref{eq:S_div_explicit}) with the ratio $m_*^2/m_4^2$.
It is also remarkable the fact that the correction in eq.~(\ref{eq:S_div_explicit}) is
independent of the elementary--composite mixings $y_{L,R}$. This implies that any
$\SO(4)$ $4$-plet below the cut-off of the effective theory would contribute to $\widehat S$
with a similar shift.~\footnote{Resonances in larger $\SO(4)$ multiplets also lead to divergent
contributions. For instance, states in the $\bf 9$ lead to a contribution $6$ times larger
than the one in eq.~(\ref{eq:S_div_explicit}).}

Notice that, in order to derive  the result in eq.~(\ref{eq:S_div_explicit}), we assumed
that the logarithmic divergence due to the fermion loops is regulated at the cut-off scale
$m_*$. This is expected to happen as a consequence of the presence of EW gauge resonances
with a mass of order $m_*$. Peculiar UV dynamics, however, could modify this picture and push
up the scale at which the divergence is regulated, resulting in
a larger contribution to $\widehat S$.

Another interesting property of the divergent contribution to $\widehat S$ is the fact that
it vanishes if $c^2 = 1/2$. As we will see later on, this choice of the parameter $c$
implies the presence of an extra symmetry in the effective Lagrangian which protects the
EW observables.

The logarithmic contribution to $\widehat S$ in eq.~(\ref{eq:S_div_explicit})
is sizable if $c^2$ is not too close to $1/2$ and is typically much larger than the
corresponding effect due to the Higgs non-linearities (eq.~(\ref{eq:S_IR})).
The correction due to fermion loops can even be comparable with the tree-level
contribution estimated in eq.~(\ref{eq:S_UV}) if the strong coupling $g_*$ is large,
$g_* \gtrsim 5$. From the point of view of our effective approach, the coefficient $c$
is just a free parameter, thus in principle the divergent fermion contribution
can have an arbitrary sign. In particular for $c^2 > 1/2$ a sizable negative shift
in $\widehat S$ would be possible, which could improve the agreement with the EW
precision measurements (see fig.~\ref{fig:ST_plane}).

\begin{figure}
\centering
\includegraphics[width=.4\textwidth]{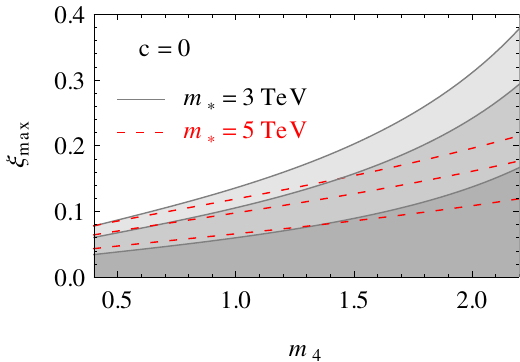}
\caption{Upper bounds on $\xi$ in the $2$-site model ($c=0$) as a function of the $4$-plet mass
parameter $m_4$ for different values of the cut-off $m_*$. The results have been obtained by
considering the shift in $\widehat S$ given in eqs.~(\ref{eq:S_UV}), (\ref{eq:S_IR})
and (\ref{eq:S_div_explicit}) and by marginalizing on $\widehat T$.
The shaded regions correspond to the points compatible
with the constraints at the $68\%$, $95\%$ and $99\%$ confidence level for $m_* = 3\ {\rm TeV}$.
The dashed red curves show how the bounds are modified for $m_* = 5\ {\rm TeV}$.}
\label{fig:xi_bound_S}
\end{figure}
It is important to notice that in explicit models that provide a partial
UV completion of our effective theory the value of $c$ is typically fixed.
A possible extension of our effective Lagrangian
is given by the $2$-site model proposed in Refs.~\cite{Panico:2011pw,Matsedonskyi:2012ym}.
In this model $c=0$, so that a sizable positive shift in $\widehat S$ seems unavoidable
if a relatively light $4$-plet is present. For example for $m_4 \simeq 700\ {\rm GeV}$
and $m_* \simeq 3\ {\rm TeV}$ a tight upper bound, $\xi \lesssim 0.1$, is obtained
if we marginalize on $\widehat T$. The limits on the compositeness scale as
a function of the $4$-plet mass taking into account only the constraints on the
$\widehat S$ parameter are shown in fig.~\ref{fig:xi_bound_S}.
Notice that the bounds become typically stronger
if the cut-off scale increases. This is due to the fact that the logarithmically enhanced
fermion contribution in eq.~(\ref{eq:S_div_explicit}) grows at larger $m_*$
and dominates over the tree-level correction in eq.~(\ref{eq:S_UV}) which instead decreases
when the gauge resonances become heavier.

The $2$-site realization of the composite models allows us also to find a connection between
the fermion corrections to $\widehat S$ and the dynamics of the gauge resonances.
In fact it turns out that the diagrams that give rise to the divergence in $\widehat S$
are closely related to the ones that determine the running of the gauge resonance coupling $g_*$.
The divergent contribution to $\widehat S$ in this picture arises from the distortion
of the mixing between the elementary and the composite gauge fields after EWSB.

A fermion contribution to $\widehat S$ similar to the one we found is in principle
present also in the extra-dimensional realization of the composite Higgs scenario.
The corrections to the oblique EW parameters due to fermion loops in this class of theories
have been considered in the literature \cite{Carena:2006bn, Panico:2010is}, however no
divergent or enhanced contribution was noticed. It is probable however that a contribution
of this kind was overlooked because of its peculiar origin. Similarly to what happens in
the $2$-site model, in extra dimensions the divergence in $\widehat S$ derives from the mixing
of the gauge zero-modes with the gauge resonances after EWSB. In the literature the computation
of $\widehat S$ has been made neglecting this mixing, thus the divergent contribution
was not found.

Notice that, in addition to the divergent contributions, which explicitly depend on the
cut-off, large finite contributions can also arise from the UV dynamics of the theory.
We can estimate the one-loop UV contributions as
\begin{equation}\label{eq:S_UV_loop}
\Delta \widehat S \sim \frac{g^2}{16 \pi^2} \xi \simeq 3\cdot 10^{-3} \xi\,.
\end{equation}
It is easy to see that these effects can in principle be sizable and
could significantly change the fit to the EW data.
The estimate in eq.~(\ref{eq:S_UV_loop}) should be considered as a lower bound on the
size of the UV corrections, valid if no accidental cancellations are present.
Larger corrections to $\widehat S$ are possible in the presence of some peculiar
UV dynamics, these however can not be predicted within our effective approach.
We will see an explicit example of non-decoupling effects in subsection~\ref{sec:singlet}.

\subsubsection*{The corrections to $\widehat S$ as a running effect}

We can understand in simple terms the origin of the large logarithmically
enhanced contributions to the $\widehat S$ parameter with an operator approach.
In the effective theory the corrections to the $\widehat S$ parameter
are induced by two dimension-$6$ operators~\cite{Giudice:2007fh}:
\begin{equation}
{\cal O}_W = i \left(H^\dagger \sigma^i \overleftrightarrow{D^\mu} H\right)
(D^\nu W_{\mu\nu})^i
\qquad {\rm and} \qquad
{\cal O}_B = i \left(H^\dagger \overleftrightarrow{D^\mu} H\right)
(D^\nu B_{\mu\nu})\,,
\end{equation}
where $H$ denotes the usual Higgs doublet and $H^\dagger \overleftrightarrow{D_\mu} H$
is the derivative $H^\dagger (D_\mu H) - (D_\mu H)^\dagger H$.

\begin{figure}
\centering
\includegraphics[width=.7\textwidth]{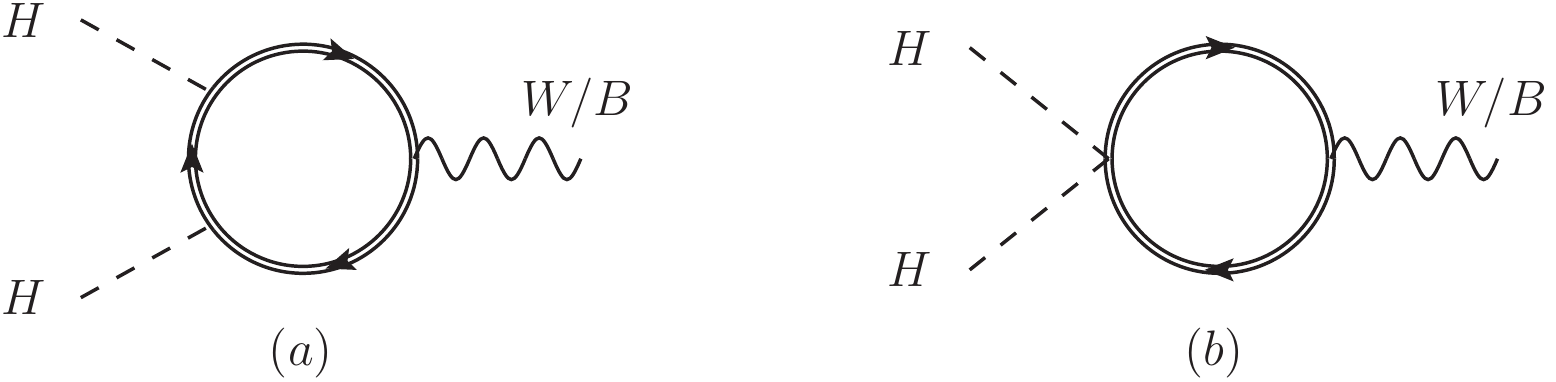}
\caption{Diagrams with resonance loops that can contribute to the ${\cal O}_{W,B}$ operators.}
\label{fig:S_effective}
\end{figure}
The corrections to the ${\cal O}_{W,B}$ operators can be connected to the diagrams with
two external Higgs states and one gauge field.
In a renormalizable theory with only standard Yukawa Higgs couplings to the fermions the
corrections from heavy resonances loops come from the $(a)$ diagrams in fig.~\ref{fig:S_effective}.
By noticing that the ${\cal O}_{W,B}$
operators contain three powers of the external momenta it is easy to realize that these
diagrams are always finite.

In a theory with a non-linear Higgs dynamics the situation is instead drastically different.
In this case non-renormalizable contact interactions with two Higgses and two composite
fermions are present.
In particular the $e_\mu$ symbol in the kinetic term of the composite $4$-plets induces
a non-renormalizable interaction $i (\vec{\Pi}^t t^a \partial_\mu \vec{\Pi}) (\overline \psi_4 \gamma^\mu \psi_4)$
(see the explicit results in appendix~\ref{app:ccwz}).
This non-linear vertex, together with the usual gauge interactions, gives rise to the new class
of diagrams denoted by $(b)$ in fig.~\ref{fig:S_effective}.
These diagrams are logarithmically divergent and induce a corresponding running
of the ${\cal O}_{W,B}$ operators leading to an enhanced contribution to $\widehat S$.
This running effect generates the $c$-independent term in the correction to $\widehat S$
(see eq.~(\ref{eq:S_div_explicit})).~\footnote{Notice that the diagrams with the new non-linear
Higgs vertex can in principle contribute also to two other dimension-$6$ operators,
${\cal O}_{HW} = i (D^\mu H)^\dagger \sigma^i (D^\nu H) W^i_{\mu\nu}$
and ${\cal O}_{HB} = i (D^\mu H)^\dagger (D^\nu H) B_{\mu\nu}$. Differently from ${\cal O}_{W,B}$,
these two operators do not contribute to $\widehat S$ and are
not minimally coupled~\cite{Giudice:2007fh}.
With an explicit computation we found that the logarithmically divergent diagrams only
generate a running of the minimally coupled operators ${\cal O}_{W,B}$
and not of ${\cal O}_{HW, HB}$.}

Non-renormalizable Higgs interactions are also generated by the $d_\mu$ symbol terms.
In particular it gives rise to a new vertex of the form $(\partial_\mu \Pi^i) \overline \psi_4^i \gamma^\mu \psi_1 + {\rm h.c.}$.
This vertex induces a logarithmically divergent contribution to ${\cal O}_{W,B}$ through
diagrams analogous to the type $(a)$ shown in fig.~\ref{fig:S_effective}.
The related contribution to the $\widehat S$ parameter corresponds to the term
proportional to $c^2$ in eq.~(\ref{eq:S_div_explicit}).

It is interesting to notice that similar contributions to the $\widehat S$
parameter are also present in technicolor models but originated from the
non-linear dynamics not of the whole Higgs doublet, as in our case,
but only of the Goldstones associated to the spontaneous breaking
$\textrm{SU}(2)_L \times \textrm{SU}(2)_R \rightarrow \textrm{SU}(2)_V$~\cite{S_technicolor}.

Before concluding the discussion on $\widehat S$ we want to comment on the relation between
our results and the ones of Refs.~\cite{Elias-Miro:2013gya,Contino:2013kra}. In Refs.~\cite{Elias-Miro:2013gya,Contino:2013kra}
an effective approach was used in which only the SM fields are retained and all the
composite resonances are integrated out. In this framework it was shown that two effective
operators ${\cal O}_{Hq} = i (\overline q_L \gamma^\mu q_L) (H^\dagger \overleftrightarrow{D_\mu}H)$
and ${\cal O}'_{Hq} = i (\overline q_L \gamma^\mu \sigma^i q_L) (H^\dagger \sigma^i \overleftrightarrow{D_\mu}H)$ induce a logarithmic running for $\widehat S$ between the
top mass, $m_t$ and the energy scale at which the effective operators are generated, $m$.
Differently from Refs.~\cite{Elias-Miro:2013gya,Contino:2013kra}, in our approach the resonances are included
in the effective theory and the effective operators ${\cal O}_{Hq}$ and ${\cal O}'_{Hq}$
are not present directly in our Lagrangian. At low energy, however, they are generated
through the exchange of resonances of mass $m$ with a coefficient $y^2/m^2$.
From the previous discussion it is easy to understand that in our approach the logarithmically
divergent corrections to $\widehat S$ found in Refs.~\cite{Elias-Miro:2013gya,Contino:2013kra} do not
appear as real divergences but rather correspond to corrections that scale as
$y^2/m^2 \log(m^2/m_t^2)$. Terms of this form can be recognized, for example, in the
explicit analytic result for $\widehat S$ given in eq.~(\ref{eq:Ssing}).~\footnote{Notice that
other effective operators with the structure ${\cal O}_t = H^\dagger H (\overline q_L H^c t_R)$
do not generate a running for $\widehat S$~\cite{Elias-Miro:2013gya}.}

\subsubsection*{The $\widehat T$ parameter}

We can now analyze the corrections to the $\widehat T$ parameter. Thanks to the custodial
symmetry $\widehat T$ does not receive correction at tree level and the only contributions come at
loop level from diagrams with insertions of the operators that break the custodial symmetry.
In our effective Lagrangian this breaking is induced by the weak gauging of the
hypercharge $\textrm{U}(1)_Y$ with coupling $g'$ and by the mixings $y_{L4,1}$
of the $q_L$ elementary doublet with the composite fermions.

The main correction due to the hypercharge coupling breaking comes from the IR contribution
associated to the Goldstone nature of the Higgs. This effect is analogous to the one we
already discussed for the $\widehat S$ parameter. The leading logarithmically enhanced
contribution is given by~\cite{Barbieri:2007bh}
\begin{equation}\label{eq:T_IR}
\Delta \widehat T = -\frac{3 g'^2}{64 \pi^2} \xi \log\left(\frac{m_*^2}{m_h^2}\right)
\simeq -3.8 \cdot 10^{-3}\, \xi\,.
\end{equation}
Differently from the analogous contribution to $\widehat S$, which was negligible due to
accidental suppression factors, the contribution in eq.~(\ref{eq:T_IR}) gives a sizable
correction to $\widehat T$. In particular, if we assume that this is the dominant correction
to $\widehat T$ and that the shift in $\widehat S$ is non negative, a very stringent
bound on $\xi$ is obtained, $\xi \lesssim 0.1$
(see fig.~\ref{fig:ST_plane}).~\footnote{A similar bound has been derived
in Ref.~\cite{Espinosa:2012im}, where the phenomenological impact of the
IR corrections to $\widehat S$ and $\widehat T$ on the fit of the Higgs couplings
has been analyzed.}

\begin{figure}
\centering
\includegraphics[width=.35\textwidth]{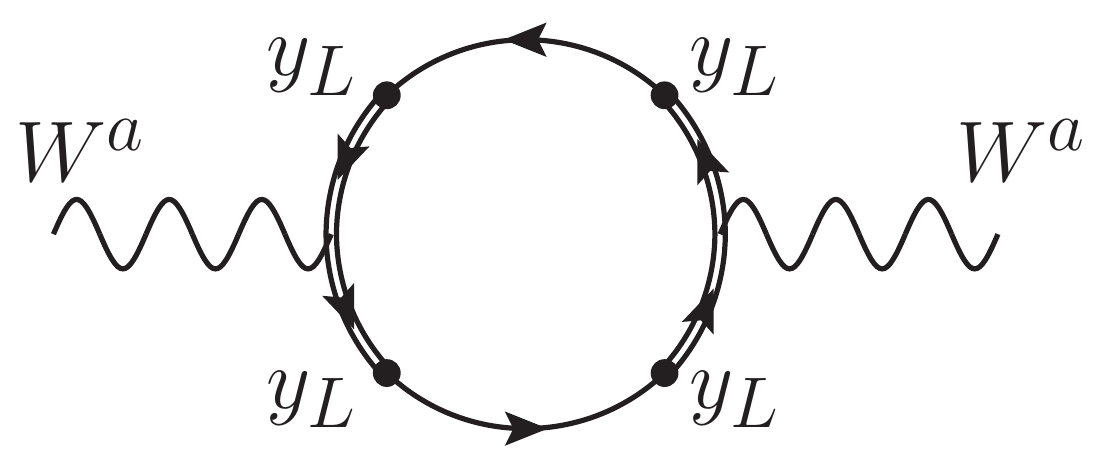}
\caption{Schematic structure of a fermion loop diagram contributing to the
$\widehat T$ parameter at leading order in the $y$ expansion.}
\label{fig:T_loop}
\end{figure}

The second correction comes from fermion loops. As already noticed, in order to induce
a contribution to $\widehat T$ the corresponding diagrams must contain some insertions of the
symmetry breaking couplings $y_{L4,1}$. Under $\textrm{SU}(2)_L \times \textrm{SU}(2)_R$
the $y_{L4,1}$ mixings transform in the $({\bf 1}, {\bf 2})$ representation,
thus at least $4$ insertions are needed to generate
a shift in $\widehat T$~\cite{Giudice:2007fh}. This minimal number of insertions guarantees
that the fermion one-loop corrections to $\widehat T$ are finite. A typical diagram
contributing at leading order in the $y$ expansion is shown in fig.~\ref{fig:T_loop}.

It is straightforward to estimate the corrections to $\widehat T$ at leading order in the
elementary--composite mixing~\cite{Giudice:2007fh}:
\begin{equation}\label{eq:T_ferm_est1}
\Delta \widehat T \simeq \frac{N_c}{16 \pi^2} \frac{y_L^4 f^2}{m^2}\, \xi\,,
\end{equation}
where we denoted by $m$ the mass scale of the lightest top partners in our effective Lagrangian.
To get a quantitative estimate we can extract the value of the $y_L$ mixing from the top mass.
If we assume that the elementary--composite mixings have comparable
sizes, $y_{L4} \simeq y_{L1} \simeq y_{R4} \simeq y_{R1} \simeq y$, the top Yukawa can be estimated as $y_t \simeq y^2 f/m$.
By using this expression we get the estimate
\begin{equation}\label{eq:T_ferm_est2}
\Delta \widehat T \simeq \frac{N_c}{16 \pi^2} y_t^2\, \xi
\simeq 2 \cdot 10^{-2}\, \xi\,.
\end{equation}
Notice that this contribution is usually dominant with respect to the one given
in eq.~(\ref{eq:T_IR}). Moreover, as we will see in the next section with an explicit calculation,
the sign of the fermion contribution can be positive, so that it can compensate
the negative shift in eq.~(\ref{eq:T_IR}). Notice that, if $\widehat S$ is not negative,
a positive correction to $\widehat T$ from the fermion loops is essential in order
to satisfy the EW constraints as can be clearly seen from the bound in fig.~\ref{fig:ST_plane}.

Notice that the finiteness of the fermion loop contribution to $\widehat T$ implies that
the correction coming from the lightest resonances is dominant with respect to the
one coming from heavier states. The contribution due to the UV dynamics
can be estimated as~\cite{Giudice:2007fh}
\begin{equation}
\Delta \widehat T \simeq \frac{N_c}{16 \pi^2} \frac{y_L^4}{g_*^2} \xi\,.
\end{equation}
This contribution is suppressed with respect to the one in eq.~(\ref{eq:T_ferm_est1})
by a factor $m^2/m_*^2$. This shows that $\widehat T$ can be predicted in a robust way
using our effective field theory approach.

\subsection{The $Z\overline b_L b_L$ vertex}\label{sec:Zbb_general}

Another observable that can be used to constrain the parameter space of new physics models
is the $Z$ boson coupling to the left-handed bottom quark. We define the $Z$ interactions
with the bottom by the formula
\begin{equation}
{\cal L}^Z = \frac{g}{c_w} Z_\mu \overline b \gamma^\mu \left[
(g^{SM}_{b_L} + \delta g_{b_L}) P_L + (g^{SM}_{b_R} + \delta g_{b_R}) P_R\right] b\,,
\end{equation}
where $g^{SM}$ denotes the SM couplings (including the loop corrections), $\delta g$
denotes the corrections due to new physics and $P_{L,R}$ are the left and right projectors.
In the following we will denote by $s_w$ and $c_w$ the sine and cosine of the weak mixing angle.
The SM tree-level values for the couplings are
\begin{equation}
g^{SM, tree}_{b_L} = -\frac{1}{2} + \frac{1}{3} s_w^2\,,
\qquad \quad g^{SM,tree}_{b_R} = \frac{1}{3} s_w^2\,,
\end{equation}
and the one-loop corrections (computed in the limit $g \rightarrow 0$) are
\begin{equation}
g^{SM, loop}_{b_L} = \frac{m_t^2}{16 \pi^2 v^2}\,,
\qquad \quad g^{SM, loop}_{b_R} = 0\,.
\end{equation}

\begin{figure}
\centering
\includegraphics[width=.4\textwidth]{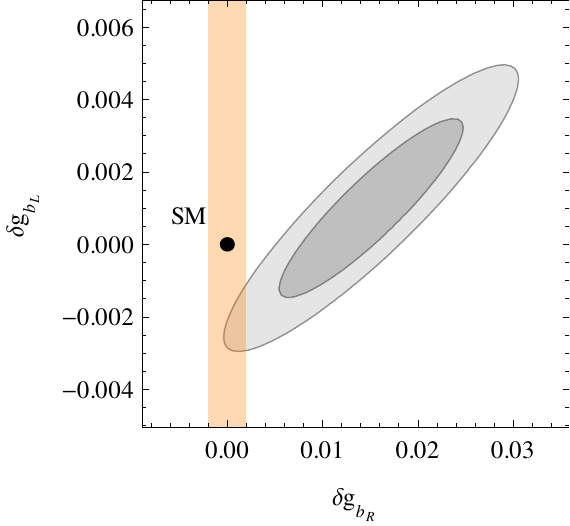}
\caption{Constraints on the corrections to the $Z$ boson couplings to the bottom quark.
The ellipses show the exclusion contours at $68\%$ and $95\%$ confidence level~\cite{constraintZbb}.
The vertical band shows the expected size of the corrections to the $g_{b_R}$ coupling.}
\label{fig:deltag_plane}
\end{figure}
As can be seen from the current bounds shown in fig.~\ref{fig:deltag_plane},
the deviation of the $Z \overline b_L b_L$ coupling are constrained to be at the level
$3\cdot 10^{-3}$, while the bounds on the coupling with the right-handed
bottom component are one order of magnitude less stringent.
In composite models the corrections to the $g_{b_R}$ coupling are typically small, at most
of the same order of the deviations in $g_{b_L}$. If we impose the constraint
$|\delta g_{b_R}| \lesssim few \cdot 10^{-3}$,
a negative value for $\delta g_{b_L}$ of order $-2\cdot 10^{-3}$ is preferred, while a
positive shift worsens the fit with respect to the SM. The region favored by the current fit
in the $(\delta g_{b_L}, \delta g_{b_R})$ plane is shown in fig.~\ref{fig:deltag_plane}
and corresponds to the intersection of the gray ellipses with the vertical band.

\subsection*{Tree-level corrections}

Let us now analyze the new physics corrections that arise in our scenario.
The presence of an automatic $P_{LR}$ symmetry in the composite sector and the fact that
the elementary $b_L$ state is invariant under this symmetry implies the absence of tree-level
corrections to the $Z\overline b_L b_L$ vertex at zero momentum~\cite{P_LR}.
The tree-level corrections induced at non-zero momentum are related to operators
of the form $D_\mu F^{\mu\nu} \overline q_L \gamma_\nu q_L$ and their
size can be estimated as
\begin{equation}\label{eq:deltag_p}
\frac{\delta g_{b_L}}{g^{SM}_{b_L}} \sim \frac{y_L^2 f^2}{m^2} \frac{m_z^2}{m_*^2}
\simeq 8 \cdot 10^{-4}\, \frac{f}{m} \left(\frac{4\pi}{g_*}\right)^2 \xi\,,
\end{equation}
where $m$ is the mass scale of the composite fields mixed with the bottom, which in our
scenario correspond to the charge $-1/3$ state inside the $4$-plet $\psi_4$.

Notice that in our effective Lagrangian we did not include an elementary $b_R$ state.
For this reason the bottom is massless in our theory. In a more complete scenario
a chiral field corresponding to the $b_R$ will be present
together extra composite fermions that are needed to generate
the bottom mass. In this case the elementary $q_L$ doublet has additional mixing terms
with the new resonances and a tree-level correction to the
$Z\overline b_L b_L$ vertex could be generated. For instance this happens in the case in which the
additional bottom partners are contained in a $5$ of $\SO(5)$ with $\textrm{U}(1)_X$
charge $-1/3$. The contribution to the $Z\overline b_L b_L$ vertex coming from these
states can be estimated as
\begin{equation}\label{eq:deltag_b_partners}
\frac{\delta g_{b_L}}{g^{SM}_{b_L}} \simeq \frac{(y_L^b f)^2}{m_B^2} \xi\,,
\end{equation}
where we denoted by $y_L^b$ the mixing of $q_L$ to the new multiplet and by $m_B$ the typical
mass scale of the new bottom partners. We can relate $y_L^b$ to the bottom Yukawa by assuming that
$y_L^b \simeq y_R^b$, in this case $(y^b_L)^2 \simeq (y_R^b)^2 \simeq y_b m_B/f$.
The correction in eq.~(\ref{eq:deltag_b_partners}) becomes
\begin{equation}\label{eq:deltag_b_partners_1}
\frac{\delta g_{b_L}}{g^{SM}_{b_L}} \simeq y_b \frac{f}{m_B} \xi
\simeq2 \cdot 10^{-2} \frac{f}{m_B} \xi\,.
\end{equation}
This correction can easily have a size comparable with the current bounds on $\delta g_{b_L}$
in the case in which the new bottom partners are relatively light. Of course this correction
can be suppressed if we relax the assumption $y_L^b \simeq y_R^b$ or if we chose $m_B \gg f$.

\subsection*{Corrections from fermion loops}

We can now consider the one-loop contributions to the $Z\overline b_L b_L$ vertex.
As a first step we will analyze the degree of divergence of the diagrams contributing
to this effect. The degree of divergence can be easily obtained by using the power-counting
method explained in Ref.~\cite{Panico:2011pw}. It is straightforward to check that the
$Z\overline b_L b_L$ operator at one loop is naively associated to a quadratic divergence.
In our set-up, however, the $P_{LR}$ symmetry implies a
reduction of the naive degree of divergence. This is an obvious consequence of the fact that
a new physics contribution to the $Z\overline b_L b_L$ vertex can be generated only if
some powers of the couplings that break the $P_{LR}$ symmetry are inserted in the diagrams.
In our Lagrangian only the $y_L$ mixings induce a breaking of this symmetry. These mixings
correspond to some mass operators, so that each insertion in
loop diagrams lowers the degree of divergence by one.~\footnote{The $y_L$ mixing
could in principle appear also in higher-dimensional operators. These operators,
which we did not include in our effective Lagrangian, are suppressed by powers of the UV cut-off
$m_*$ as can be inferred from our power-counting rule in eq.~(\ref{eq:power-counting}).
For this reason their insertions also lead to a reduction of
the degree of divergence in agreement with the power counting expectation.}
Let us now count how many insertions of the $y_L$ mixing are necessary to generate a
distortion of the $Z\overline b_L b_L$ vertex. Each external $b_L$ is of course associated
to a power of $y_L$. However, due to the fact that the $b_L$ fields are external legs and
they are invariant under $P_{LR}$, these insertions do not lead to a breaking of the
symmetry. As a consequence at least four insertions of $y_L$ are needed to generate a
non-vanishing contribution.~\footnote{A more rigorous proof of this statement can be
obtained by using an operator analysis. For simplicity we do not present this analysis
in the main text and postpone it to appendix~\ref{app:Zbb_operator_analysis}.}

\begin{figure}
\centering
\includegraphics[height=.25\textwidth]{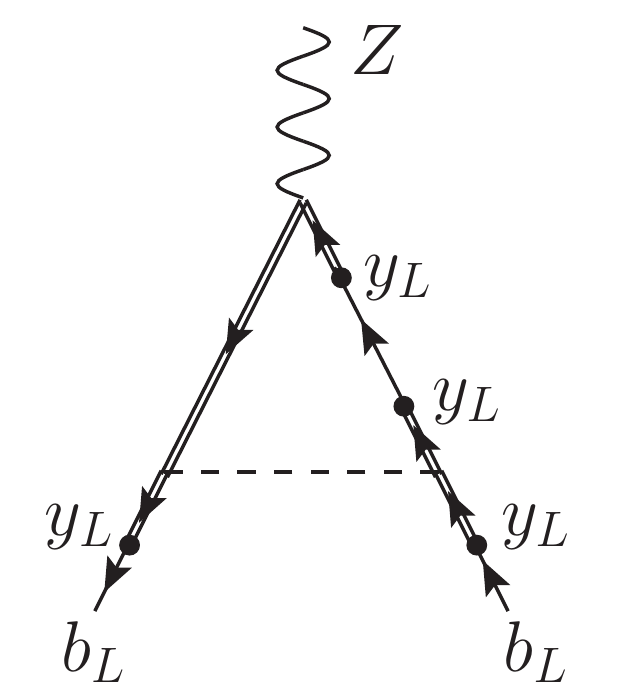}
\hspace{5em}
\includegraphics[height=.25\textwidth]{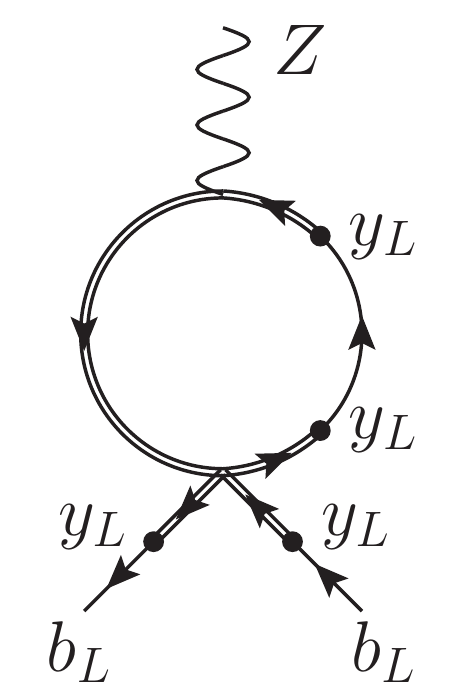}
\caption{Schematic structure of fermion loop diagrams contributing to the
$Z\overline b_L b_L$ vertex with insertions of the $y_L$ couplings on the external fermion legs.}
\label{fig:Zbb_y_insertions}
\end{figure}

If the four $y_L$ insertions are all inside the loop the corresponding contribution to
the $Z\overline b_L b_L$ vertex is finite. This necessarily happens in the
case in which only a singlet is present in the effective theory. Instead, if
a $4$-plet is also present, two $y_L$ insertions can be on the external legs. In this case the
two ``external'' insertions do not influence the degree of divergence and a logarithmically
divergent contribution can be present. Examples of diagrams that could lead to
this kind of corrections are shown in fig.~\ref{fig:Zbb_y_insertions}.

In our effective theory a further subtlety is present which partially protects the
$Z\overline b_L b_L$ vertex. The structure of the elementary--composite mixings implies
the presence of a selection rule that forbids logarithmically divergent corrections coming
from a large class of diagrams.
As we will see the only diagrams that can lead to a divergent contribution are a subset
of the ``bubble''-type diagrams (see the diagram on the right of fig.~\ref{fig:Zbb_y_insertions}),
so that this kind of correction is necessarily related to the presence of $4$-fermion operators.

To understand the origin of the selection rule we can analyze the ``triangle''-type diagrams
with $y_L$ insertions on the external legs shown on the left of fig.~\ref{fig:Zbb_y_insertions}.
The external $b_L$'s are both mixed with the $B_L$ state coming from $\psi_4$.
In order to generate a divergence the vertices containing a Goldstone boson must
also contain a power of the momentum, that is they must be of the type
$\partial_\mu \phi \overline \psi_L \gamma^\mu \psi_L$, where we generically denote by $\phi$ the
Goldstone field and by $\psi$ the composite fermions.~\footnote{In our effective Lagrangian
vertices of this kind are generated by the $d_\mu$ symbol term.} The structure of the vertex
implies that the composite fermions that enter in the loop must be necessarily left-handed.
But the left-handed composite fermions in the leading order Lagrangian mix with the
elementary states only through $y_R$. As a consequence in order to generate
a triangle diagram of this type some $y_R$ or some composite mass insertions
are needed in addition to the $y_L$
mixings and this lowers the degree of divergence making the diagrams finite.

The only diagrams that can give rise to a logarithmic divergence are the ``bubble'' ones
shown on the right of fig.~\ref{fig:Zbb_y_insertions}. They of course crucially depend
on the presence of $4$-fermion operators in the effective Lagrangian. Two types of $4$-fermion
vertices can generate a diagram that contributes to $\delta g_{b_L}$. The first type of vertex
has the form
\begin{equation}\label{eq:4-ferm_l}
{\cal O}^{4-ferm}_L = \frac{e_L}{f^2} (\overline B_L \gamma^\mu B_L)(\overline {\cal T}_L \gamma_\mu {\cal T}_L)\,,
\end{equation}
where by ${\cal T}$ we denote any composite state with charge $2/3$. For shortness
in eq.~(\ref{eq:4-ferm_l}) we did not specify the color structure which is not relevant for
the present discussion. By adapting the previous analysis of the ``triangle'' diagrams,
it is straightforward to show that the ``bubble'' diagrams
with the vertex in eq.~(\ref{eq:4-ferm_l}) are also protected by the selection rule,
so that they are finite.
The second type of $4$-fermion vertex is of the form
\begin{equation}\label{eq:4-ferm_r}
{\cal O}^{4-ferm}_R = \frac{e_R}{f^2} (\overline B_L \gamma^\mu B_L)(\overline {\cal T}_R \gamma_\mu {\cal T}_R)\,.
\end{equation}
In this case the selection rule is violated because the ${\cal T}_R$ fields can clearly
mix with the $q_L$ doublet through $y_L$. This class of vertices, as we will show with
an explicit calculation, gives rise to a logarithmically divergent contribution to
the $Z\overline b_L b_L$ vertex.

Of course in our effective Lagrangian higher-order mixing terms between the elementary
and the composite states can in general be present. An example of such operators is
a kinetic mixing between the $q_L$ doublet and the composite $4$-plet:
$y_{L} f/m_* \left(\overline q_L^{\bf 5}\right)^I U_{Ii} \Dslash \psi^i_{4L} + \textrm{h.c.}$.
A term like this would induce a correction to the $Z\overline b_L b_L$ vertex through diagrams
analogous to the ``triangle'' ones we considered before. Such a diagram would be
superficially quadratically divergent (the kinetic higher-order mixing gives an extra power
of the momentum). However the coefficient of the kinetic mixing, following our
power counting in eq.~(\ref{eq:power-counting}), is suppressed by the UV cut-off, $m_*$, so that
the final contribution is finite. Even though these diagrams can not give
a logarithmically divergent contribution, they induce a correction that is not
suppressed by powers of the cut-off, thus they can contribute at leading order to
the $Z\overline b_L b_L$ vertex.

Notice that the presence of unsuppressed contributions of this kind also implies a
non-decoupling of the fermionic resonances. Even if we send the mass of a resonance to
the cut-off, it can generate a higher-order effective operator in the low-energy Lagrangian
that breaks the selection rule and gives a sizable contribution to the $Z\overline b_L b_L$ vertex.
We will discuss an example of this effect in the next section.

The above discussion clearly shows that, even in the absence of logarithmically
divergent contributions, the $Z\overline b_L b_L$ vertex is highly sensitive to the UV
dynamics of the theory and can be reliably computed in a low-energy effective approach
only if the logarithmically divergent contributions dominate or
if we assume that the contributions coming from the UV dynamics are (accidentally) suppressed.

To conclude the general analysis of the $Z\overline b_L b_L$ vertex corrections we
derive an estimate of the size of the contribution due to the fermion loops.
The logarithmically divergent contribution can be estimated as
\begin{equation}\label{eq:div_Zbb}
\frac{\delta g_{b_L}}{g^{SM}_{b_L}} \simeq \frac{y_L^2}{16 \pi^2}
\frac{y_{L4}^2 f^2}{m_4^2 + y_{L4}^2 f^2} \xi \log\left(\frac{m_*^2}{m_4^2}\right)\,.
\end{equation}
Notice that we explicitly included a factor $y_{L4}^2 f^2/(m_4^2 + y_{L4}^2 f^2)$, which corresponds
to the mixings between the $b_L$ and the $B_L$ that appears in the external legs of the
logarithmically divergent diagrams. Using the relation between $y_{L,R}$ and the top Yukawa we get
\begin{equation}\label{eq:div_Zbb_est}
\frac{\delta g_{b_L}}{g^{SM}_{b_L}} \simeq \frac{y_t^2}{16 \pi^2} \xi
\log\left(\frac{m_*^2}{m_4^2}\right)
\simeq 2 \cdot 10^{-2}\, \xi\,,
\end{equation}
where for the numerical estimate we set $m_* \simeq 3\ {\rm TeV}$ and $m_4 \simeq 700\ {\rm GeV}$.
In the case in which the logarithmically divergent contribution is not present or is suppressed
the estimate becomes
\begin{equation}\label{eq:UV_Zbb}
\frac{\delta g_{b_L}}{g^{SM}_{b_L}} \simeq \frac{y_L^2}{16 \pi^2}
\frac{y_L^2 f^2}{m^2} \xi
\simeq \frac{y_t^2}{16 \pi^2} \xi
\simeq 6 \cdot 10^{-3}\, \xi\,,
\end{equation}
with $m$ the mass of the lightest top partner.

The corrections in eqs.~(\ref{eq:div_Zbb}) and (\ref{eq:UV_Zbb}) are typically
larger than the tree-level contribution generated at non zero momentum
given in eq.~(\ref{eq:deltag_p}). This is especially true if the mass of the resonances
is not too small, $m \gtrsim f$, and the strong coupling is large, $g_* \gtrsim 5$.
The corrections due to the bottom partners estimated in eq.~(\ref{eq:deltag_b_partners_1})
can in principle be comparable to the ones coming from fermion loops if the scale of
the bottom partner is relatively small $m_B \sim f$. These corrections crucially depend on the
quantum numbers of the bottom partners. In minimal scenarios (bottom partners in
the fundamental representation of $\SO(5)$) they are positive and some cancellation
seems required to pass the present bounds. For simplicity, in our explicit analysis
we will neglect both tree-level corrections.

\subsection{Symmetries in the effective Lagrangian}\label{sec:symm_eff_lagr}

As we saw in the analysis of the $\widehat S$ parameter the divergent contributions
coming from fermion loops are finite if the relation $c^2 = 1/2$ holds.
We want now to study our effective Lagrangian in this case and understand the
origin of the protection of the EW parameters. For definiteness we will focus on
the case $c = 1/\sqrt{2}$ and we will comment at the end on the other possibility $c= -1/\sqrt{2}$.

Let us start with the Lagrangian for the composite fields given in eq.~(\ref{eq:lagr_comp}).
A straightforward computation shows that the leading order terms in the case $c = 1/\sqrt{2}$ can
be simply rewritten as
\begin{equation}\label{eq:Lagr_comp_c}
{\cal L}_{comp}^{c=1/\sqrt{2}} = i \overline (\Psi U^\dagger) \gamma^\mu (\partial_\mu - i g A_\mu)
(U \Psi) - m_4 \overline \Psi \Psi - (m_1 - m_4) \overline \Psi_5 \Psi_5\,,
\end{equation}
where we introduced the $5$-plet
\begin{equation}
\Psi =
\left(
\begin{array}{c}
\psi_4\\
\psi_1
\end{array}
\right)
\end{equation}
and we denoted by $\Psi_5$ the fifth component of $\Psi$, namely $\Psi_5 = \psi_1$,
while $A_\mu$ represents the elementary gauge fields in a compact notation. A simple field
redefinition, $\Psi \rightarrow \Psi' \equiv U^\dagger \Psi$, shows that the only dependence on the
Goldstone fields in the composite fermion Lagrangian is associated to the mass term
\begin{equation}
{\cal L}_{comp}^{c=1/\sqrt{2}} \supset -(m_1 - m_4) (\overline \Psi' U)_5 (U^\dagger \Psi')_5\,,
\end{equation}
which gives the mass splitting between the $4$-plet and the singlet. Notice that this property
is a consequence of our choice of $c$, in the general Lagrangian the dependence on the Goldstones
in the kinetic terms of the composite fields can not be removed. It is clear that, if $m_1 = m_4$,
in the composite sector Lagrangian an additional $\SO(5)$ symmetry is present, which allows
us to remove the Higgs VEV.

With the same redefinition of the composite fields the Lagrangian for the elementary
states in eq.~(\ref{eq:lagr_elem}) becomes
\begin{eqnarray}
{\cal L}_{elem}^{c=1/\sqrt{2}} &=& i \overline q_L \Dslash q_L + i \overline t_R \Dslash t_R\nonumber\\
&&+\, y_{L4} f \overline q_L^{{\bf 5}} \Psi'
+ (y_{L1} - y_{L4}) f \left(\overline q_L^{{\bf 5}} U\right)_5 (U^\dagger \Psi')_5\nonumber\\
&&+\, y_{R4} f \overline t_R^{{\bf 5}} \Psi'
+ (y_{R1} - y_{R4}) f \left(\overline t_R^{{\bf 5}} U\right)_5 (U^\dagger \Psi')_5 + \textrm{h.c.}\,.
\label{eq:Lagr_elem_c}
\end{eqnarray}
The Goldstones in this case appear only in association with the $(y_{L1} - y_{L4})f$
and $(y_{R1} - y_{R4})f$ mass mixings.

From the structure of the Lagrangian in eqs.~(\ref{eq:Lagr_comp_c}) and (\ref{eq:Lagr_elem_c})
we can simply understand why no divergence arises in the fermion contribution to $\widehat S$.
In order to generate an effect that feels EWSB the corresponding operator must necessarily
include some insertions of the Lagrangian terms containing the Goldstones.
For our choice of $c$ the Goldstones are always associated to mass operators
and any insertion leads to a reduction of the degree of divergence. The $\widehat S$
parameter is naively logarithmically divergent at one loop, thus the extra mass
insertions make it finite.

A similar protection mechanism is also present for the fermion corrections to the
$Z\overline b_L b_L$ vertex. In the case in which $y_{L1} = y_{L4}$ the remaining
$y_{L4} f \overline q^{\bf 5}_L \Psi'$ mixing is independent of the Goldstones.
The only operators containing the $U$ matrix
are the $(m_1 - m_4)$ mass term and the $(y_{R1} - y_{R4})f$ mixing.
In order to generate a correction to $g_{b_L}$ some insertions of these operators are needed
in addition to the four insertions of $y_{L4}$. These extra mass insertions make the
corrections to the $Z\overline b_L b_L$ vertex finite.

A similar structure of the effective Lagrangian is also present if $c = -1/\sqrt{2}$.
This case can be connected to the one we discussed with the redefinitions
$\psi_1 \rightarrow - \psi_1$, $y_{L,R1} \rightarrow -y_{L,R1}$, which just reverse the sign of $c$.

A particular implementation of our effective Lagrangian with $c = 1/\sqrt{2}$ has been studied in
Ref.~\cite{Anastasiou:2009rv}. In this work the additional relations $y_{L4} = y_{L1}$
and $y_{R4} = y_{R1}$ are assumed. In this particular case the only dependence on the Goldstones
comes from the mass splitting term between the composite $4$-plet and the singlet.
The explicit computation of the fermion corrections to the $Z\overline b_L b_L$ vertex presented
in Ref.~\cite{Anastasiou:2009rv} shows that the new physics contributions are finite,
in agreement with the results of our analysis.

\section{Results in explicit models}\label{sec:models}

After the general analysis presented in the previous section,
we now focus on a more detailed study of the corrections to the EW precision parameters
in some explicit scenarios. First of all we will consider the simplified set-ups in which
only one light composite multiplet is present in the effective theory. Afterwards
we will study two more complete models containing a composite $4$-plet as well as a singlet.

The analysis of explicit scenarios is of course essential to obtain a reliable quantitative
determination of the constraints coming from the EW precision data. Moreover
it allows to check the validity of the general results derived in the
previous section.

In all our numerical results we fix the top mass to the value $m_t = m_t^{\overline{MS}}(2\ {\rm TeV}) = 150\ {\rm GeV}$,
which corresponds to the pole mass $m_t^{pole} = 173\ {\rm GeV}$. Moreover, to estimate the constraints
from the oblique parameters, we chose a cut-off scale $m_* = 3\ {\rm TeV}$.

\subsection{The case of a light singlet}\label{sec:singlet}

As a first example we consider the case in which only a light composite singlet
is present in the effective theory. The effective Lagrangian for this set-up can be
easily read from the general one of section~\ref{sec:the_model} by removing
the terms containing $\psi_4$. In this configuration the resonance spectrum contains
only one composite state, the $\widetilde T$, which has the same electric
charge as the top and a mass
\begin{equation}
m_{\widetilde T}^2 = m_1^2 + y_{R1}^2 f^2\,.
\end{equation}

We start our analysis by considering the corrections to the $\widehat S$ parameter.
In the general analysis we saw that the fermion contributions to $\widehat S$ can
diverge only if the spectrum contains a light $4$-plet, thus in our present set-up
we expect a finite result.
In fact at leading order in the $v/f$ expansion we find that the one-loop fermion contribution is given by
\begin{equation}\label{eq:Ssing}
\Delta \widehat S_{ferm} = \frac{g^2}{192 \pi^2} \xi \frac{m_1^2 y_{L1}^2 f^2}{(m_1^2 + y_{R1}^2 f^2)^2}
\left[-5 + 2 \log\left(\frac{2 (m_1^2 + y_{R1}^2 f^2)^2}{v^2 y_{L1}^2 y_{R1}^2 f^2}\right)\right]\,.
\end{equation}
Notice that the argument of the logarithm can be identified with the ratio between
the mass of the heavy fermion resonance $m_{\widetilde T}$ and the top mass.
\begin{equation}
m_t^2 \simeq \frac{v^2\, y_{L1}^2 y_{R1}^2 f^2}{2 (m_1^2 + y_{R1}^2 f^2)}\,.
\end{equation}
For typical values of the parameters,
$y_{L1} \sim y_{R1} \sim 1$, $m_1 \lesssim 1\ {\rm TeV}$ and $\xi \lesssim 0.2$,
the contribution in eq.~(\ref{eq:Ssing}) is positive and small,
$\Delta \widehat S_{ferm} \lesssim 10^{-4}$.

As we discussed in section~\ref{sec:general_analysis}, although the correction to $\widehat S$
coming from the low-energy dynamics is calculable,
large uncalculable UV contributions can be present. Even if we assume that the tree-level
effects given in eq.~(\ref{eq:S_UV}) are negligible, the loop contributions
coming from the UV dynamics (see the estimate in eq.~(\ref{eq:S_UV_loop}))
are typically dominant with respect to the corrections in eq.~(\ref{eq:Ssing}).
We can check that the UV effects can be important by slightly
modifying our explicit computation.
We consider an effective theory in which a composite $4$-plet is present as well as a singlet.
In order to recover the case with only a light singlet, we then take the limit in which the $4$-plet
mass is sent to the cut-off $m_*$.
To ensure that $\widehat S$ is calculable in the effective theory we set $c^2 = 1/2$.
The explicit computation of $\Delta \widehat S$ leads to the result in
eq.~(\ref{eq:Ssing}) plus an additional shift which,
at the leading order in an expansion in the cut-off, is given by
\begin{equation}\label{eq:S_4plet_UV}
\Delta \widehat S_{ferm}^{UV} = - \frac{g^2}{24 \pi^2} \xi \simeq -1.8 \cdot 10^{-3}\, \xi\,.
\end{equation}
As expected, the $4$-plet does not decouple in the limit in which it becomes heavy.
The UV corrections in eq.~(\ref{eq:S_4plet_UV})
have a size compatible with our estimate in eq.~(\ref{eq:S_UV_loop}) and are typically larger than
the singlet contribution in eq.~(\ref{eq:Ssing}).
Notice that the result in eq.~(\ref{eq:S_4plet_UV}) gives only an example of possible UV
effects and should not be thought as a complete determination of the UV contributions.
In order to properly compute the total shift in $\widehat S$ the whole UV completion of the
model should be taken into account.

\begin{figure}
\centering
\includegraphics[width=.4\textwidth]{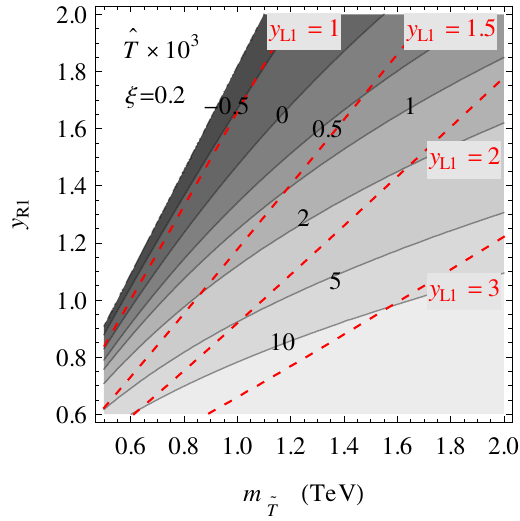}
\caption{Corrections to the $\widehat T$ parameter as a function of the singlet mass
$m_{\widetilde T}$ and of the $y_{R1}$ mixing. The result corresponds to
the case with only a light singlet and includes the contribution due to the Higgs
non-linear dynamics in eq.~(\ref{eq:T_IR}) and the exact fermion one-loop correction.
The compositeness scale has been fixed to the value $\xi = 0.2$.
The red dashed lines correspond to the contours with fixed $y_{L1}$.}
\label{fig:T_hat_singlet}
\end{figure}

Let us now consider the $\widehat T$ parameter. As shown in the general analysis,
the fermion corrections are finite and saturated by the low-energy contributions.
The explicit calculation gives the following result at leading order in $v/f$:
\begin{equation}\label{eq:Tsing}
\Delta \widehat T_{ferm} = \frac{3\, \xi}{64 \pi^2} \frac{y_{L1}^4 m_1^2 f^2}{(m_1^2 + y_{R1}^2 f^2)^3}
\left\{m_1^2 + 2 y_{R1}^2 f^2 \left[
\log\left(\frac{2 (m_1^2 + y_{R1}^2 f^2)^2}{v^2 y_{L1}^2 y_{R1}^2 f^2}\right) - 1\right]\right\}\,.
\end{equation}
This contribution is positive and, in a large part of the parameter space, can compensate the
negative shift that comes from the non-linear Higgs dynamics (see eq.~(\ref{eq:T_IR})).
In the points in which $y_{L1} \sim y_{R1} \sim 1$, the estimate given in eq.~(\ref{eq:T_ferm_est1})
is approximately valid. The total shift in $\widehat T$ is shown in fig.~\ref{fig:T_hat_singlet}
for the reference value $\xi = 0.2$, corresponding to $f = 550\ {\rm GeV}$. It can be seen
that sizable positive values of $\Delta \widehat T$ can easily be obtained for
reasonable values of the singlet mass and of the elementary--composite mixings.

Finally we analyze the corrections to the $Z \overline b_L b_L$ vertex. We showed in
section~\ref{sec:general_analysis} that in the case with only a light singlet
the one-loop fermion corrections to this observable are finite. The absence of a $4$-plet
also implies that additional contributions coming from $4$-fermion operators and from the
UV dynamics are suppressed by the cut-off scale and can be expected to be negligible. At leading order in
$v/f$ we find that the shift in $g_{b_L}$ is given by
\begin{equation}\label{eq:Zbbsing}
\delta g_{b_L} = \frac{\xi}{64 \pi^2} \frac{y_{L1}^4 m_1^2 f^2}{(m_1^2 + y_{R1}^2 f^2)^3}
\left\{m_1^2 + 2 y_{R1}^2 f^2 \left[\log\left(\frac{2 (m_1^2 + y_{R1}^2 f^2)^2}{v^2 y_{L1}^2 y_{R1}^2 f^2}\right) - 1\right]\right\}\,.
\end{equation}
Comparing this result with the fermion contribution to $\widehat T$ in eq.~(\ref{eq:Tsing})
we can notice that a strict relation exists between the two quantities
$\Delta \widehat T_{ferm} = 3 \delta g_{b_L}$.~\footnote{This relation was already
noticed in Refs.~\cite{Barbieri:2007bh,Gillioz:2008hs}.} In particular the positive
correction to $\widehat T$ is related to a corresponding positive shift in $g_{b_L}$.
For the typical size of the fermion contribution to $\widehat T$ needed
to satisfy the experimental bounds, $1 \cdot 10^{-3} < \Delta \widehat T < 2 \cdot 10^{-3}$,
a moderate contribution to $\delta g_{b_L}$ is found: $g_{b_L}$: $0.33 \cdot 10^{-3} < \delta g_{b_L} < 0.66 \cdot 10^{-3}$.
As we already discussed (see fig.~\ref{fig:deltag_plane}),
the experimental measurements disfavor a positive contribution to the $Z\overline b_L b_L$
coupling. Thus the scenario with only a light singlet
tends to be in worse agreement with the EW precision data than the SM.

On the other hand, if we neglect the constraints on $\delta g_{b_L}$ and only consider the bounds on the oblique EW
parameters, it is not hard to satisfy the experimental constraints even for sizable values of $\xi$.

\subsection{The case of a light $4$-plet}\label{sec:4plet}

As a second simplified scenario we consider the case in which the resonance spectrum contains
only a light $4$-plet. The general analysis of section~\ref{sec:general_analysis} showed
that in this case only $\widehat T$ receives a finite contribution from fermion loops, whereas
the corrections to the $\widehat S$ parameter and to the $Z \overline b_L b_L$ vertex
are logarithmically divergent.~\footnote{The corrections to the $\widehat T$ parameter
and to the $Z \overline b_L b_L$ vertex in this set-up have been studied also in
Ref.~\cite{Gillioz:2008hs}. The results for $\widehat T$ are similar to the ones we find.
The results for the $Z \overline b_L b_L$ corrections are also in agreement with ours if we
exclude the contributions from $4$-fermion operators
which are not included in the analysis of Ref.~\cite{Gillioz:2008hs}.}

Before discussing in details the contributions to the EW parameter, we analyze the spectrum of the
resonances. The $4$-plet gives rise to two ${\rm SU}(2)_L$ doublets with hypercharges
$1/6$ and $7/6$. The ${\bf 2}_{1/6}$ doublet contains a top partner $T$ and a bottom partner
$B$, while the ${\bf 2}_{7/6}$ doublet contains an exotic state with charge $5/3$ ($X_{5/3}$)
and a top resonance ($X_{2/3}$).
The mixing with the elementary states induces a mass splitting between the two doublets.
The states inside each doublet, instead, receive only a small splitting due to EWSB effects
and are nearly degenerate in mass.
In particular the $B$ and $X_{5/3}$ states are not coupled
to the Higgs and their masses do not receive corrections after EWSB. The masses of the
composite resonances are given by
\begin{equation}
m^2_{X_{2/3}} \simeq m_{X_{5/3}}^2 = m_4^2\qquad {\rm and}
\qquad m^2_{T} \simeq m_{B}^2 = m_4^2 + y_{L4}^2 f^2\,.
\end{equation}
The top mass at the leading order in $v/f$ is given by
\begin{equation}\label{eq:topmass_4plet}
m_t^2 \simeq \frac{v^2\, y_{L4}^2 y_{R4}^2 f^2}{2 (m_4^2 + y_{L4}^2 f^2)}\,.
\end{equation}

The dominant contribution to the $\widehat S$ parameter comes from the logarithmically enhanced
corrections due to loops of fermion resonances. The explicit result
can be obtained from eq.~(\ref{eq:S_div_explicit}) by setting $c=0$:~\footnote{The same
result can be obtained with the following equivalent procedure.
We consider an effective theory containing a $4$-plet and a singlet with $c^2 = 1/2$.
In this case the fermion contribution to $\widehat S$ is finite and calculable.
The explicit computation shows that a contribution of the form $g^2/(8 \pi^2) \xi \log(m_1^2/m_4^2)$
is present. In the limit in which the singlet becomes heavy, $m_1 \rightarrow m_*$,
we recover, as expected, the contribution in eq.~(\ref{eq:S_div_4plet}).}
\begin{equation}\label{eq:S_div_4plet}
\Delta \widehat S_{ferm} = \frac{g^2}{8 \pi^2} \xi \log\left(\frac{m_*^2}{m_4^2}\right)
\simeq 1.6 \cdot 10^{-2}\,\xi\,,
\end{equation}
where the numerical estimate has been obtained by setting $m_4 \simeq 700\ {\rm GeV}$
and $m_* \simeq 3\ {\rm TeV}$. If the gauge resonances are heavy, $m_*/f = g_* \gtrsim 4$,
the correction in eq.~(\ref{eq:S_div_4plet}) is
comparable or even larger than the tree-level one in eq.~(\ref{eq:S_UV}).

\begin{figure}
\centering
\includegraphics[width=.4\textwidth]{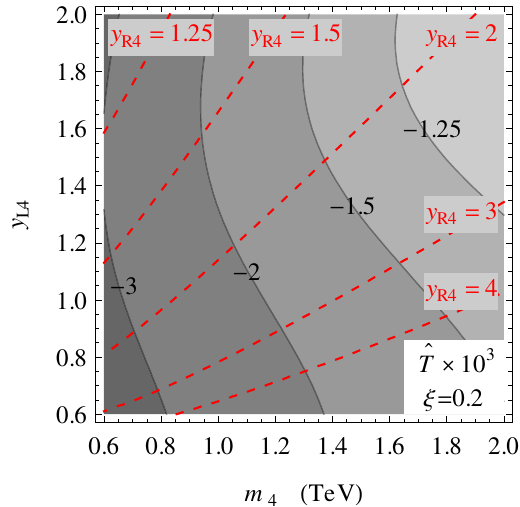}
\caption{Corrections to the $\widehat T$ parameter as a function of the mass
parameter $m_4$ and of the $y_{L4}$ mixing. The result corresponds to
the case with only a light $4$-plet and includes the contribution due to the Higgs
non-linear dynamics in eq.~(\ref{eq:T_IR}) and the exact fermion one-loop correction.
The compositeness scale has been fixed to the value $\xi = 0.2$.
The red dashed lines correspond to the contours with fixed $y_{R4}$.}
\label{fig:T_hat_4plet}
\end{figure}
The sizable positive contribution to the $\widehat S$ parameter implies
a quite stringent bound on the compositeness scale, $\xi \lesssim 0.1$ (see fig.~\ref{fig:ST_plane}).
An even stronger constraint is obtained if we also
consider the corrections to the $\widehat T$ parameter. The full expression of the
fermion contributions at leading order in $v/f$ is in this case too involved and
does not give useful insights, so we only report here the leading term in the $y$ expansion:
\begin{equation}\label{eq:T_4plet_leading}
\Delta \widehat T_{ferm} = - \frac{\xi}{32 \pi^2} \frac{y_{L4}^4 f^2}{m_4^2}\,.
\end{equation}
The approximate result suggests that the shift in $\widehat T$ is negative.
This conclusion is typically correct and has been explicitly verified
with a numerical computation. The main contributions to $\widehat T$ coming from the
non-linear Higgs dynamics (see eq.~(\ref{eq:T_IR})) and from fermion loops are shown in fig.~\ref{fig:T_hat_4plet} for $\xi = 0.2$.
Similar results are obtained for different values of $\xi$. Notice that
the leading order expression in eq.~(\ref{eq:T_4plet_leading}) capture only the
overall size of the fermion contributions. The exact result can deviate from the estimate
at order one especially in the parameter space region in which $y_{R4}$ becomes large.

The fact that the shift in $\widehat T$ is necessarily negative
makes the constraints coming from the oblique parameters extremely severe.
Using the results in fig.~\ref{fig:ST_plane} an upper bound
$\xi \lesssim 0.02$ at the $99\%$ confidence level is obtained, which corresponds
to a lower bound $f \gtrsim 1.7\ {\rm TeV}$.

Although the configuration with only a light $4$-plet is strongly disfavored by the
large corrections to the oblique parameters, it is still worth discussing the
form of the corrections to the $Z \overline b_L b_L$ vertex. The explicit computation
will be useful to verify the results obtained in our general analysis in section~\ref{sec:general_analysis}.

We start by considering the contributions related to the leading-order terms in the
effective Lagrangian. If we neglect the effects coming from higher-dimensional operators
and from $4$-fermion contact interactions, we get the following
corrections to the $Z \overline b_L b_L$ vertex at the leading order in the $v/f$ expansion:
\begin{eqnarray}
\delta g^{4-plet}_{b_L} &=& -\frac{\xi}{32 \pi^2} \frac{y_{L4}^2 y_{R4}^2 f^2}{m_4^2 + y_{L4}^2 f^2}\Bigg[\frac{y_{L4}^2 f^2}{m_4^2 + y_{L4}^2 f^2} + \left(1 - \frac{y_{R4}^2 f^2}{4 m_4^2}\right)
\log\left(1 + \frac{y_{L4}^2f^2}{m_4^2}\right)\nonumber\\
&& - y_{L4}^2 f^2\frac{4 m_4^2(m_4^2 + y_{L4}^2 f^2) - (2 m_4^2 + y_{L4}^2 f^2) y_{R4}^2 f^2}{4 m_4^2 (m_4^2 + y_{L4}^2 f^2)^2}
\log\left(\frac{2 (m_4^2 + y_{L4}^2 f^2)^2}{v^2 y_{L4}^2 y_{R4}^2 f^2}\right)\Bigg]\,.\label{eq:Zbb4plet}
\end{eqnarray}
As expected, due to the selection rule discussed in subsection~\ref{sec:Zbb_general},
the fermion contribution to the $g_{b_L}$ coupling is finite.

If higher-order operators and in particular higher-order mixings between the elementary
and the composite states are present in the effective Lagrangian, the selection rule
can be violated and sizable corrections to the result in eq.~(\ref{eq:Zbb4plet}) can arise.
This is a signal of the fact that the $Z \overline b_L b_L$ vertex is sensitive to
the UV dynamics of the theory. To explicitly verify this property we can use a procedure
analogous to the one we adopted for the $\widehat S$ parameter in the case with only
a light singlet. We consider a theory with a $4$-plet as well as a singlet and then we recover
the configuration with only a light $4$-plet by taking the limit in which the singlet mass
goes to the cut-off $m_*$. Using this procedure we find that the fermion correction
to the $Z \overline b_L b_L$ vertex contains an additional contribution with respect to
the result in eq.~(\ref{eq:Zbb4plet}):
\begin{equation}\label{eq:Zbb_4plet_add}
\delta g_{b_L} = \delta g^{4-plet}_{b_L} + \frac{\xi}{32 \pi^2} \frac{y_{L4}^2 f^2}{m_4^2 + y_{L4}^2 f^2} c^2y_{L1}\left(y_{L1} - \sqrt{2} c y_{L4}\right)\,.
\end{equation}
The additional contribution arises at leading order in the $y$ expansion and is independent
of the singlet mass, it only depends on the mixing of the singlet with the elementary
states $y_{L1}$.

An equivalent way to understand the non-decoupling of the singlet is the
following. In the limit in which the singlet becomes heavy we can integrate it out from
the effective theory. This procedure generates a set of higher-order operators, in particular
it gives rise to a term of the form $(y_{L1} c/m_{*}) (q_L^{\bf 5} U)_5 \gamma^\mu d_\mu^i \psi_4^i
+ {\rm h.c.}$, where we replaced the singlet mass by the cut-off $m_*$.
This higher-order mixing couples the $q_L$ doublet with the left-handed component of the
composite $4$-plet and induces a breaking of the $Z \overline b_L b_L$ selection rule, as can
be easily inferred from the discussion in subsection~\ref{sec:Zbb_general}.

Notice that in the case in which $c=0$ the higher-dimension operators are not generated
by integrating out the singlet, thus the selection rule is still unbroken and the additional
correction to the $Z \overline b_L b_L$ vertex in eq.~(\ref{eq:Zbb_4plet_add}) vanishes.
There is also a second case in which the additional corrections are not there. As we saw
in subsection~\ref{sec:symm_eff_lagr}, if $c = \pm 1/\sqrt{2}$ and $y_{L1} = \pm y_{L4}$
the low-energy theory acquires an extra symmetry which protects the EW observables.
In this case we expect the decoupling of the heavy dynamics to occur and, in fact, the
extra correction in eq.~(\ref{eq:Zbb_4plet_add}) exactly cancels.

To conclude the analysis of the case with only a light $4$-plet we now consider the effects
due to the $4$-fermion contact operators. As expected, vertices of the form
given in eq.~(\ref{eq:4-ferm_l}) induce a finite correction to the $Z \overline b_L b_L$
vertex:
\begin{eqnarray}
\delta g_{b_L}^{4-ferm}  &=& \frac{3 e_{L4} \xi y_{L4}^2 f^2}{64 \pi^2 (m_4^2 + y_{L4}^2 f^2)^3}
\Bigg\{m_4^2 y_{L4}^2 (m_4^2 + y_{L4}^2 f^2 - 4 y_{R4}^2 f^2)\nonumber\\
&&+\,2 y_{R4}^2 \Bigg[ (m_4^2 + y_{L4}^2 f^2)^2 \log\left(\frac{m_4^2 + y_{L4}^2 f^2}{m_4^2}\right)
+ y_{L4}^4 f^4 \log\left(\frac{v^2 y_{L4}^2 y_{R4}^2 f^2}{2 (m_4^2 + y_{L4}^2 f^2)^2}\right)\Bigg] \Bigg\}\,.\label{eq:4ferm_1}
\end{eqnarray}
On the other hand, the vertex in eq.~(\ref{eq:4-ferm_r}) induces a logarithmically divergent
contribution:
\begin{equation}\label{eq:4ferm_2}
\delta g_{b_L}^{4-ferm} = \frac{3\, e_{R4}}{32 \pi^2}\xi \frac{y_{L4}^2 f^2}{m_4^2 + y_{L4}^2 f^2}
y_{L4}^2 \log\left(\frac{m_*^2}{m_4^2}\right)\,.
\end{equation}
Notice that the results in eqs.~(\ref{eq:4ferm_1}) and (\ref{eq:4ferm_2})
correspond to the case in which the $4$-fermion vertex has the structure
$(\overline B_L^a \gamma^\mu B^a_L)(\overline{T}^b \gamma_\mu T^b
+ \overline{X}_{2/3}^b \gamma_\mu X_{2/3}^b)$, where
$a$ and $b$ are color indices. Different color structures lead to results that only differ
by group theory factors.~\footnote{The combination of $T$ and $X_{2/3}$ is dictated by the
$P_{LR}$ symmetry which is unbroken in the composite sector.}

The sign of the $4$-fermion contribution crucially depends on the sign of the coefficients $e_{L,R}$.
In our low-energy effective theory $e_{L,R}$ are completely free parameters, thus their sign is
not fixed. From the UV perspective, instead, the operators in eqs.~(\ref{eq:4-ferm_l}) and
(\ref{eq:4-ferm_r}) arise from the exchange of heavy bosonic resonances and the sign of their
coefficients is usually fixed by the quantum numbers of the resonances. It can be
checked that the $e_{L,R}$ coefficients can be generated with arbitrary sign by considering
resonances in different representations of $\SO(4)$.

\subsection{Two complete models}\label{sec:complete_models}

In this subsection we finally consider two more complete models that include both
a $4$-plet and a singlet. In order to reduce the number of parameters
we choose a common value for the left and right elementary mixings: $y_{L4} = y_{L1} = y_L$
and $y_{R4} = y_{R1} = y_R$. In this case the fermion Lagrangian (excluding the
interactions with the gauge fields) becomes equal to the one of the $2$-site model
proposed in Refs.~\cite{Panico:2011pw,Matsedonskyi:2012ym}.

An interesting byproduct of this
choice is the fact that the fermion contribution, which dominates the Higgs potential,
becomes only logarithmically divergent. One renormalization condition is enough to regulate
the divergence and one can fix it by choosing the compositeness scale $f$. In this way the
Higgs mass becomes calculable and an interesting relation between $m_h$
and the masses of the top partners holds~\cite{Matsedonskyi:2012ym}:
\begin{equation}\label{eq:mh_toppartners}
\displaystyle \frac{m_h}{m_t} \simeq \frac{\sqrt{2 N_c}}{\pi} \frac{m_T m_{\widetilde T}}{f}
\sqrt{\frac{\log(m_T/m_{\widetilde T})}{m_T^2 - m_{\widetilde T}^2}}\,,
\end{equation}
where $m_T$ is the mass of the states
in the ${\bf 2}_{1/6}$ doublet coming from the $4$-plet and
$m_{\widetilde T}$ is the mass of the heavy singlet
after the mixing with the elementary states. The complete spectrum of the
composite resonances is a combination of the ones described in the cases
with only one light multiplet considered in the previous subsections.
The complete mass matrix for the charge $2/3$ states is given by
\begin{equation}
M = \left(
\begin{array}{c@{\hspace{.75em}}c@{\hspace{.75em}}c@{\hspace{.75em}}c}
0 & - \frac{1}{2}y_{L4} f (c_h + 1) & \frac{1}{2}y_{L4} f (c_h - 1) & \frac{1}{\sqrt{2}}y_{L1}f s_h\\
\rule{0pt}{1.15em}-\frac{1}{\sqrt{2}}y_{R4} f s_h & m_4 & 0 & 0\\
\rule{0pt}{1.15em}\frac{1}{\sqrt{2}}y_{R4} f s_h & 0 &m_4 & 0\\
\rule{0pt}{1.15em}- y_{R1} f c_h & 0 & 0 & m_1
\end{array}
\right)\,,
\end{equation}
where $c_h \equiv \cos(\langle h \rangle/f)$ and $s_h \equiv \sin(\langle h \rangle/f)$.
The relation in eq.~(\ref{eq:mh_toppartners}) allows us to fix the mass of
one heavy multiplet as a function of the other parameters of the effective Lagrangian.
Another mass parameter can be fixed by the requirement of reproducing the top mass.
At the leading order in the $v/f$ expansion we find that $m_t$ is given by
\begin{equation}
m_t^2 = \frac{v^2 (m_4 - m_1)^2 y_L^2 y_R^2 f^2}{2 (m_4^2 + y_L^2 f^2)(m_1^2 + y_R^2 f^2)}\,.
\end{equation}

Apart from the masses of the composite multiplets and the elementary mixings, only one free
parameter appears in the effective Lagrangian: the coefficient of the $d$-symbol term, $c$.
In the following we will analyze the models obtained for two particular choices of $c$.
The first one is the case $c=0$ which exactly corresponds to the $2$-site model
of Refs.~\cite{Panico:2011pw,Matsedonskyi:2012ym}. The second case corresponds to the
choice $c = 1/\sqrt{2}$ which, as explained in subsection~\ref{sec:symm_eff_lagr},
implies the presence of an additional protection for the EW parameters. This second choice
reproduces the model studied in Ref.~\cite{Anastasiou:2009rv}.

\subsubsection*{The case $c=0$}

We start by considering the $2$-site model ($c=0$). In this case the
leading corrections to the $\widehat S$ parameter are the same as in the case
with only one light $4$-plet. As shown in section~\ref{sec:general_analysis},
the constraints on $\widehat S$ alone are
strong enough to put an absolute upper bound on the compositeness scale $\xi \lesssim 0.1$,
as can be seen from fig.~\ref{fig:xi_bound_S}.

Let us now consider the $\widehat T$ parameter. We can reduce the number of free
parameters by fixing the top and Higgs masses.
The requirement of reproducing the correct Higgs mass gives a relation between
$m_T$ and $m_{\widetilde T}$ (see eq.~(\ref{eq:mh_toppartners})), while fixing
the top mass allows us to determine the right mixing $y_R$ as a function of the
other parameters. With this procedure we are left with only two free parameters,
which we choose to be $m_T$ and the $q_L$ compositeness angle $\phi_L$ defined as
\begin{equation}
\sin \phi_L \equiv \frac{y_L f}{\sqrt{m_4^2 + y_L^2 f^2}}\,.
\end{equation}
Notice that with this procedure the right mixing $y_R$ is determined up to a twofold
ambiguity. In the figures that show the numerical results we will thus include
two plots that correspond to the two choices of $y_R$.

\begin{figure}
\centering
\includegraphics[width=.4\textwidth]{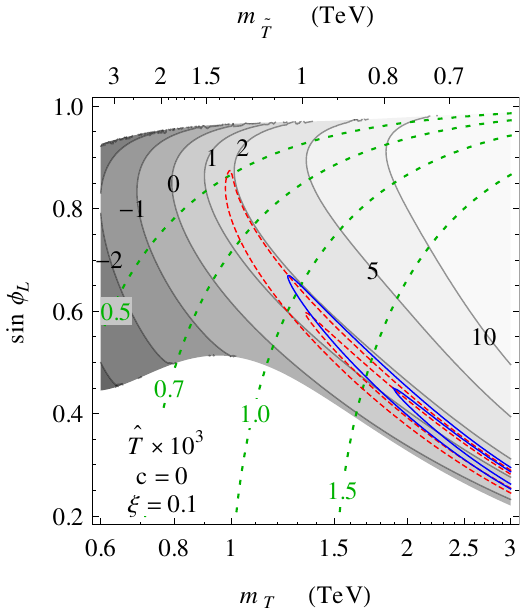}
\hspace{3.em}
\includegraphics[width=.4\textwidth]{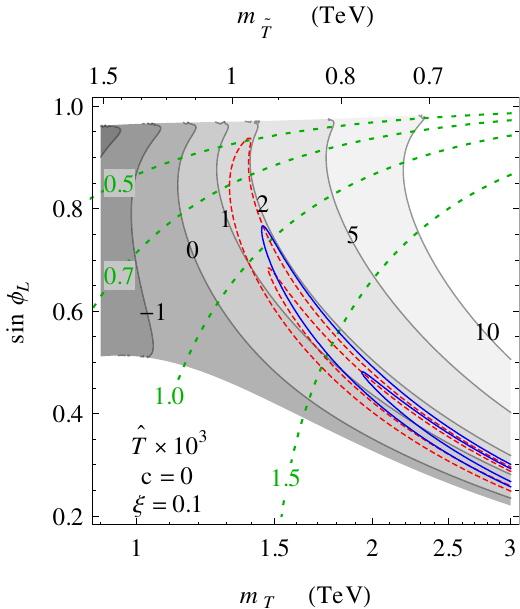}
\caption{Corrections to the $\widehat T$ parameter as a function of the mass
of the top partners and of the $q_L$ compositeness in the model with $c=0$ for $\xi = 0.1$.
The two plots correspond to the two different choices of $y_R$ that allow to obtain the
correct Higgs and top masses at fixed $m_T$ and $\phi_L$ (see the main text for further details).
In the white regions at the top and at the bottom of the plots the Higgs and top masses
can not be reproduced.
The dashed green contours show the mass (in TeV) of the exotic composite state $X_{5/3}$.
The solid blue contours give the regions that pass the
constraints on the oblique parameters at the $68\%$ and $95\%$ confidence level, while
the dashed red lines show how the bounds are modified if we assume a $25\%$
reduction of $\widehat S$.}
\label{fig:T_hat_2site}
\end{figure}

The corrections to the $\widehat T$ parameter are shown in fig.~\ref{fig:T_hat_2site}
for $\xi = 0.1$. To obtain the numerical results we fixed the Higgs
mass to the value $m_h = 126\ {\rm GeV}$.~\footnote{For simplicity we do not take into account
the running of the Higgs mass.}
As expected from the results we discussed in the previous simplified cases,
in the region in which the $4$-plet is the lightest multiplet the
corrections to $\widehat T$ are negative, whereas a light singlet typically implies
a positive shift. The fit of the oblique parameters can put strong bounds on the
parameter space of the model. In the plots we showed the allowed regions for
$68\%$ and $95\%$ confidence level. To obtain the constraints we estimated
$\widehat S$ by adding the leading corrections in eqs.~(\ref{eq:S_UV}), (\ref{eq:S_IR})
and (\ref{eq:S_div_explicit}) for the choice $m_* = 3\ {\rm TeV}$.

The numerical results show that the oblique parameters can be used to set
some lower bounds on the masses of the resonances coming from the composite $4$-plet.
At the $95\%$ confidence level one finds
$m_{X_{2/3}} \simeq m_{X_{5/3}} \gtrsim 0.95\ {\rm TeV}$
for the masses of the exotic doublet ${\bf 2}_{7/6}$ and
$m_{T} \simeq m_{B} \gtrsim 1.2\ {\rm TeV}$ for the ${\bf 2}_{1/6}$ states.
If we assume a $25\%$ cancellation in the corrections to the $\widehat S$ parameter
the bounds are significantly relaxed: $m_{X_{2/3}} \simeq m_{X_{5/3}} \gtrsim 0.5\ {\rm TeV}$
and $m_{T} \simeq m_{B} \gtrsim 1\ {\rm TeV}$.
Notice that these bounds are competitive or even stronger than the ones obtained from direct searches.
For instance the current bounds on the exotic top partners is
$m_{X_{5/3}} \gtrsim 700\ {\rm GeV}$~\cite{Bounds53_CMS,Bounds53_ATLAS}.

Let us finally discuss the corrections to the $Z \overline b_L b_L$ vertex.
The presence of a $4$-plet in the low-energy spectrum makes this observable sensitive
to the UV dynamics of the theory and to possible $4$-fermion interactions present in
the effective Lagrangian. In particular, as discussed in the general analysis of
section~\ref{sec:general_analysis}, logarithmically divergent contributions can arise
from a set of $4$-fermion interactions.

\begin{figure}
\centering
\includegraphics[width=.4\textwidth]{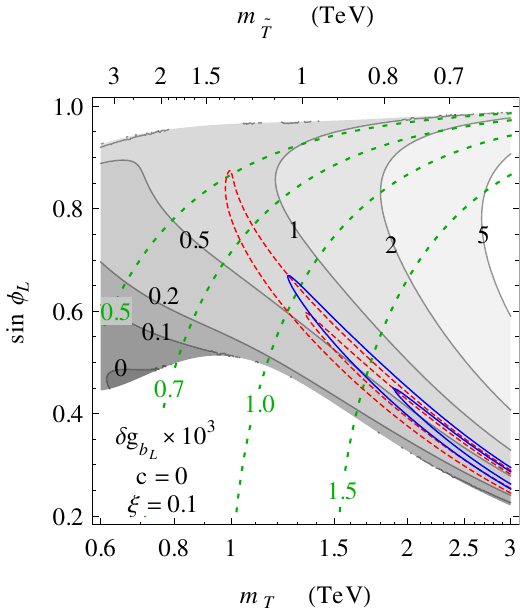}
\hspace{3.em}
\includegraphics[width=.4\textwidth]{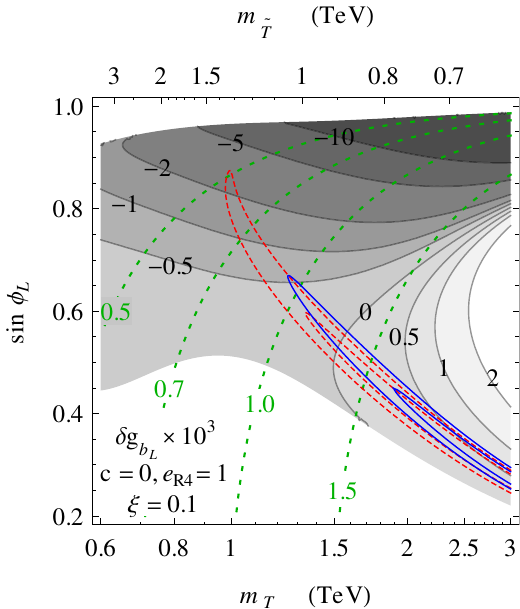}
\caption{Corrections to the $Z\overline b_L b_L$ vertex in the model with $c=0$
for $\xi = 0.1$.
The results on the left panel are obtained by neglecting the UV effects
and the contributions from $4$-fermion operators. On the right panel we added
the logarithmically enhanced contribution induced by the
operator in eq.~(\ref{eq:4ferm_2site}) with $e_{R4} = 1$.
The configurations correspond to the ones chosen for the left plot in
fig.~\ref{fig:T_hat_2site}.}
\label{fig:Zbb_2site}
\end{figure}

If we neglect the UV contributions and set to zero the $4$-fermion operators
we find that the shift in the $Z \overline b_L b_L$ vertex
is positive and somewhat correlated with the corrections to $\widehat T$.
As an example we show in the left panel of fig.~\ref{fig:Zbb_2site} the shift in $g_{b_L}$
for the configurations corresponding to the left plot in fig.~\ref{fig:T_hat_2site}.
One can see that the corrections become typically large and positive in the presence
of a light singlet. The points that pass the constraints on the oblique
parameters have a small positive shift in the $Z \overline b_L b_L$ vertex:
$0.2 \cdot 10^{-3} \lesssim \delta g_{b_L} \lesssim 0.8 \cdot 10^{-3}$.

The UV contributions and the effects of $4$-fermion operators can however drastically
change the above result. In the right panel of fig.~\ref{fig:Zbb_2site} we show
how the previous result changes if we add to the low-energy Lagrangian the interaction
\begin{equation}\label{eq:4ferm_2site}
\frac{e_{R4}}{f^2} \big(\overline B_L^a \gamma^\mu B^a_L\big)\Big(\overline{T}^b_R \gamma_\mu T^b_R
+ \overline{X}_{2/3R}^b \gamma_\mu X_{2/3R}^b\Big)\,,
\end{equation}
with $e_{R4} = 1$. To obtain the numerical result we only included the leading logarithmically
enhanced contribution to $\delta g_{b_L}$ and we set the cut-off to the value $m_* = 3\ {\rm TeV}$.
As expected, the new correction strongly changes the result in the configurations
with large $q_L$ compositeness, whereas the points with small $\phi_L$ are only
marginally affected.

\subsubsection*{The case $c=1/\sqrt{2}$}

The second complete model we consider corresponds to the case $c=1/\sqrt{2}$. In this set-up the
EW observables are finite. In particular the main corrections to the $\widehat S$ parameter
are given by the tree-level UV contributions and by the logarithmically enhanced corrections
due to the non-linear Higgs dynamics. These corrections, for a reasonably high cut-off
($m_* \gtrsim 3\ {\rm TeV}$) are well below the absolute upper bound on $\widehat S$.

\begin{figure}
\centering
\includegraphics[width=.4\textwidth]{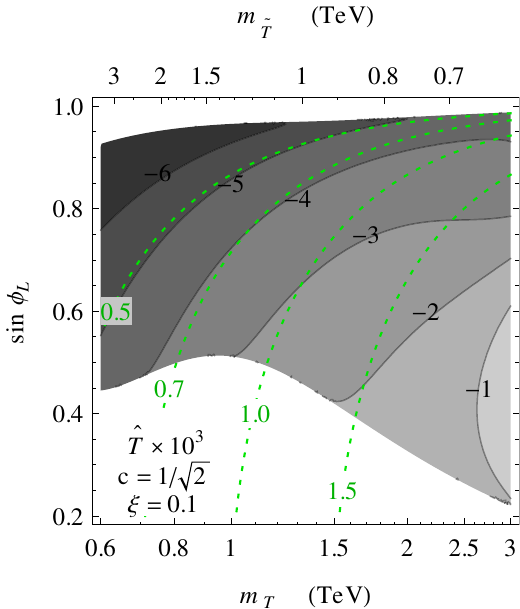}
\hspace{3.em}
\includegraphics[width=.4\textwidth]{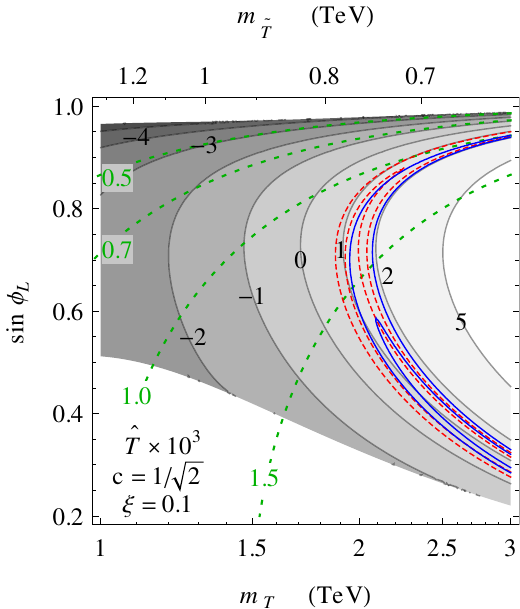}
\caption{Corrections to the $\widehat T$ parameter as a function of the mass
of the top partners and of the $q_L$ compositeness in the model with $c=1/\sqrt{2}$
for $\xi = 0.1$.}
\label{fig:T_hat_c}
\end{figure}

The corrections to the $\widehat T$ parameter are shown in fig.~\ref{fig:T_hat_c}.
The configurations chosen for the plots correspond to the ones we used for the
analogous plots in the case $c=0$ (see fig.~\ref{fig:T_hat_2site}). The results, however,
significantly differ in the two cases. In the case $c=1/\sqrt{2}$ the corrections to $\widehat T$
tend to be more negative and a much lighter singlet is needed in order to pass
the constraints on the oblique parameters ($m_{\widetilde T} \lesssim 0.8\ {\rm TeV}$).
Notice that in this case the constraints are not significantly modified if we assume
that some amount of cancellation in $\widehat S$ is present. Differently from the
case $c=0$, the corrections to $\widehat S$ are small and are typically much below
the absolute upper bound $\widehat S \lesssim 2.5 \cdot 10^{-3}$.

\begin{figure}
\centering
\includegraphics[width=.4\textwidth]{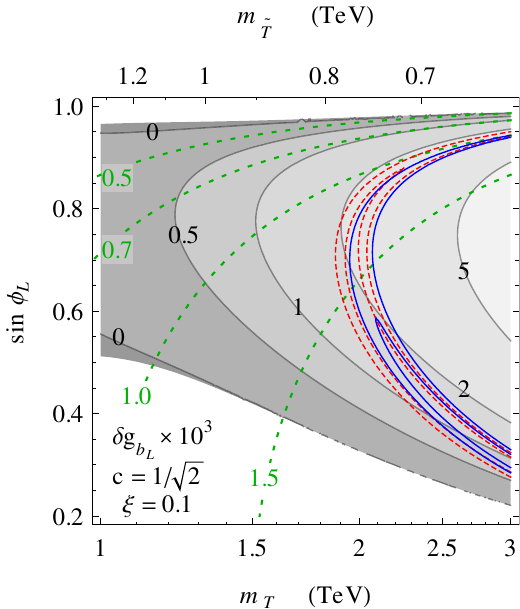}
\hspace{3.em}
\includegraphics[width=.4\textwidth]{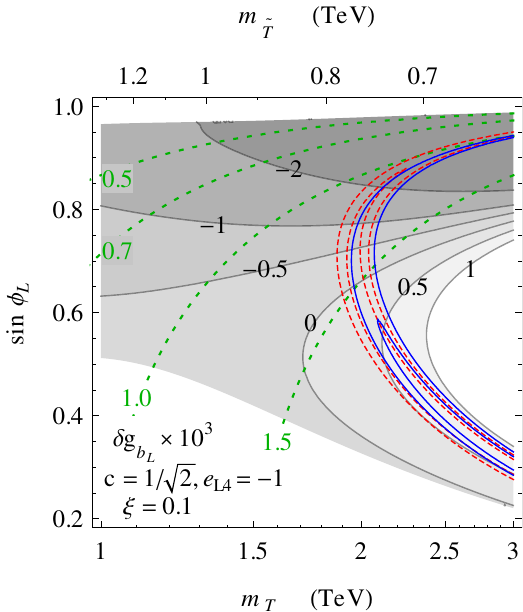}
\caption{Corrections to the $Z\overline b_L b_L$ vertex in the model with $c=1/\sqrt{2}$
for $\xi = 0.1$.
In the left plot we neglected the UV effects
and the contributions from $4$-fermion operators. On the right panel we added
the shift induced by the operator in eq.~(\ref{eq:4ferm_c}) with $e_{L4} = -1$.
The configurations correspond to the one chosen for the right plot in
fig.~\ref{fig:T_hat_c}.}
\label{fig:Zbb_c}
\end{figure}
As in the case $c=0$, if we neglect the contributions
from the UV dynamics and from the $4$-fermion operators,
the corrections to the $Z \overline b_L b_L$ parameter tend to be
positive and correlated to the shift in $\widehat T$. The numerical results in the
plane corresponding to the right plot in fig.~\ref{fig:T_hat_c} are shown in the left panel of
fig.~\ref{fig:Zbb_c}. Due to the protection of the EW observables, the presence of $4$-fermion
operators can not induce logarithmically divergent contributions to the $Z \overline b_L b_L$
vertex. However sizable finite corrections are still possible.
In the right panel of fig.~\ref{fig:Zbb_c} we show how $\delta g_{b_L}$ is modified if we
add the contributions due to the vertex
\begin{equation}\label{eq:4ferm_c}
\frac{e_{L4}}{f^2} \big(\overline B_L^a \gamma^\mu B^a_L\big)\Big(\overline{T}^b_L \gamma_\mu T^b_L
+ \overline{X}_{2/3L}^b \gamma_\mu X_{2/3L}^b\Big)\,,
\end{equation}
with $e_{L4} = -1$. As expected, the corrections are large only in the parameter space region
in which the $q_L$ has a large degree of compositeness. In this region the additional
correction can easily induce a negative value for $\delta g_{b_L}$. Notice however that the
sign of the corrections crucially depends on the sign of the coefficient of the
$4$-fermion operators. In our effective approach this coefficient is a free parameter,
but in a theory including a UV completion of our Lagrangian some constraints on the
size and on the sign of the $4$-fermion operators could be present.

\section{The case of a totally composite $t_R$}\label{sec:compositetR}

So far we analyzed a class of models based on the standard implementation
of partial compositeness in which all the SM fermions have a corresponding elementary counterpart.
Of course, due to the quantum numbers of the left-handed SM fermions, including them in the
effective Lagrangian via some elementary fields is the only reasonable option if we want
to preserve the global $\SO(5)$ invariance in the composite sector. The situation
is different for the right-handed fermions. They are singlets under the $\SO(4)$
symmetry and can be embedded in the theory as elementary fields or, alternatively,
as chiral fermions coming from the strong dynamics.
In this case the right-handed fermions are part of the composite sector and are total singlets
under the global $\SO(5)$ invariance.

This alternative implementation of partial compositeness is particularly appealing for the
right-handed top component. As shown in Ref.~\cite{Panico:2012uw} models with a totally
composite $t_R$ can lead to minimally tuned implementations of the composite Higgs idea
and can give rise to an interesting collider phenomenology~\cite{DeSimone:2012fs}.

In this section we analyze the corrections to the EW observables that are present in this
alternative scenario. Our strategy will be similar to the one followed in the previous sections.
We will use an effective Lagrangian approach to parametrize the low-energy dynamics of the
models and we will analyze the EW parameters with particular attention to the corrections
coming from the light composite fermions.

\subsection{The effective Lagrangian}

As we did for the models in section~\ref{sec:the_model},
we will concentrate on a minimal scenario in which the elementary top component is mixed
with a composite operator that transforms in
the fundamental representation of the global $\SO(5)$ symmetry. For simplicity we only
include one level of composite resonances which transform as a $4$-plet
($\psi_4$) and a singlet ($\psi_1$) under the $\SO(4)$ subgroup.
The elementary sector of the theory contains
the left-handed doublet $q_L$, while the $t_R$ is now an $\SO(5)$ chiral singlet
belonging to the composite sector.

The effective Lagrangian for the composite states is given by~\footnote{The presence of chiral states
coming from the strong dynamics does not allow us to impose a
parity symmetry in the strong sector. For this reason in eq.~(\ref{eq:Lagr_comp_tR})
we wrote independent $d$-symbol interactions for the left- and right-handed chiralities.}
\begin{eqnarray}
{\cal L}_{comp} &=& i \overline \psi_4 \Dslash \psi_4 + i \overline \psi_1 \Dslash \psi_1
+ i \overline t_R \Dslash t_R - m_4 \overline \psi_4 \psi_4 - m_1 \overline \psi_1 \psi_1
\label{eq:Lagr_comp_tR}\\
&& +\, \left(i c_L \overline \psi_{4L}^i \gamma^\mu d_\mu^i \psi_{1L}
+ i c_R \overline \psi_{4R}^i \gamma^\mu d_\mu^i \psi_{1R} +{\rm h.c.}\right)
+ \left(i c_t \overline \psi_{4R}^i \gamma^\mu d_\mu^i t_R + {\rm h.c.}\right)
+ \frac{1}{f^2} (\overline \psi \psi)^2\,.\nonumber
\end{eqnarray}
As in eq.~(\ref{eq:lagr_comp}), the covariant derivative for the $4$-plet $\psi_4$ contains
the CCWZ $e_\mu$ symbol: $D_\mu \psi_4 = (\partial_\mu - 2/3 i g' X_\mu + i e_\mu)\psi_4$.
Notice that a mass term of the form $m_R \overline t_R \psi_{1L} + {\rm h.c.}$ can be
added to the effective Lagrangian in eq.~(\ref{eq:Lagr_comp_tR}). This term can however be
removed by a redefinition of the $\psi_{1R}$ and $t_R$ fields.
The Lagrangian containing the kinetic terms for the elementary fields and the mixings is
\begin{equation}\label{eq:Lagr_elem_tR}
{\cal L}_{elem + mixing} = i \overline q_L \Dslash q_L
+ \left(y_{Lt} f \left(\overline q_L^{\bf 5}\right)^I U_{I5} t_R
+ y_{L4} f \left(\overline q_L^{\bf 5}\right)^I U_{Ii} \psi_4^i
+ y_{L1} f \left(\overline q_L^{\bf 5}\right)^I U_{I5} \psi_1 + {\rm h.c.}\right)\,.
\end{equation}
Differently from the case with an elementary right-handed top, in the present scenario a direct
mass mixing between the $q_L$ doublet and the $t_R$ singlet appears in the
effective Lagrangian.
The parameters in our effective Lagrangian are in general complex and some of the complex
phases can not be removed by field redefinitions. For simplicity we assume that our
theory is invariant under $CP$, in this way all the parameters in eqs.~(\ref{eq:Lagr_comp_tR})
and (\ref{eq:Lagr_elem_tR}) are real.

An interesting question is whether the scenarios with totally composite $t_R$ can
correspond to a particular limit of the case with an elementary $t_R$.
To address this question we can notice that a property of the scenario
with a totally composite right-handed top
is the fact that the couplings and mixing of the $t_R$ field with the other composite
resonances respect the $\SO(5)$ symmetry. The only breaking of the global invariance
in the fermion sector comes from the mixings of the elementary doublet $q_L$
in eq.~(\ref{eq:Lagr_elem_tR}).
In the case with an elementary $t_R$, instead, the $y_R$ mixings induce an extra
source of $\SO(5)$ breaking. The different symmetry structure of the two implementations
of partial compositeness clearly points out that the two scenarios are independent and
can not be simply connected by a limiting procedure.

\subsection{Results}

We can now discuss the explicit results for the scenarios with a totally composite $t_R$.
The analysis presented in section~\ref{sec:general_analysis}
can be straightforwardly adapted to the present set-up,
in particular all the general results are still valid. Before presenting the numerical results
for some simplified models, we briefly summarize the main differences with respect to the
results of section~\ref{sec:general_analysis}.

The contributions to the oblique parameters due to the non-linear Higgs dynamics
(sse eqs~.(\ref{eq:S_IR}) and (\ref{eq:T_IR})) and the tree-level corrections to the
$\widehat S$ parameter due to the gauge resonances (eq.~(\ref{eq:S_UV})) are universal
and do not depend on the assumptions on fermion compositeness.
The presence of a light $4$-plet of composite resonances still induces a logarithmically divergent
contribution to the $\widehat S$ parameter, which is now given by
\begin{equation}
\Delta \widehat S^{div}_{ferm} = \frac{g^2}{8 \pi^2}\left(1 - c_L^2 - c_R^2 - c_t^2\right)
\xi \log\left(\frac{m_*^2}{m_4^2}\right)\,.
\end{equation}
Notice that in this case the $d$-symbol involving the $t_R$ and the $4$-plet can
lead to a cancellation of the divergent contributions even if no light singlet is
present in the spectrum. This cancellation happens for $c_t = 1$.

As in the case with a partially composite $t_R$, the only couplings that break
the custodial invariance and the $P_{LR}$ symmetry are the mixings of the elementary $q_L$.
In the present case, however, we can write three mixings of this kind, $y_{L4}$,
$y_{L1}$ and $y_{Lt}$.
The fermion contribution to the $\widehat T$ parameter is generated at order $y_L^4$,
thus it is finite and dominated by the contributions coming from the lightest resonances.

The corrections to the $Z\overline b_L b_L$ vertex are in general logarithmically divergent.
We can extend to the present set-up the discussion of subsection~\ref{sec:Zbb_general}
and show that a selection rule exists also in this case. In particular a
logarithmically divergent correction can be generated only by specific $4$-fermion operators
and requires the presence of a light composite $4$-plet.
If the elementary $q_L$ is significantly composite
non-decoupling effects can arise and the contribution from the UV dynamics can be
sizable making the corrections to $g_{b_L}$ non predictable in the effective theory.

Notice that in the present set-up the top Yukawa is mainly determined by the $y_{Lt}$ mixing.
At the leading order in the $v/f$ expansion we find
\begin{equation}
m_t^2 = \frac{m_4^2}{m_4^2 + y_{L4}^2 f^2} \frac{y_{Lt}^2 v^2}{2}\,.
\end{equation}
The presence of a direct mixing between the elementary doublet $q_L$ and
the singlet $t_R$, allows to get the correct top mass even if we set to zero the $y_{L4}$
and $y_{L1}$ mixings. In this limit the composite $4$-plet and singlet do not feel directly
the breaking of the custodial and $P_{LR}$ symmetries and their corrections to the
$\widehat T$ parameter and to the $Z \overline b_L b_L$ vertex are totally negligible.
The contributions to $\widehat S$, instead, can still be sizable.

In the following we will consider in details two simplified scenarios, namely the
cases in which only a light composite singlet or a light composite $4$-plet are present
in the effective theory.

\subsubsection*{The case of a light singlet}

As a first simplified model we consider the case with only a light composite singlet.
As we will see, in this limit the model with a totally composite $t_R$ has
many properties in common with the case of a partially composite $t_R$ discussed
in subsection~\ref{sec:singlet}.

The deviations in $\widehat S$ are dominated by the tree-level UV contribution
and by the corrections due to the non-linear Higgs dynamics. For a high enough cut-off
($m_* \gtrsim 3\ {\rm TeV}$) the corrections to the $\widehat S$ parameter are well below
the maximal value allowed by the EW precision tests.

The fermion contributions to the $\widehat T$ parameter can be sizable and are typically positive.
At the leading order in $v/f$ they are given by
\begin{equation}\label{eq:T_comptR_sing}
\Delta \widehat T_{ferm} = \frac{3}{64 \pi^2} \xi \frac{y_{L1}^2 f^2}{m_1^2}
\left\{y_{L1}^2 + 2 y_{Lt}^2\left[\log\left(\frac{2 m_1^2}{v^2 y_{Lt}^2}\right) - 1\right]
\right\}\,.
\end{equation}
\begin{figure}
\centering
\includegraphics[width=.4\textwidth]{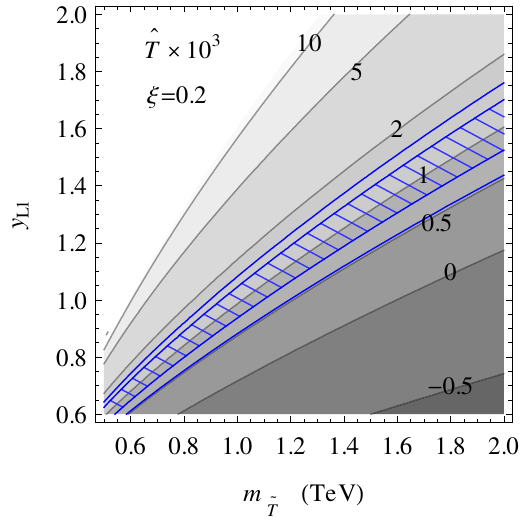}
\hspace{3.em}
\includegraphics[width=.4\textwidth]{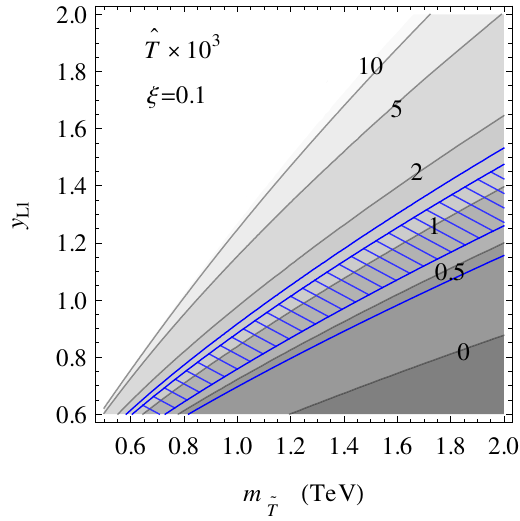}
\caption{Corrections to the $\widehat T$ parameter as a function of the mass
of the top partners and of the $q_L$ compositeness. The result corresponds to the
scenario with a totally composite $t_R$ with only a light singlet. The compositeness
scale has been fixed to $\xi = 0.2$ in the left panel and $\xi = 0.1$ in the right one.
The solid blue contours give the regions that pass the
constraints on the oblique parameters at the $68\%$ and $95\%$ confidence level.}
\label{fig:T_hat_comptR_sing}
\end{figure}
In fig.~\ref{fig:T_hat_comptR_sing} we show the total correction to $\widehat T$ including
the leading IR effects given in eq.~(\ref{eq:T_IR}).

As in the analogous case with a partially composite $t_R$, the fermion contributions to
the $Z \overline b_L b_L$ vertex are strongly correlated with the corrections to $\widehat T$.
At leading order in $v/f$ we find
\begin{equation}
\delta g_{b_L} = \frac{1}{64 \pi^2} \xi \frac{y_{L1}^2 f^2}{m_1^2}
\left\{y_{L1}^2 + 2 y_{Lt}^2\left[\log\left(\frac{2 m_1^2}{v^2 y_{Lt}^2}\right) - 1\right]
\right\}\,.
\end{equation}
By comparing this expression with the result in eq.~(\ref{eq:T_comptR_sing}) we
find the same relation we obtained in subsection~\ref{sec:singlet}:
$\Delta \widehat T_{ferm} = 3 \delta g_{b_L}$. The values of $\widehat T$
compatible with the bounds ($0 \lesssim \widehat T \lesssim 2 \cdot 10^{-3}$) imply
a moderate positive shift in $\delta g_{b_L}$. This shift slightly worsens the
agreement with the experimental data with respect to the SM.

\subsubsection*{The case of a light $4$-plet}

The second simplified model we consider is the effective theory with only a light $4$-plet.
As can be seen from eqs.~(\ref{eq:Lagr_comp_tR}) and (\ref{eq:Lagr_elem_tR}),
in this case the low-energy Lagrangian contains $4$ free parameters:
the elementary--composite mixings, the $4$-plet mass and the coefficient of the $d$-symbol
term, $c_t$. As we will see, the $d$-symbol term can sizably affect the
corrections to the EW observables. Its presence makes the properties of the model
quite different from the ones found in the case with an elementary $t_R$ (compare
subsection~\ref{sec:4plet}).
Moreover, as was pointed out in the analysis of Ref.~\cite{DeSimone:2012fs}, the $d$-symbol
term can also play an important role for collider phenomenology.

In addition to the corrections from the Higgs non-linear dynamics and the UV tree-level
shift, the $\widehat S$ parameter receives a logarithmically enhanced
contributions from fermion loops:
\begin{equation}
\Delta \widehat S^{div}_{ferm} = \frac{g^2}{8 \pi^2}\left(1 - c_t^2\right)
\xi \log\left(\frac{m_*^2}{m_4^2}\right)\,.
\end{equation}
If $c_t$ is not close to $1$, this shift can be sizable and can induce stringent
constraints on the compositeness scale $\xi$.

\begin{figure}
\centering
\includegraphics[width=.4\textwidth]{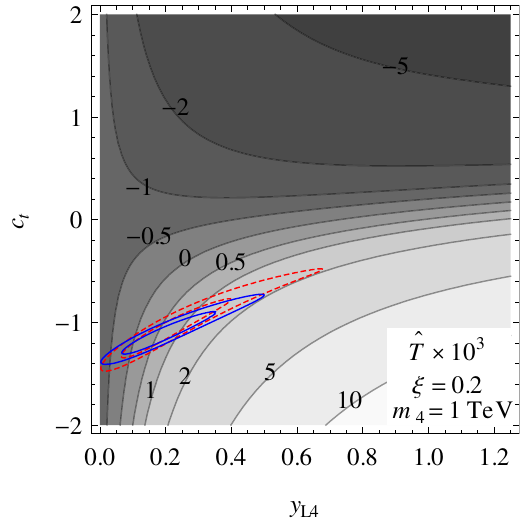}
\hspace{3.em}
\includegraphics[width=.4\textwidth]{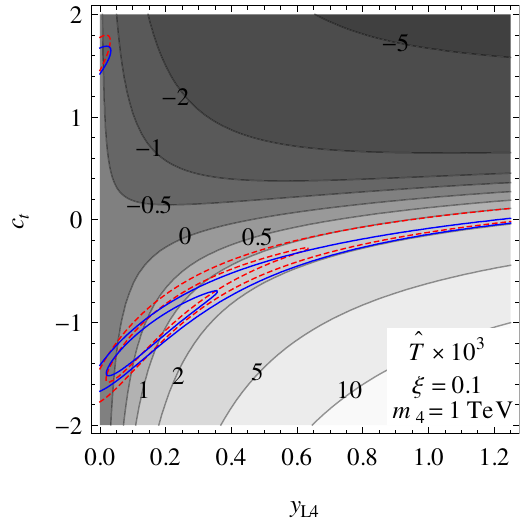}
\caption{Corrections to the $\widehat T$ parameter as a function of the $y_{L4}$ mixing
and of $c_t$. The result corresponds to the scenario with a totally composite $t_R$
with only a light $4$-plet with mass $m_4 = 1\ {\rm TeV}$. The compositeness
scale has been fixed to $\xi = 0.2$ in the left panel and $\xi = 0.1$ in the right one.
The solid blue contours give the regions that pass the
constraints on the oblique parameters at the $68\%$ and $95\%$ confidence level.
The dashed red lines show how the bounds are modified if we assume a $25\%$
reduction in $\widehat S$.
}
\label{fig:T_hat_comptR_4plet}
\end{figure}

The contributions to the $\widehat T$ parameter coming from fermion loops at
leading order in $v/f$ are given by
\begin{eqnarray}
\Delta \widehat T_{ferm} &=& -\frac{\xi}{32 \pi^2} \frac{y_{L4} f^2}{m_4^2}\Bigg\{
3 c_t^2 y_{L4} (y_{L4}^2 - 4 y_{Lt}^2) + y_{L4}^2 (y_{L4} - 3 \sqrt{2} c_t y_{Lt})\nonumber\\
&& -\, 3 y_{Lt}^2(y_{L4} - 4 \sqrt{2} c_t y_{Lt}) \left[\log\left(\frac{2 m_4^2}{v^2 y_{Lt}^2}\right)
- 1 \right] \Bigg\}\,.
\end{eqnarray}
Notice that the terms related to the $d$-symbol operator come with accidentally
large coefficients, thus even a relatively small value of $c_t$ can drastically modify
the result. In fig.~\ref{fig:T_hat_comptR_4plet} we show the total correction to
$\widehat T$ as a function of $y_{L4}$ and $c_t$ for a fixed value of the $4$-plet
mass, $m_4 = 1\ {\rm TeV}$. One can see that a positive correction to the $\widehat T$
parameter is possible, but requires a sign correlation between $y_{L4}$
and $c_t$.~\footnote{Notice that the Lagrangian is invariant under the transformation
$y_{L4} \rightarrow - y_{L4}$ and $c_t \rightarrow - c_t$.} In the plots we also show the
regions compatible with the constraints on the oblique parameters.
The parameter space regions with better agreement with the EW data are the ones with
$c_t \sim -1$, in which the logarithmically enhanced shift in $\widehat S$
is partially cancelled.

\begin{figure}
\centering
\includegraphics[width=.4\textwidth]{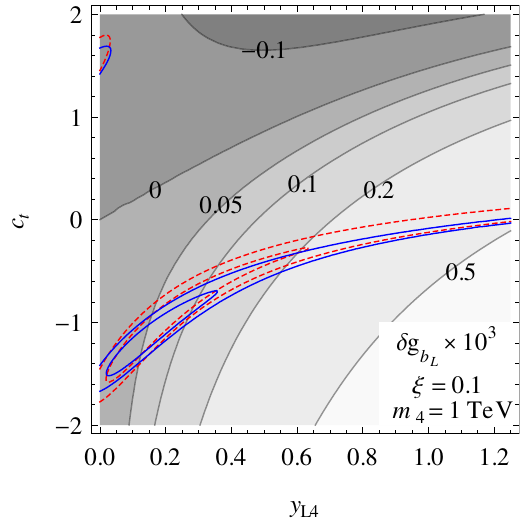}
\hspace{3.em}
\includegraphics[width=.4\textwidth]{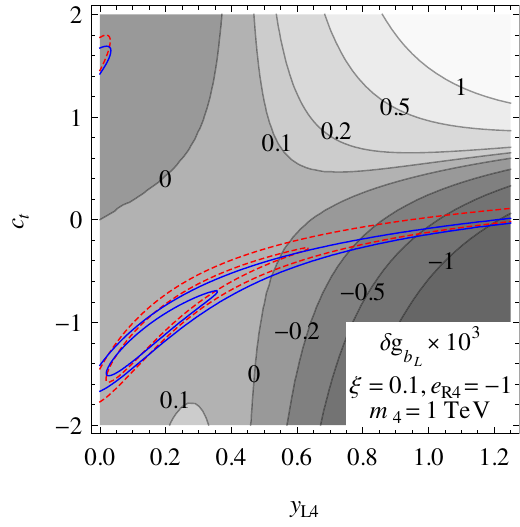}
\caption{Corrections to the $Z \overline b_L b_L$ vertex as a function of the $y_{L4}$ mixing
and of $c_t$. The results correspond to the scenario with a totally composite $t_R$
with only a light $4$-plet with mass $m_4 = 1\ {\rm TeV}$. The compositeness
scale has been fixed by $\xi = 0.1$. In the left panel we neglected the contributions
from $4$-fermion operators, while in the right panel we included the corrections due to the
operator in eq.~(\ref{eq:4ferm_2site}) with $e_{R4} = -1$.}
\label{fig:Zbb_comptR_4plet}
\end{figure}

The corrections to the $Z \overline b_L b_L$ vertex are given at the leading order in $v/f$ by
\begin{eqnarray}
\delta g_{b_L} &=& -\frac{\xi}{64 \pi^2}
\frac{m_4^2 y_{L4} y_{Lt}^2 f^2}{(m_4^2 + y_{L4}^2)^2} \Bigg[
2 y_{L4} - \sqrt{2} c_t y_{Lt}\nonumber\\
&& +\, \left(2 y_{L4} - \sqrt{2} c_t y_{Lt}
+ \frac{y_{L4} y_{Lt}^2 f^2}{2 (m_4^2 + y_{L4}^2 f^2)}\right)
\log\left(\frac{v^2 m_4^2 y_{Lt}^2}{2 (m_4^2 + y_{L4}^2 f^2)^2}\right)
\Bigg]\,.
\end{eqnarray}
The above formula contains only the corrections coming from the lowest order terms in the
effective Lagrangian without the contributions from $4$-fermion operators.
As can be seen from the numerical result in the left panel of fig.~\ref{fig:Zbb_comptR_4plet},
the sign of $\delta g_{b_L}$ has some correlation with the sign of $\widehat T$.
The size of the corrections to the $Z\overline b_L b_L$ vertex is however typically
one order of magnitude smaller than the one in $\widehat T$.
The points compatible with the constraints on the oblique EW parameters have $\delta g_{b_L}$
in the range $0 \lesssim \delta g_{b_L} \lesssim 0.5 \cdot 10^{-3}$.

The corrections to the $Z\overline b_L b_L$ vertex can of course be modified if $4$-fermion
interactions are present in the effective Lagrangian. In particular logarithmically divergent
contributions can be induced by operators of the form given in eq.~(\ref{eq:4-ferm_r}).
As an example we will show how the previous result for $\delta g_{b_L}$ is modified
by the operator given in eq.~(\ref{eq:4ferm_2site}). In this case the following additional
contribution arises:
\begin{equation}\label{eq:zbb_comptr_div}
\delta g_{b_L} = \frac{e_{R4}}{32\pi^2}\xi \frac{y_{L4}^2 f^2}{m_4^2 + y_{L4}^2 f^2}
y_{L4} \left(y_{L4} - \sqrt{2} c_t y_{Lt}\right) \log\left(\frac{m_*^2}{m_4^2}\right)\,,
\end{equation}
In the right panel of fig.~\ref{fig:Zbb_comptR_4plet} we show the numerical result
for $\delta g_{b_L}$ including the extra contribution in eq.~(\ref{eq:zbb_comptr_div})
for $e_{R4} = -1$.
In the region with sizable values for $y_{L4}$ the new contribution dominates and can
induce a negative shift in $\delta g_{b_L}$, which would improve the compatibility
with the experimental measurements.

\section{Corrections to the top couplings}\label{sec:topcouplings}

So far we devoted our attention to the oblique EW parameters and the bottom couplings.
The tight experimental bounds on these observables do not allow for
large deviations from the SM predictions and lead to strong bounds on the new
physics effects. Another class of observables, in particular the ones related to
the top quark, are instead less constrained from the present data which allow
sizable deviation from the SM. Large corrections to the top couplings are
naturally predicted in the scenarios with partial compositeness due to the strong
mixing of the third generation quarks with the composite dynamics.
Notice that the $P_{LR}$ invariance, which suppresses the corrections to the $Z \overline b_L b_L$
vertex, does not protect the couplings of the top quark. Thus big tree-level contributions can
be generated which could be eventually tested at the LHC.
The aim of this section is to determine the size of the
distortion of the top couplings to the $Z$ and to the $W$ bosons.

The top coupling to the $Z$ boson are described by the following effective Lagrangian
\begin{equation}
{\cal L}^Z = \frac{g}{c_w} Z_\mu \overline t \gamma^\mu \left[
(g_{t_L}^{SM} + \delta g_{t_L}) P_L + (g_{t_R}^{SM} + \delta g_{t_R}) P_R\right] t\,,
\end{equation}
where $g^{SM}$ denote the SM couplings and $\delta g$ correspond to the new physics
contributions. In the above formula $P_{L,R}$ are the left and right chiral projectors.
The tree-level values of the SM couplings are given by
\begin{equation}
g_{t_L}^{SM} = \frac{1}{2} - \frac{2}{3} s_w^2\,,
\qquad \quad g_{t_R}^{SM} = - \frac{2}{3} s_w^2\,.
\end{equation}

The couplings of the left-handed top component with the charged $W$ boson
are related to the $V_{tb}$ element of the CKM matrix.
We will parametrize the new physics contributions as $V_{tb} = 1 + \delta V_{tb}$.
The current LHC results already put a constraint on the new physics contribution at the $10\%$ level:
$V_{tb} = 1.020 \pm 0.046\, {\rm (meas.)} \pm 0.017\, {\rm (theor.)}$~\cite{Chatrchyan:2012ep}.
As we will see, the bounds on the models coming from this measurement are still
weaker than the ones coming from the EW precision data.

\subsection{A relation between $\delta g_{t_L}$ and $\delta V_{tb}$}

Before discussing the results in the explicit models we considered in this paper,
we rederive a general relation that links the deviations in the $Z\overline t_L t_L$
vertex to the corrections to $V_{tb}$ as already noticed in Refs.~\cite{delAguila:2000aa, delAguila:2000rc, Aguilar-Saavedra:2013pxa}. In the effective Lagrangian describing
the Higgs doublet and the SM fields
only two dimension-six operators contribute to the corrections to the $t_L$ couplings~
\cite{delAguila:2000aa, Grzadkowski:2010es,Giudice:2007fh,Contino:2013kra}:
\begin{equation}\label{eq:operators_tL}
{\cal L} = i \frac{c_{Hq}}{f^2} (\overline q_L \gamma^\mu q_L)
\left(H^\dagger \overleftrightarrow{D_\mu} H\right)
+ i \frac{c'_{Hq}}{f^2} (\overline q_L \sigma^i \gamma^\mu q_L)
\left(H^\dagger \sigma^i \overleftrightarrow{D_\mu} H\right)\,.
\end{equation}

A combination of the two operators in eq.~(\ref{eq:operators_tL}) is strongly
constrained by the experimental bound on the corrections to the $Z \overline b_L b_L$ vertex.
Notice that, in the models we considered in our analysis, the corrections to $g_{b_L}$
exactly vanish at tree level thanks to the $P_{LR}$ symmetry. The condition of vanishing
corrections to the $Z \overline b_L b_L$ coupling implies the relation
$c'_{Hq} = - c_{Hq}$~\cite{P_LR, AguilarSaavedra:2012vh}. Using this relation we find that the operators in eq.~(\ref{eq:operators_tL})
give rise to the following interactions of the top quark with the EW gauge bosons:
\begin{equation}
{\cal L} \supset 2 c_{Hq}\, v^2 \left[\frac{g}{c_w}\overline t_L Z^\mu \gamma_\mu t_L
+ \frac{g}{2} \left(\overline t_L \left(W^1_\mu - i W^2_\mu\right) \gamma^\mu b_L
+ {\rm h.c.}\right)
\right]\,.
\end{equation}
From this equation we can easily conclude that the leading corrections to the $Z \overline t_L t_L$
vertex and to the $V_{tb}$ matrix element satisfy the relation
\begin{equation}\label{eq:Ztt_Vtb}
\delta g_{t_L} = \delta V_{tb}\,.
\end{equation}
Notice that the above result holds only at order $v^2/f^2$. The subleading terms,
as for instance the dimension-eight operators, can generate independent corrections
to $g_{t_L}$ and $V_{tb}$.

It is important to stress that this analysis is valid as far as we can neglect the
corrections to the $Z \overline b_L b_L$ vertex with respect to the corrections to the
top couplings. Thus the result in eq.~(\ref{eq:Ztt_Vtb}) is true in general
and not only in the composite Higgs scenarios.

\subsection{The case of an elementary $t_R$}

As a first class of models we consider the scenarios with an elementary $t_R$.
The corrections to the $t_L$ couplings at leading order in $v/f$ are given by
\begin{equation}
\delta g_{t_L} = \delta V_{tb} = -\frac{\xi}{4} \frac{f^2}{m_4^2 + y_{L4}^2 f^2}\left[
\left(\frac{m_4 m_1 y_{L1} + y_{L4} y_{R4} y_{R1} f^2}{m_1^2 + y_{R1}^2 f^2} - \sqrt{2} c y_{L4}\right)^2
+ (1 - 2 c^2) y_{L4}^2\right]\,.
\end{equation}
This explicit result is in agreement with the relation derived in the previous subsection
(see eq.~(\ref{eq:Ztt_Vtb})). We also verified that at order $(v/f)^4$ the corrections
to $g_{t_L}$ and $V_{tb}$ do not coincide.

The coupling of the $t_R$ with the $Z$ boson is modified as well. The leading corrections
take the form
\begin{equation}
\delta g_{t_R} = \frac{\xi}{4} \frac{f^2}{m_1^2 + y_{R1}^2 f^2}
\left[\left(\frac{m_4 m_1 y_{R4} + y_{L4} y_{L1} y_{R1} f^2}{m_4^2 + y_{L4}^2 f^2}
- \sqrt{2} c y_{R1}\right)^2
- \left(\frac{m_1 y_{R4}}{m_4} - \sqrt{2} c y_{R1}\right)^2
\right]\,.
\end{equation}

\begin{figure}
\centering
\includegraphics[width=.4\textwidth]{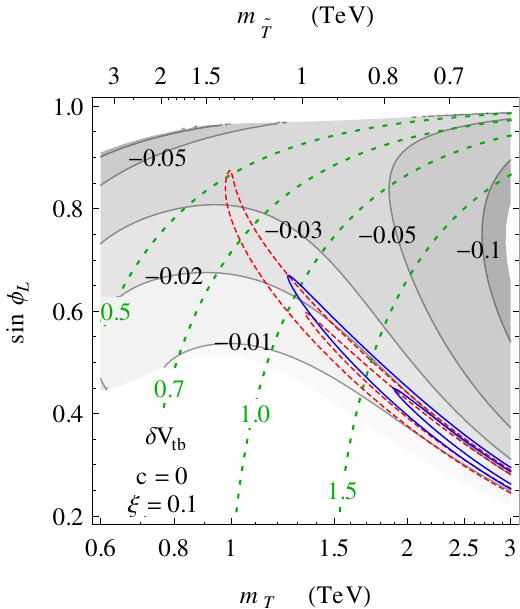}
\hspace{3.em}
\includegraphics[width=.4\textwidth]{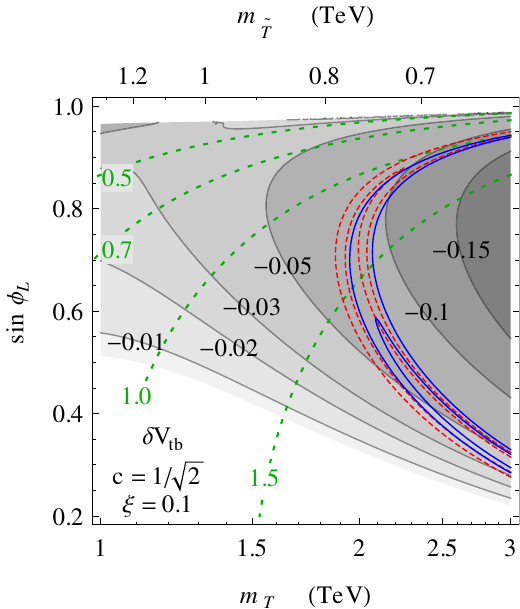}
\caption{Corrections to the $V_{tb}$ matrix element in the complete models with
$c=0$ (left panel) and $c=1/\sqrt{2}$ (right panel) for $\xi = 0.1$.
The configurations correspond to the ones of the left plot of fig.~\ref{fig:T_hat_2site}
for the case $c=0$ and of the right plot of fig.~\ref{fig:T_hat_c} for the case $c=1/\sqrt{2}$.
}
\label{fig:t_couplings}
\end{figure}
As explicit numerical examples we show in fig.~\ref{fig:t_couplings}
the distortion of the $V_{tb}$ matrix element in
the complete models with $c=0$ and $c=1/\sqrt{2}$ (see subsection~\ref{sec:complete_models}).
In the case with $c=0$, the configurations allowed by the constraints
on the oblique EW parameters have small corrections to $V_{tb}$,
$-0.03 \lesssim \delta V_{tb} \lesssim 0$, which are below the present
experimental sensitivity.
On the contrary, in the model with $c=1/\sqrt{2}$, the corrections to $V_{tb}$ can be
sizable, $-0.12 \lesssim \delta V_{tb} \lesssim -0.03$, and the current bounds can
already exclude a corner of the parameter space allowed by the EW precision data.
In our numerical analysis we also found that, in the realistic regions of the parameter space,
the deviations in the $t_R$ couplings are always small, $\delta g_{t_R} \lesssim 0.01$.
Moreover we checked numerically that the correlation between $\delta g_{t_L}$ and $\delta V_{tb}$
is always well verified and the deviations from eq.~(\ref{eq:Ztt_Vtb}) are of order $\xi$ as expected.

To conclude the analysis of the top couplings in the models with an elementary $t_R$,
it is interesting to consider the simplified cases with only one light composite multiplet.
In the limit with only a light singlet we find
\begin{equation}
\delta g_{t_L} = \delta V_{tb} = -\frac{\xi}{4} \frac{m_1^2 y_{L1}^2 f^2}{(m_1^2 + y_{R1}^2 f^2)^2}\,,
\qquad \quad
\delta g_{t_R} = 0\,.
\end{equation}
This shows that the corrections to the $t_L$ couplings are suppressed in the parameter space region
with a sizable $t_R$ compositeness ($y_{R1} f > m_1$ and $y_{R1} > y_{L1}$). The corrections
to $g_{t_R}$ vanish in this case because the $t_R$ can only mix with composite states
with the same coupling to the $Z$ boson.

In the case with only a light $4$-plet we obtain the following results
\begin{equation}
\delta g_{t_L} = \delta V_{tb} = -\frac{\xi}{4} \frac{y_{L4}^2 f^2}{m_4^2 + y_{L4}^2 f^2}\,,
\qquad \quad
\delta g_{t_R} = -\frac{\xi}{4} \frac{y_{L4}^2 y_{R4}^2 f^2}{m_4^2 + y_{L4}^2 f^2}
\left(\frac{f^2}{m_4^2} + \frac{f^2}{m_4^2 + y_{L4}^2 f^2}\right)\,.
\end{equation}
In this case the experimental bounds on $V_{tb}$ can be used to put
an upper bound on the $t_L$ compositeness.
Notice that the mixing of the $t_R$ does not break the $P_{LR}$ symmetry.
The $g_{t_R}$ coupling, however, can receive tree-level corrections through
the mixing between the elementary $t_R$ and composite resonances with
different quantum numbers, which is induced by the non-zero top mass. This origin explains
why the prefactor in the expression for $\delta g_{t_R}$ is proportional to the square
of the top Yukawa (see eq.~(\ref{eq:topmass_4plet})). The correction to $g_{t_R}$ is
enhanced if the top partners are light.

\subsection{The case of a composite $t_R$}

We now consider the scenarios with a totally composite $t_R$.
The leading corrections to the $V_{tb}$ matrix element and to the
top couplings to the $Z$ boson are given by
\begin{equation}\label{eq:Vtb_comptR}
\delta g_{t_L} = \delta V_{tb} = -\frac{\xi}{4} \frac{f^2}{m_4^2 + y_{L4}^2 f^2}\left[
\left(\frac{m_4 y_{L1}}{m_1} - \sqrt{2} c_L y_{L4}\right)^2
+ (1 - 2 c_L^2) y_{L4}^2\right]\,,
\end{equation}
and
\begin{equation}\label{eq:gtR_comptR}
\delta g_{t_R} = \frac{\xi}{4} \frac{y_{L4} y_{Lt} f^2}{(m_4^2 + y_{L4}^2 f^2)^2}
\left[y_{L4} y_{Lt} f^2 - 2 \sqrt{2} c_t (m_4^2 + y_{L4}^2 f^2)\right]\,.
\end{equation}

In the limits with only one light multiplet the expressions in eqs.~(\ref{eq:Vtb_comptR})
and (\ref{eq:gtR_comptR}) can be drastically simplified.
If only a light singlet is present in the effective theory we find:
\begin{equation}
\delta g_{t_L} = \delta V_{tb} = -\frac{\xi}{4} \frac{y_{L1}^2 f^2}{m_1^2}\,,
\qquad \quad
\delta g_{t_R} = 0\,.
\end{equation}
In this case the corrections to the $Z \overline t_R t_R$ coupling are negligible, while the $V_{tb}$
matrix element and the $Z \overline t_L t_L$ vertex can become large if the
composite singlet is light.

In the model with only a light composite $4$-plet the corrections to the top
couplings become
\begin{equation}
\delta g_{t_L} = \delta V_{tb} = -\frac{\xi}{4} \frac{y_{L4}^2 f^2}{m_4^2 + y_{L4}^2 f^2}\,,
\qquad
\delta g_{t_R} = \frac{\xi}{4} \frac{y_{L4} y_{Lt} f^2}{(m_4^2 + y_{L4}^2 f^2)^2}
\left[y_{L4} y_{Lt} f^2 - 2 \sqrt{2} c_t (m_4^2 + y_{L4}^2 f^2)\right]\,.
\end{equation}
Analogously to the case with an elementary $t_R$, the corrections to the $V_{tb}$ matrix
element can be used to put an upper bound on the degree of compositeness of the elementary
doublet $q_L$.

\section{Conclusions}\label{sec:conclusions}

In this work we studied the corrections to the EW observables that arise in
composite Higgs scenarios due to the presence of fermionic resonances.
In realistic models light composite fermions are typically predicted and this
motivated the use of an effective field theory approach for our analysis.
For definiteness we focused our attention on the minimal composite Higgs realization based
on the symmetry structure $\SO(5)/\SO(4)$. Within this framework we considered
a general parametrization of the case in which the elementary SM fermions are mixed
with operators in the fundamental representation of the global $\SO(5)$ group.
We included in our effective Lagrangian one level of composite fermionic resonances
which correspond to a $4$-plet and a singlet under the unbroken $\SO(4)$ symmetry.

We quantified the relevance of the fermionic contribution to the deviation of
the precision electroweak observables. In particular we focused on the oblique electroweak
parameters, $\widehat S$ and $\widehat T$,
and on the $Z \overline b_L b_L$ coupling, which are very well determined experimentally
and can be used to put tight constraints on new physics effects.

One interesting result is the identification of a new parametrically enhanced contribution
to the $\widehat S$ parameter. This effect is entirely generated by the composite dynamics
and appears if light composite fermions (in particular $\SO(4)$ $4$-plets) are present
in the spectrum. The origin of the new enhanced contribution can easily be understood
from an effective field theory point of view.
The non-renormalizable Higgs interactions due to the non-linear $\sigma$-model dynamics
induce new logarithmically divergent diagrams and generate a running of the two dimension-$6$
operators, ${\cal O}_{W,B}$, which contribute to the $\widehat S$ parameter.
This effect is calculable in the effective theory and its size turns out to be comparable
or even larger than the tree-level shift given by the heavy gauge resonances.

In minimal scenarios with a light $4$-plet ($m_4 \lesssim 1\ {\rm TeV}$), the constraints on the
$\widehat S$ parameter imply a tight bound on the compositeness scale $\xi \lesssim 0.1$,
which corresponds to $f \gtrsim 750\ {\rm GeV}$ (see fig.~\ref{fig:xi_bound_S}).
This bound can be relaxed if additional light states are present in the spectrum
(for instance a singlet). Cancelling the $4$-plet contribution, however, seems possible only
at the price of some additional tuning.

Another consequence of the presence of logarithmic divergence in $\widehat S$ is the fact that
the UV dynamics does not necessarily decouple and can generate non-negligible finite
corrections. We discussed an example of this effect in one explicit model,
but we did not systematically investigate this aspect. We leave this analysis for future work.

Differently from $\widehat S$, the $\widehat T$ parameter is finite in our scenario
thanks to the protection coming from the custodial symmetry. The corrections coming from the
composite sector are thus dominated by the contributions of the lightest composite states and
can be reliably computed in our effective field theory. This allows us to use the $\widehat T$
parameter to put robust bounds on the parameter space of the composite Higgs models.

We found that a positive shift in $\widehat T$, which is typically needed to satisfy the
constraints on the oblique parameters, can be easily generated by the fermion loops.
In the standard scenarios, in which the $t_R$ is a partially elementary state,
obtaining a positive correction to $\widehat T$ requires the presence of a relatively light
singlet. In configurations with only a light $4$-plet the corrections are instead always negative.
On the contrary, in the alternative scenarios in which the $t_R$ is a completely composite state,
a positive contribution to $\widehat T$ can be obtained also in the configurations
with only a light $4$-plet. This can be done at the price of a mild correlation among
the parameters (see fig.~\ref{fig:T_hat_comptR_4plet}).

The third precision observable we considered is the $Z \overline b_L b_L$ coupling.
In this case, power-counting arguments show that the composite resonances contributions can
be logarithmically divergent. We found however that, if only the operators
with the  lowest dimension are included in the effective Lagrangian, a selection
rule forbids the appearance of divergent contributions
and makes the corrections to the $g_{b_L}$ coupling finite.
This is no longer true if higher dimensional operators and in particular $4$-fermion interactions
are present in the effective theory. In this case, if a light $4$-plet is included in the
theory, a logarithmically enhanced correction to $g_{b_L}$ can be generated.
Moreover, as in the case of the $\widehat S$ parameter, the UV dynamics typically does not
decouple and can generate sizable corrections.

If only the lowest-dimensional operators are included in the effective Lagrangian, the corrections
to the $Z \overline b_L b_L$ vertex tend to be correlated to the corrections to $\widehat T$.
In particular a sizable positive shift in $\widehat T$ usually corresponds to a positive
contribution to $\delta g_{b_L}$, which is disfavored by the current experimental bounds.
Higher-dimensional operators, which are typically generated by the composite dynamics,
can however induce large contributions to the $Z \overline b_L b_L$ coupling and
remove the correlation with $\widehat T$.

Finally we analyzed the corrections to the top EW gauge couplings. In the composite Higgs
scenarios we considered these couplings can receive large tree-level distortions
due to the sizable degree of compositeness of the top. We found that the deviations
of the $Z \overline t_L t_L$ vertex are strongly correlated with the corrections to the $W t_L b_L$ coupling.
Stringent bounds on the deviations of the $V_{tb}$ matrix element would therefore strongly
disfavor the presence of large corrections to the $Z$ coupling.

The constraints on the model coming from the current measurement of the $V_{tb}$ matrix element
are typically weaker than the ones from the EW precision data and can become competitive with them
only in a small region of the parameter space. For a moderate amount of tuning, $\xi = 0.1$,
the corrections to the $V_{tb}$ matrix element can be of order $5\%$ and the
corrections to the $Z \overline t_L t_L$ of order $10\%$.

\section*{Acknowledgments}

We thank E.~Furlan, R.~Rattazzi, S.~Rychkov, M.~Serone and M.~Trott for useful discussions.
We also thank A.~Wulzer for many suggestions and comments about the manuscript.
O.~M. and G.~P. acknowledge the Galileo Galilei Institute in Florence for hospitality
during the completion of this work.
The work of C.~G. and O.~M. is supported by the European Programme Unification in the LHC Era,
contract PITN-GA-2009-237920 (UNILHC). C.~G. is  also supported by the
Spanish Ministry MICNN under contract FPA2010-17747. 
This research has been partly funded by the European Commission under the
ERC Advanced Grant 226371 MassTeV.

\appendix

\section{The CCWZ notation}\label{app:ccwz}

In this appendix we define our notation for the $\SO(5)$ algebra and for the CCWZ operators.
For most of our definitions we follow the notation of Ref.~\cite{DeSimone:2012fs}.

\subsection*{The $\SO(5)$ algebra and the Goldstones}

A useful basis for the $\SO(5)$ generators, which shows explicitly the $\SO(4)$ subgroup, is given by
\begin{equation}
(T^\alpha_{L,R})_{IJ} = -\frac{i}{2}\left[\frac{1}{2}\varepsilon^{\alpha\beta\gamma}
\left(\delta_I^\beta \delta_J^\gamma - \delta_J^\beta \delta_I^\gamma\right) \pm
\left(\delta_I^\alpha \delta_J^4 - \delta_J^\alpha \delta_I^4\right)\right]\,,
\label{eq:SO4_gen}
\end{equation}
\begin{equation}
T^{i}_{IJ} = -\frac{i}{\sqrt{2}}\left(\delta_I^{i} \delta_J^5 - \delta_J^{i} \delta_I^5\right)\,,
\label{eq:SO5/SO4_gen}
\end{equation}
where $T^{\alpha}_{L,R}$ ($\alpha = 1,2,3$) correspond to the
$\SO(4) \simeq \textrm{SU}(2)_L \times \textrm{SU}(2)_R$ generators,
$T^i$ ($i = 1,\ldots,4$) are the generators of the coset $\SO(5)/\SO(4)$ and
the indices $I,J$ take the values $1,\ldots,5$. We chose to normalize the
generators in eqs.~(\ref{eq:SO4_gen}) and (\ref{eq:SO5/SO4_gen}) such that
${\rm Tr}[T^A, T^B] = \delta^{AB}$. With this normalization the $\textrm{SU}(2)_{L,R}$ generators
satisfy the usual commutation relations
\begin{equation}
\left[T^{\alpha}_{L,R}, T^{\beta}_{L,R}\right] = i \varepsilon^{\alpha\beta\gamma}\, T^{\gamma}_{L,R}\,.
\end{equation}

The Goldstone matrix for the coset $\SO(5)/\SO(4)$ is given by
\begin{equation}
U = \exp\left[i \frac{\sqrt{2}}{f_\pi} \Pi_{i} T^{i}\right]
= \left(
\begin{array}{c@{\hspace{1.5em}}c}
\displaystyle I_{4 \times 4} - \frac{\vec{\Pi} \vec{\Pi}^t}{\Pi^2} \left(1 - \cos \frac{\Pi}{f}\right)
& \displaystyle \frac{\vec{\Pi}}{\Pi} \sin \frac{\Pi}{f}\\
\rule{0pt}{2em}\displaystyle - \frac{\vec{\Pi}^t}{\Pi} \sin \frac{\Pi}{f} & \displaystyle \cos \frac{\Pi}{f}
\end{array}
\right)\,,
\end{equation}
where we defined $\Pi^2 \equiv \vec{\Pi}^t \vec{\Pi}$. Under an $\SO(5)$ transformation, $g$, the Goldstones
transform according to the standard relation
\begin{equation}\label{eq:U_transformation}
U(\Pi) \rightarrow U(\Pi^{(g)}) = g \cdot U(\Pi)\cdot h^t(\Pi; g)\,,
\end{equation}
where $h(\Pi; g)$ is an element of the $\textrm{SO(4)}$ subgroup:
\begin{equation}
h =
\left(
\begin{array}{cc}
h_4 & 0\\
0 & 1
\end{array}
\right)
\end{equation}
Under the unbroken $\textrm{SO(4)}$ symmetry the Goldstones transform
linearly: $\Pi^i \rightarrow (h_4)^i_j \Pi^j$.

The standard Higgs doublet $H = (h_u, h_d)$ is related to the $\Pi$
$4$-plet as
\begin{equation}
\Pi = \frac{1}{\sqrt{2}}\left(
\begin{array}{c}
-i (h_u - h_u^\dagger)\\
h_u + h_u^\dagger\\
i (h_d - h_d^\dagger)\\
h_d + h_d^\dagger
\end{array}
\right)\,.
\label{ferem}
\end{equation}
The physical Higgs $\rho$ can be obtained adopting the unitary gauge in which the
Higgs doublet reads
\begin{equation}
h_d \equiv \frac{h}{\sqrt{2}}
= \frac{\langle h \rangle + \rho}{\sqrt{2}}\,, \qquad \quad h_u = 0\,.
\end{equation}
In this gauge the Goldstone matrix takes the simple form
\begin{equation}
U =
\left(
\begin{array}{ccccc}
1 & 0 & 0 & 0 & 0\\
0 & 1 & 0 & 0 & 0\\
0 & 0 & 1 & 0 & 0\\
0 & 0 & 0 & \cos \frac{h}{f} & \sin \frac{h}{f}\\
0 & 0 & 0 & -\sin \frac{h}{f} & \cos \frac{h}{f}
\end{array}
\right)\,.
\end{equation}

\subsection*{The CCWZ operators}

In order to define the $e_\mu$ and $d_\mu$ CCWZ symbols it is useful to describe the
elementary gauge bosons in an $\SO(5)$ notation.
The SM vector fields are introduced in the theory by weakly gauging
the $\textrm{SU}(2)_L \times \textrm{U}(1)_{R3}$ subgroup of $\SO(4)$
and their embedding is given explicitly by
\begin{equation}\label{eq:gauge_embedding}
A_\mu = \frac{g}{\sqrt{2}} W^+_\mu (T^1_L + i T^2_L)
+ \frac{g}{\sqrt{2}} W^-_\mu (T^1_L - i T^2_L)
+ g (c_w Z_\mu + s_w A_\mu) T^3_L
+ g' (c_w A_\mu - s_w Z_\mu) T^3_R\,,
\end{equation}
where $g$ and $g'$ are the gauge coupling corresponding to the $\textrm{SU}(2)_L$
and $\textrm{U}(1)_Y$ subgroups, while $c_w$ and $s_w$ are the cosine and sine of
the weak mixing angle, $\tan \theta_w = g'/g$.

To define the CCWZ symbols we can start from the following quantity
\begin{equation}
\overline A_\mu \equiv A_\mu^{(U^t)} = U^t [A_\mu + i \partial_\mu ] U\,,
\end{equation}
and we can define $e_\mu$ and $d_\mu$ as the coefficient of the
decomposition of $\overline A_\mu$ in terms of broken and unbroken $\SO(5)$ generators:
\begin{equation}
\overline A_\mu = -d_\mu^{i} T^{i} - e^a_\mu T^a\,.
\end{equation}
It is not difficult to prove that the $e$ and $d$ symbols transform
under $\SO(5)$ as
\begin{equation}
e_\mu \equiv e_\mu^a t^a \rightarrow h_4 [e_\mu - i \partial_\mu] h^t_4
\quad \textrm{and} \quad
d_\mu^{i} \rightarrow (h_4)^{i}_{j} d^{j}_\mu\,,
\end{equation}
where we denoted by $t^a$ the $\SO(4)$ generators in a $4 \times 4$ matrix form.

Using the embedding of the gauge fields given in eq.~(\ref{eq:gauge_embedding}) we get
the explicit expressions
\begin{eqnarray}
d^{i}_\mu &=& \sqrt{2} \left(\frac{1}{f} - \frac{\sin \Pi/f}{\Pi}\right)
\frac{\vec \Pi \cdot \nabla_\mu \vec \Pi}{\Pi^2}\Pi^{i}
+ \sqrt{2} \frac{\sin \Pi/f}{\Pi}\nabla_\mu \Pi^{i}\\
e^a_\mu &=& -A^a_\mu + 4 i \frac{\sin^2 (\Pi/2f)}{\Pi^2}\vec \Pi^t t^a \nabla_\mu \vec \Pi
\end{eqnarray}
where $\nabla_\mu \Pi$ is defined as
\begin{equation}
\nabla_\mu \Pi^{i} = \partial_\mu \Pi^{i} - i A^a_\mu (t^a)^i_j \Pi^{j}\,.
\end{equation}
The expressions for the $d_\mu$ and $e_\mu$ symbols in the unitary gauge
are given by
\begin{equation}
d_\mu = -\frac{g}{\sqrt{2}} \sin\frac{h}{f}
\left(
\begin{array}{c}
W^1_\mu\\
W^2_\mu\\
\frac{1}{c_w} Z_\mu\\[1mm]
-\frac{2}{gf \sin h/f} \partial_\mu h
\end{array}
\right)\,,
\end{equation}
and
\begin{equation}
e_\mu = \frac{i g}{2}
\left(
\begin{array}{cccc}
0 & 2 s_w A_\mu + \frac{1 - 2 s_w^2}{c_w} Z_\mu & -W^2_\mu & W^1_\mu \cos \frac{h}{f}\\
-2 s_w A_\mu - \frac{1 - 2 s_w^2}{c_w} Z_\mu & 0 & W^1_\mu & W^2_\mu \cos \frac{h}{f}\\
W^2_\mu & - W^1_\mu & 0 & \frac{1}{c_w} Z_\mu \cos \frac{h}{f}\\
- W^1_\mu \cos \frac{h}{f} & - W^2_\mu \cos \frac{h}{f} &
- \frac{1}{c_w} Z_\mu \cos \frac{h}{f} & 0
\end{array}
\right)\,.
\end{equation}

Using the $d_\mu$ symbol we can write the kinetic term for the Goldstones in the form
\begin{equation}
{\cal L}_\pi = \frac{f^2}{4} d_\mu^i d^\mu_i\,.
\end{equation}
In the unitary gauge the above expression becomes
\begin{equation}
{\cal L}_\pi = \frac{1}{2} (\partial h)^2 + \frac{g^2}{4} f^2 \sin^2 \frac{h}{f}
\left(|W|^2 + \frac{1}{2 c_w^2} Z^2\right)\,.
\end{equation}
From this expression we can extract the mass of the $W$ boson, $m_w = (g/2) f \sin(\langle h \rangle)$
and derive the exact relation between the Higgs VEV and the EW scale $v = 246\ \textrm{GeV}$:
\begin{equation}
v = f \sin \frac{\langle h \rangle}{f}\,.
\end{equation}
When a gap between the EW scale $v$ and the compositeness scale $f$ exists, such that $v \ll f$,
the Higgs VEV and the EW scale can be identified $v \simeq \langle h \rangle$.
As it is clear from our analysis the condition $(v/f)^2 \ll 1$ is required by the EW
constraints and we can safely replace the Higgs VEV with $v$ as we did in this paper.

Finally we discuss the introduction of fermions in the CCWZ notation.
We included in our effective theory two possible composite multiplets: $\psi_4$
which transforms as a $4$-plet of $\SO(4)$ and $\psi_1$ which is a singlet.
Under the non-linearly realized $\SO(5)$ transformations $\psi_1$ is invariant, while $\psi_4$ transforms as
\begin{equation}
\psi_4 \rightarrow h_4 \cdot \psi_4\,.
\end{equation}
The covariant derivative for the singlet is the standard one
\begin{equation}
D_\mu \psi_1 = [\partial_\mu - i g' X (c_w A_\mu - s_w Z_\mu)] \psi_1\,,
\end{equation}
where $X$ denotes the charge under ${\textrm U}(1)_X$.
The covariant derivative of the $4$-plet, instead, contains an extra term
given by the $e_\mu$ symbol:
\begin{equation}
D_\mu \psi_4 = [\partial_\mu + i e_\mu - i g' X (c_w A_\mu - s_w Z_\mu)] \psi_4\,.
\end{equation}
The presence of the extra term is essential to restore the full $\SO(5)$ invariance.

\section{Operator analysis for the $Z\overline b_L b_L$ vertex}\label{app:Zbb_operator_analysis}

In section~\ref{sec:general_analysis} we presented a general analysis of the one-loop corrections
to the $Z\overline b_L b_L$ vertex that are induced by the presence of composite
fermion resonances. We found that logarithmically divergent contributions can be
present if a light composite $4$-plet is present in the spectrum.
For simplicity in the main text we did not report rigorous proofs
of our statements and we only gave some partial justifications. The aim of this appendix is
to present a more rigorous and systematic study based on an operator analysis.

\subsection*{General considerations}

An important feature of our effective Lagrangian
is the presence of a $P_{LR}$ symmetry, which is exact in the composite
sector and is only broken by the mixing with the elementary states (in particular with
the doublet $q_L$). The $P_{LR}$ symmetry plays an essential
role in protecting the $Z\overline b_L b_L$ vertex from large tree-level corrections
and it also leads to a reduction of the degree of divergence of the loop contributions.
In the following we will take into account the consequences of the $P_{LR}$
invariance through the method of spurions.

As a first step we need to formally restore the global $\SO(5)$ invariance in our effective
Lagrangian. For this purpose we assume that the elementary fields transform only under
an ``elementary'' $\textrm{SU}(2)_L \times \textrm{U}(1)_Y$ global group which is
independent with respect to the global $\SO(5)$ invariance of the composite sector.
In this picture the SM group corresponds to the diagonal combination of the ``elementary''
and the ``composite'' groups. The mixing between the elementary and the composite states
clearly induces a breaking of the extended global invariance. We can however formally
restore the complete global symmetry by promoting the couplings to spurions with
non-trivial transformation properties under the ``elementary'' and the ``composite'' groups.
In our set-up we need two spurions:
\begin{itemize}
\item[i)] $(\tilde y_L)^\alpha_A$, which transforms as a doublet (${\bf \overline 2}_{-1/6}$)
under the ``elementary'' symmetry (index $\alpha$) and belongs to the fundamental
representation of $\SO(5)$ with $U(1)_X$ charge $2/3$ (index $A$). Its physical
value is given by
\begin{equation}
\langle \tilde y_L \rangle =
\frac{1}{\sqrt{2}}
\left(
\begin{array}{cc}
0 & i\\
0 & 1\\
i & 0\\
-1 & 0\\
0 & 0
\end{array}
\right)\,.
\end{equation}

\item[ii)] $(\tilde y_R)_A$, which is a singlet under the ``elementary'' group (${\bf 1}_{-2/3}$)
and transforms in the fundamental representation of the ``composite'' group (${\bf 5}_{2/3}$).
Its physical value is given by
\begin{equation}
\langle \tilde y_R \rangle =
\left(
\begin{array}{c}
0\\
0\\
0\\
0\\
1
\end{array}
\right)\,.
\end{equation}
\end{itemize}
It is important to remark that in our definition the two spurions transform linearly under
the $\SO(5)$ ``composite'' group.

Using the spurions we can rewrite the elementary--composite mixings in a fully invariant form
\begin{eqnarray}
{\cal L}_{mix} &=& y_{L4}\, \overline q_L^\alpha \big(\tilde y_L^\dagger\big)^\alpha_A U_{Ai} \psi_4^i
+ y_{L1}\, \overline q_L^\alpha \big(\tilde y_L^\dagger\big)^\alpha_A U_{A5} \psi_1\nonumber\\
&& + y_{R4}\, \overline t_R \big(\tilde y_R^\dagger\big)_A U_{Ai} \psi_4^i
+ y_{R1}\, \overline t_R \big(\tilde y_R^\dagger\big)_A U_{A5} \psi_1
+ \textrm{h.c.}\,.\label{eq:Lagr_spurions}
\end{eqnarray}
Notice that the two mixings of the $q_L$ doublet are associated to the same spurion
$\tilde y_L$ and analogously the $t_R$ mixings correspond to the spurion $\tilde y_R$.
From the Lagrangian in eq.~(\ref{eq:Lagr_spurions}) we can recover the original
mixing terms in eq.~(\ref{eq:lagr_elem}) by replacing the spurions with their
physical values $\langle \tilde y_{L,R} \rangle$.

We can now identify the building blocks that can be used to construct the operators in
our effective theory. One key element is of course the Goldstone matrix $U$.
As shown in eq.~(\ref{eq:U_transformation}), under the $\SO(5)$ group
$U$ transforms linearly on one side and non-linearly on the other. We can thus split
the Goldstone matrix in two components: $U_{Ai}$ whose index $i$ transforms as a CCWZ
$4$-plet and $U_{A5}$ which is a singlet. In both cases the index $A$ corresponds to a
linear realization of the fundamental representation of $\SO(5)$.

It is also useful to introduce a slight generalization of the covariant derivative.
We define it in such a way that it acts on all the indices of a given object,
for instance the covariant derivative of the $4$-plet Goldstone component is
\begin{equation}
(D_\mu U)_{Ai} \equiv \partial_\mu U_{Ai} - i (A_\mu U)_{Ai} - i (U e_\mu)_{Ai}\,.
\end{equation}
For the elementary fermions and the composite resonances the convariant derivative
coincides with the one we used so far. It is useful to notice
that the covariant derivative of the Goldstone matrix can always be expressed in terms
of the $d_\mu$ symbol:
\begin{equation}
(D_\mu U)_{Ai} = - U_{A5} d_\mu^i
\qquad {\rm and} \qquad (D_\mu U)_{A5} = - U_{Ai} d_\mu^i\,.
\end{equation}
Moreover it is easy to check that the covariant derivative of the spurions vanishes
when it is computed on the spurion physical values, $\langle D_\mu y_{L,R} \rangle = 0$.

In our analysis, for simplicity, we will consider the limit in which the gauge
couplings are sent to zero. This limit is justified by the fact that the largest
corrections to the $Z\overline b_L b_L$ vertex come from loops containing the
Goldstones and not the transverse gauge field components. Within this approximation,
the elementary fermion interactions are necessarily mediated by the elementary--composite
mixings. This implies that, in classifying the operators that contribute to the
$Z\overline b_L b_L$ coupling, we can assume that the elementary fields are always
contracted with the $\tilde y_{L,R}$ spurions.

To construct the operators that can appear in the effective Lagrangian we
can use the following building blocks:
\footnote{Multiple covariant derivatives can be also used ({\it e.g.}
$D_\mu D_\nu \psi$) but they are not relevant for our analysis.}
\begin{center}
\begin{tabular}{rl}
elementary fields: & $q_L^\alpha$ and $t_R$\\
\rule{0pt}{1.25em}composite fields: & $\psi_{4}^i$ and $\psi_1$\\
\rule{0pt}{1.25em}cov. der. of the fermions: & $(D_\mu q_L)^\alpha$,
$D_\mu t_R$, $(D_\mu \psi_4)^i$ and $D_\mu \psi_1$\\
\rule{0pt}{1.25em}$d_\mu$ symbol: & $d_\mu^i$\\
\rule{0pt}{1.25em}mixings:  & $(U^\dagger \tilde y_L)^\alpha_{i,5}$ and $(U^\dagger \tilde y_R)_{i,5}$
\end{tabular}
\end{center}
Notice that, thanks to the unitarity of the Goldstone matrix, we can always write the
spurions in the combinations $U^\dagger \tilde y_{L,R}$.

\subsection*{Classification of the operators}

We can now analyze the operators that can modify the coupling of the $Z$ boson
to the $b_L$ with the aim of determining their degree of divergence.
This can be easily achieved by classifying the operators in an expansion in the
elementary--composite mixings.

To simplify the analysis it is more convenient to work in the basis of the
elementary and composite fields and not in the one of the mass eigenstates.
The mass eigenstate corresponding to the physical $b_L$,
which we will denote here by $\widetilde b_L$,
is given by a combination of the elementary
$b_L$ and of the composite state $B$ contained in the $4$-plet $\psi_4$:
\begin{eqnarray}
b_L &=& \frac{m_4}{\sqrt{m_4^2 + y_{L4}^2 f^2}} \widetilde b_L -
\frac{y_{L4} f}{\sqrt{m_4^2 + y_{L4}^2 f^2}} \widetilde B_L\,,\label{eq:b_decomp}\\
B_L &=& \frac{y_{L4} f}{\sqrt{m_4^2 + y_{L4}^2 f^2}} \widetilde b_L
+ \frac{m_4}{\sqrt{m_4^2 + y_{L4}^2 f^2}} \widetilde B_L\,,\label{eq:B_decomp}
\end{eqnarray}
where we denoted by $\widetilde B$ the heavy mass eigenstate.
The operators that induce a distortion of the $g_{b_L}$ coupling
are trivially related to the ones that give the couplings of the
$Z$ boson to the elementary $b_L$ and the composite $B_L$.

Notice that under the SM gauge group the $b_L$ and the  $B_L$ fields have exactly the same charges
as the physical $\widetilde b_L$, thus operators containing the covariant derivatives
$D_\mu b_L$ and $D_\mu B_L$ do not give any distortion of the couplings. They only
induce a rescaling of the canonical kinetic terms.

We start by analyzing the operators containing only $q_L$. As we said before,
the elementary $q_L$ must necessarily be contracted with the spurion $\tilde y_L$,
thus the relevant operators contain at least two spurion insertions.
The $q_L$ field appears in the combination
\begin{equation}
(U^\dagger \tilde y_L q_L)_{i,5}
\end{equation}
where $i$ and $5$ denote the uncontracted index of $U^\dagger$. The singlet
component (index $5$) does not contain the $b_L$ field, thus only the $4$-plet
part is relevant for our analysis. To get the $Z$ boson we must use the covariant
derivative or the $d_\mu^i$ symbol. The index structure, however, does not allow us
to construct an operator with $d_\mu^i$. The only possibility is
\begin{equation}
i\, \overline q_L \tilde y_L^\dagger \gamma^\mu \tilde y_L D_\mu q_L\,,
\end{equation}
which gives a renormalization of the usual $b_L$ kinetic term and does not
induce a correction to the $g_{b_L}$ coupling.
At order $y_L^4$ we get one operator that contributes to the distortion of
the $Z\overline b_L b_L$ vertex:
\begin{equation}
{\cal O} = i (\overline q_L y_L^\dagger \gamma^\mu y_L q_L)
\left(U^\dagger_{5A} (y_L)^\alpha_A (y_L^\dagger)^\alpha_B U_{Bi} d_\mu^i\right)
+ \textrm{h.c.}\,.
\end{equation}
In this case the $4$ insertions of the $\tilde y_L$ spurion ensure that
the corrections are finite at one loop.

We can now consider the operators containing only the composite $4$-plet $\psi_4$.
At least two spurion insertions are needed to generate an operator that breaks the $P_{LR}$ symmetry
and corrects the $Z\overline b_L b_L$ vertex. Notice that if more than two spurions
are present the operator corresponds to a finite one-loop contribution.
If we want to classify possible divergent corrections, we can focus on the case with only two $\tilde y_L$
insertions.

From the previous discussion it follows that the only way to contract the $\tilde y_L$
spurions is
\begin{equation}
U^\dagger_{*A} (y_L)^\alpha_A (y_L^\dagger)^\alpha_B U_{B*}\,,
\end{equation}
where each $*$ denotes a free index which can correspond to a $4$-plet or
a singlet of $\SO(4)$.
As we noticed before, operators containing $D_\mu \psi_4$ can only induce a
rescaling of the canonical kinetic term for the $B$. Thus in order to obtain a distortion
of the coupling with the $Z$ boson we need to include the $d_\mu^i$ symbol.
It is easy to show that the expression $d_\mu^i \psi_4^i$ does not
contain a term of the form $Z_\mu B$. This term can only be generated if the $d$-symbol index
is contracted with the Goldstone matrix $U$. We are left with only one possibility:
\begin{equation}\label{eq:op_Zbb_B}
{\cal O} = i (\overline \psi_4 \gamma^\mu \psi_4)
\left(U^\dagger_{5A} (y_L)^\alpha_A (y_L^\dagger)^\alpha_B U_{Bi} d_\mu^i\right) + {\rm h.c.}\,.
\end{equation}
With an explicit computation we find that this operator contains
a coupling of the $B$ with the $Z$ boson:
\begin{equation}
{\cal O} \supset
\left(\sqrt{2} \sin^2 \left(\frac{\langle h \rangle}{f}\right)\right) \frac{g}{c_w} Z_\mu
\overline B \gamma^\mu B\,.
\end{equation}

The operator in eq.~(\ref{eq:op_Zbb_B}) contains only two spurion insertions and
corresponds to a logarithmically divergent contribution at one loop. After the
rotation to the mass eigenstates a correction to the $Z\overline b_L b_L$ vertex
is induced. Using eq.~(\ref{eq:B_decomp}) we find that this correction arises
at order $y_L^4$, as expected.

Finally we can consider the mixed operators containing one elementary and one composite field.
The elementary $b_L$ must necessarily be contracted with a $\tilde y_L$ spurion.
It is straightforward to show that at least two other spurion insertions are needed
to construct an operator that can contribute to $\delta g_{b_L}$ and the associated
one-loop corrections are finite.

To conclude we summarize the results of this section. We found that the one-loop
corrections to the $Z\overline b_L b_L$ can be logarithmically divergent.
Moreover we showed that the divergence can only come from diagrams with two composite
$B$'s as external states. The contributions related to the elementary $b_L$
fields are instead always finite.

\section{Computation of the loop corrections to the $Z\overline b_L b_L$ vertex}

In this appendix we compute the one-loop corrections to the $Z \overline b_L b_L$
vertex. For simplicity we consider the limit in which the gauge couplings are
sent to zero. This approximation is justified by the fact that, as in the SM,
the most relevant contributions are related to the Yukawa interactions and not to the
gauge couplings.~\footnote{We
verified numerically in the model of Ref.~\cite{Anastasiou:2009rv} that the corrections
due to non-vanishing gauge couplings are small and can be safely neglected.}

The computation can be significantly simplified by using a consequence of the
operator analysis presented in appendix~\ref{app:Zbb_operator_analysis}.
We saw that, an operator can contribute to the distortion
of the $Z \overline b_L b_L$ interaction only if it contains the
CCWZ $d_\mu^i$ symbol. Moreover we found that the $4$-plet index of
$d_\mu$ must be necessarily contracted with the Goldstone matrix.
By an explicit computation one easily finds that the combination $U_{Ai} d_\mu^i$
contains the $Z$ boson always in association with the neutral Goldstone $\phi^0$:
\begin{equation}
U_{Ai} d_\mu^i \supset -\frac{1}{\sqrt{2}} \left(\frac{g}{c_w}
\sin\left(\frac{\langle h \rangle}{f}\right) Z_\mu
+ 2 \partial_\mu \phi^0\right)\,,
\end{equation}
where $\phi^0$ denotes the canonically normalized neutral
Goldstone, $\phi^0 = - (f/\langle h \rangle) \sin(\langle h \rangle/f) \Pi_3$ (see appendix~\ref{app:ccwz}).
It is also straightforward to check that the covariant derivatives $D_\mu b_L$
and $D_\mu \psi_4$ do not contain any term of the form $(\partial_\mu \phi^0) b_L$.
From these results it follows that we can extract the corrections to the $g_{b_L}$ coupling
by computing the one loop contributions to the $(\partial_\mu \phi^0) \overline b_L \gamma^\mu b_L$
interaction.~\footnote{Another proof of the correctness of this procedure
was given in Ref.~\cite{Barbieri:1992dq}, in which the
two loop corrections to the $Z\overline b_L b_L$ vertex in the SM are computed.}

Notice that, thanks to the $P_{LR}$ symmetry under which $\phi^0$ is odd, the vertex
$(\partial_\mu \phi^0) \overline b_L \gamma^\mu b_L$ is not present at tree level and
this makes the computation of the $(\partial_\mu \phi^0) \overline b_L \gamma^\mu b_L$
one-loop corrections even simpler. Due to the presence of a tree-level $Z \overline b_L b_L$
vertex, the one loop renormalization of the $b_L$ must be taken into account to compute $\delta g_{b_L}$
in the standard way. In the case of the $(\partial_\mu \phi^0) \overline b_L \gamma^\mu b_L$
interaction, instead, the wave function renormalization does not induce a one-loop
contribution, thus we only need to compute the vertex correction.

We parametrize the relevant Goldstone couplings in the following way:
\begin{eqnarray}
{\cal L} &=& \overline T_i (A_i\, \phi^+ + i\, B_i \partialslash\, \phi^+) b_L + {\rm h.c.}\nonumber\\
&& +\,\left(i\, C_{ij}\, \phi^0\, \overline T_i P_L T_j + {\rm h.c.}\right)
+ \partial_\mu \phi^0\, \overline T_i \gamma^\mu \left(D_{ij}^L P_L + D_{ij}^R P_R\right) T_j\nonumber\\
&& +\,\overline T_i\left(i E_i \phi^+ \phi^0 + F_i^+ \phi^0 \partialslash\, \phi^+
+ F_i^0 \phi^+ \partialslash\, \phi^0\right)b_L + {\rm h.c.}\,,
\end{eqnarray}
where we denoted by $T_i$ the charge $2/3$ states in the mass eigenbasis and $P_{L,R}$
are the left and right projectors. $\phi^+$ and $\phi^0$ are the canonically normalized
Goldstone fields, in particular the charged Goldstone is given by $\phi^+ = (f/\langle h \rangle) \sin(\langle h \rangle/f) h_u$
(see appendix~\ref{app:ccwz}). Notice that, in the effective theory we considered in this paper,
the $\phi^0$ Goldstone has no vertex that involves only charge $-1/3$ states.
As a consequence the diagrams that give a correction to the $Z \overline b_L b_L$ vertex
only contain charge $2/3$ fermions inside the loop.

As we discussed in the main text, corrections to the $g_{b_L}$ coupling can also be induced
by $4$-fermion effective interactions. We parametrized them by the Lagrangian:
\begin{equation}
{\cal L}^{4-ferm.} = G_{ij}^L [\overline b_L^a \gamma_\mu b_L^a][\overline T_i^b \gamma^\mu P_L T_j^b]
+ G_{ij}^R [\overline b_L^a \gamma_\mu b_L^a][\overline T_i^b \gamma^\mu P_R T_j^b]\,,
\end{equation}
where $a$ and $b$ are color indices. For simplicity we consider only the color structure given in the previous
formula. The results for different color structures only differ by an overall group theory
factor.

\begin{figure}
\centering
\includegraphics[height=.2\textwidth]{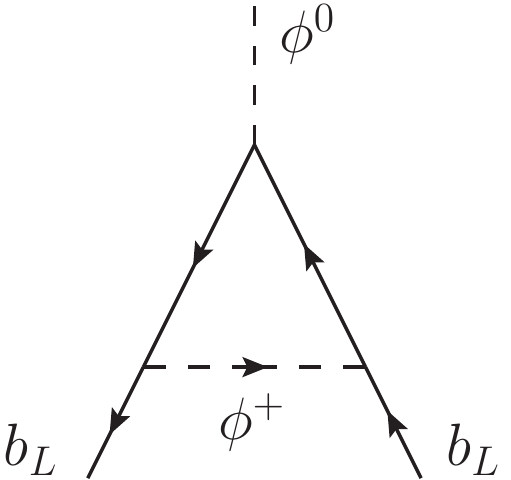}
\hspace{2.em}
\includegraphics[height=.2\textwidth]{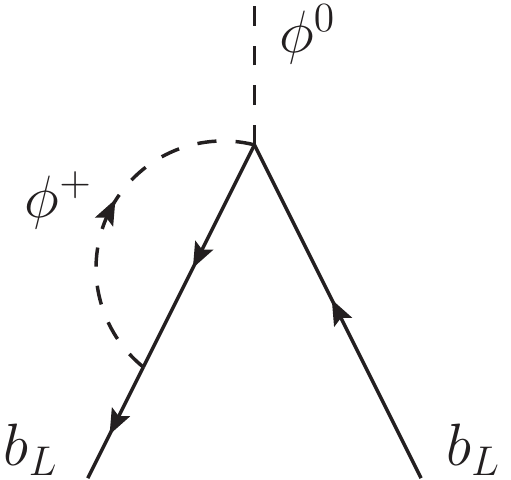}
\hspace{2.em}
\includegraphics[height=.2\textwidth]{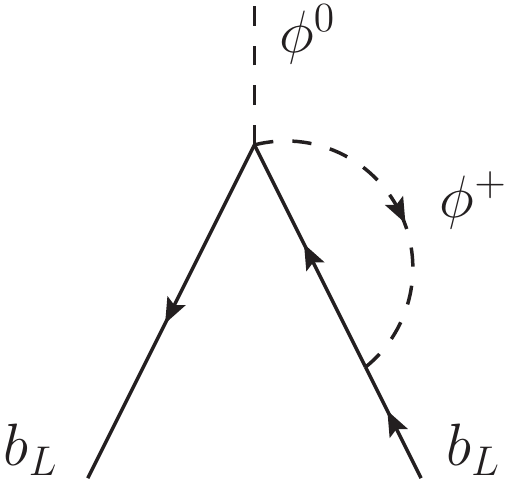}
\hspace{2.em}
\includegraphics[height=.2\textwidth]{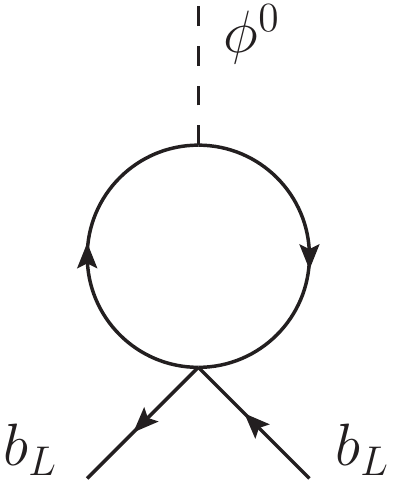}
\caption{Topologies of the diagrams contributing to the
$(\partial_\mu \phi^0) \overline b_L \gamma^\mu b_L$ interaction.
The internal fermion lines are fields with electric charge $2/3$.}
\label{fig:phi0bb}
\end{figure}

The topologies of the diagrams that contribute to the $(\partial_\mu \phi^0) \overline b_L \gamma^\mu b_L$
interaction are shown in fig.~\ref{fig:phi0bb}.
The ``triangle'' topology and the diagrams with a loop on the external legs arise from
the leading order terms in the composite Higgs effective Lagrangian. The $4$-fermion
interactions, instead, generate the diagrams with a ``bubble'' topology.
For our explicit computation we use dimensional regularization and we encode the divergent part
in the parameter $\Delta \equiv 1/\epsilon - \gamma + \log(4\pi)$, where $\epsilon$ is
defined by $d = 4 - 2 \epsilon$. We denote the renormalization scale by $\mu$.

The correction to the $Z \overline b_L b_L$ vertex coming from the ``triangle''
diagrams is given by
\begin{eqnarray}
\delta g_{b_L}^{\rm triangle} &=& \frac{f \sin(\langle h \rangle/f)}{64 \pi^2}
\sum_{i,j} \Bigg\{A_j A_i^* \left[D_{ij}^R I_1^{ij} + 2 D_{ij}^L m_i m_j I_2^{ij}
- C_{ij} m_j (I_2^{ij} - I_4^{ij}) - C^\dagger_{ij} m_i (I_2^{ij} + I_4^{ij})\right]\nonumber\\
&&+ B_j B_i^* \left[D_{ij}^R m_i m_j I_1^{ij} - 2 D_{ij}^L I_3^{ij}
+ \frac{1}{2} C_{ij} m_i (I_1^{ij} + I_5^{ij}) + \frac{1}{2} C^\dagger_{ij} m_j (I_1^{ij} - I_5^{ij})\right]\\
&&+\,{\rm Re}\left[A_j B_i^* \left(C^\dagger_{ij} (3 I_1^{ij} - I_5^{ij} + 1)
+ 2 C_{ij} m_i m_j I_4^{ij} + 2 D_{ij}^R m_i I_1^{ij} - 2 D_{ij}^L m_j (2 I_1^{ij} + 1)\right)\right]\Bigg\}\,,\nonumber
\end{eqnarray}
where we defined the $I_{1, \ldots, 5}$ functions as
\begin{eqnarray}
I_1^{ij} &=& \displaystyle \Delta + \frac{1}{2} - \frac{1}{m_i^2 - m_j^2}\left[m_i^2 \log\left(\frac{m_i^2}{\mu^2}\right)
- m_j^2 \log\left(\frac{m_j^2}{\mu^2}\right)\right]\,,\nonumber\\
I_2^{ij} &=& \displaystyle \frac{1}{m_i^2 - m_j^2} \log\left(\frac{m_i^2}{m_j^2}\right)\,,\nonumber\\
I_3^{ij} &=& \displaystyle (m_i^2 + m_j^2) (\Delta + 1) - \frac{1}{m_i^2 - m_j^2}\left[m_i^4 \log\left(\frac{m_i^2}{\mu^2}\right)
- m_j^4 \log\left(\frac{m_j^2}{\mu^2}\right)\right]\,,\\
I_4^{ij} &=& \displaystyle \frac{1}{m_i^2 - m_j^2} - \frac{m_i^2 + m_j^2}{2(m_i^2 - m_j^2)^2}\log\left(\frac{m_i^2}{m_j^2}\right)\,,\nonumber\\
I_5^{ij} &=& \displaystyle \frac{m_i^2 + m_j^2}{m_i^2 - m_j^2} - \frac{2 m_i^2 m_j^2}{(m_i^2 - m_j^2)^2} \log\left(\frac{m_i^2}{m_j^2}\right)\,.
\nonumber
\end{eqnarray}
The contribution from the diagrams with loops on the external legs is given by
\begin{eqnarray}
\delta g_{b_L}^{\rm legs} &=& \frac{f \sin(\langle h \rangle/f)}{128 \pi^2} \sum_i
{\rm Re}\Bigg[4 F_i^0 m_i \left(A_i^* + B_i^* m_i\right) I_6^i
- E_i \left(A_i^* (I_6^i + 1) - B_i^* m_i (I_6^i - 1)\right)\nonumber\\
&& -\, F_i^+ m_i \left(A_i^* (I_6^i - 1) + B_i^* m_i (3 I_6^i -1)\right)\Bigg]\,,
\end{eqnarray}
where $I_6$ is given by
\begin{equation}
I_6^i = 2 \Delta + 2 - 2 \log\left(\frac{m_i^2}{\mu^2}\right)\,.
\end{equation}
Notice that in the effective theory we considered in this paper
the two contributions $\delta g_{b_L}^{\rm triangle}$ and $\delta b_{b_L}^{\rm legs}$ are
always finite.

Finally the contribution induced by the $4$-fermion interactions is given by
\begin{eqnarray}
\delta g_{b_L}^{\rm bubble} &=& N_c \frac{f \sin(\langle h \rangle/f)}{32 \pi^2} \sum_{i,j}
\Bigg\{\left(D_{ij}^L G_{ji}^L + D_{ij}^R G_{ji}^R\right) \left(I_3^{ij} - (m_i^2 + m_j^2)/2\right)\nonumber\\
&& -\, \left(D_{ij}^R G_{ji}^L + D_{ij}^L G_{ji}^R\right) m_i m_j \left(2 I_1^{ij} + 1\right)
+ {\rm Re}\left[C_{ij} G_{ji}^L - C^\dagger_{ij} G^R_{ji}\right] m_i I_7^{ij}\Bigg\}\,,
\end{eqnarray}
where
\begin{equation}
I_7^{ij} = 2 \Delta + 3 - 2 \frac{m_i^2}{m_i^2 - m_j^2} - 2 \frac{1}{(m_i^2 - m_j^2)^2}
\left[(m_i^4 - 2 m_i^2 m_j^2) \log\left(\frac{m_i^2}{\mu^2}\right)
+ m_j^4 \log\left(\frac{m_j^2}{\mu^2}\right)\right]\,.
\end{equation}
Differently from the first two classes of diagrams, in our effective theory the ``bubble'' diagrams
can give a divergent contribution. This can happen if the $G_{ij}^R$ couplings are non-vanishing.
The $G_{ij}^L$ couplings, instead, give rise only to finite corrections.


\end{document}